%% file: ANA-TRIG-2019-01-PAPER.tex
\pdfoutput=1
\documentclass[PAPER,atlasdraft=false,texlive=2016,USenglish,cernpreprint,texmf]{atlasdoc}
 
\usepackage{multirow}
 
\usepackage{atlaspackage}
\usepackage{atlasbiblatex}
 
\usepackage{atlasphysics}

\graphicspath{{logos/}{figures/}}
 
\usepackage{ANA-TRIG-2019-01-PAPER-defs}

\AtlasTitle{Performance of the missing transverse momentum triggers for the ATLAS detector during Run-2 data taking}
 
\AtlasJournal{JHEP}
\AtlasJournalRef{JHEP 08 (2020) 80}
\AtlasDOI{https://doi.org/10.1007/JHEP08(2020)080}
\AtlasAbstract{
The factor of four increase in the LHC luminosity, from
$0.5\times 10^{34}\,\textrm{cm}^{-2}\textrm{s}^{-1}$ to $2.0\times 10^{34}\textrm{cm}^{-2}\textrm{s}^{-1}$,
and the corresponding increase in pile-up collisions during the 2015--2018 data-taking period,
presented a challenge for the ATLAS trigger,
particularly for those algorithms that select events with missing transverse momentum.
The output data rate at fixed threshold typically increases exponentially with the number of pile-up collisions,
so the legacy algorithms from previous LHC data-taking periods had to be tuned and
new approaches developed to maintain the high trigger efficiency achieved in earlier operations.
A study of the trigger performance and comparisons with simulations show that these changes resulted
in event selection efficiencies of >98\% for this period,
meeting and in some cases exceeding the performance of similar triggers in earlier run periods,
while at the same time keeping the necessary bandwidth within acceptable limits.
}
 
\author{The ATLAS Collaboration}
 
\AtlasRefCode{TRIG-2019-01}
\PreprintIdNumber{CERN-EP-2020-050}

\hypersetup{pdftitle={ATLAS document},pdfauthor={The ATLAS Collaboration}}
 
\begin{document}
 
\maketitle
 
\tableofcontents

\section{Introduction}

The trigger system~\cite{TRIG-2016-01} of the ATLAS experiment~\cite{PERF-2007-01}
is responsible for deciding which  proton--proton ($pp$) bunch-crossing events are kept for later analysis.
Storage and processing requirements limit the fraction of events that can be retained to the order of
$10^{-5}$, with the rest being discarded and hence unavailable for further physics analysis.
 
Particles that interact via neither the strong nor the electromagnetic force,
and that escape the experiment without decaying, leave no visible signature.
Efficient trigger selection of events that contain such invisible particles
is nevertheless essential for much of the ATLAS physics programme.
Examples include searches for decays of the Higgs boson into invisible final states~\cite{HinvComb,EXOT-2016-37},
searches for new charged Higgs bosons decaying into $\tau\nu$~\cite{HIGG-2016-11},
searches for dark matter based on, for example, events in which invisible particles recoil against a single energetic jet~\cite{EXOT-2016-27},
supersymmetry searches that involve a stable and invisible neutralino~\cite{SUSY-2016-30,EWKSusy}, top-quark scalar partner searches~\cite{SUSY-2016-33} and searches for final states with stable long-lived particles~\cite{LLP}.
Another recent example
is the Standard Model (SM) Higgs boson decay into
$b$-quarks~\cite{HIGG-2018-04}, a process first observed in events in which
the Higgs boson was produced in association with a $\Zboson$ boson which itself decayed into unobserved neutrinos.
 
Selecting events that contain invisible particles
is particularly difficult, precisely because such particles do not register in the detector.
The strategy employed is to deduce the presence of these invisible particles
from the apparent imbalance of the momentum calculated from the visible particles.
In practice the imbalance in the direction parallel to the
proton beams is not sensitive since the fraction of each proton's momentum that participates in the
collision is unknown, and much of the outgoing momentum in the beam direction is not observed.
Instead, the momentum imbalance in the plane perpendicular
to the proton beams is the quantity of most interest;
it is known as the missing transverse momentum, and its magnitude is conventionally denoted by $\met{}$.
 
The $\met{}$ triggers used by ATLAS are based on transverse momentum imbalance within the calorimeter only.
Muons are approximately invisible in the calorimeter~\cite{TCAL-2010-01},
and so are treated in these calculations much like neutrinos.
Neglecting muons results in a negligible cost in terms of additional trigger rate
since events containing muons with large transverse momentum are rare.
These calorimeter-only algorithms also have the advantage that
they can efficiently select events that contain high-\pt{} muons.
For example, the \met{} trigger is also used to select events in which the Higgs boson is
produced in association with a $\Zboson$ boson decaying into muons,
or events containing a $\Wboson$ boson decaying into $\mu\nu$~\cite{HIGG-2018-04}.
 
Given that the selection of events by the \met{} trigger is based on
energy deposited throughout the calorimeter,
there are particular reconstruction challenges.
ATLAS employs a trigger system that
uses a region-of-interest trigger strategy~\cite{TRIG-2016-01} where
the lowest-level trigger identifies potentially interesting objects
in each event and, for those events that satisfy the selection criteria,
it provides regions of interest to be further analysed by the higher-level trigger.
This technique of reconstructing objects only in particular regions of the detector is useful
for simplifying the computational task, but generally unsuited to $\met{}$ triggers
which must sum momenta over the full solid angle that is instrumented.
 
The most significant challenge to the $\met{}$ triggers during the
13\,\TeV{} Run-2 data-taking period (2015--2018)
was the factor of four increase in the number of proton--proton collisions occurring within
each bunch crossing.
The additional collisions, known as pile-up,
were a consequence of the corresponding increase in LHC luminosity from
$0.5\times 10^{34}\,\textrm{cm}^{-2}\textrm{s}^{-1}$ in 2015 to
$2.0\times 10^{34}\,\textrm{cm}^{-2}\textrm{s}^{-1}$ in 2017 and 2018.
The peak luminosity of $2.0\times 10^{34}\,\textrm{cm}^{-2}\textrm{s}^{-1}$ was achieved
with 2544 bunches of circulating protons, a mean number of $pp$ interactions per bunch crossing
$\langle\mu\rangle$ = 56, and a peak pile-up of 70 interactions.
The energy from the additional pile-up collisions
is deposited throughout the detector.
Due to the shaping time of the front-end electronics,
the calorimeter response is affected by pile-up from several preceding bunch
crossings~\cite{PERF-2014-07}.
The overall effect of both forms of pile-up is to degrade the \met resolution of the detector.
With the existing Run-1 algorithms, this rise in pile-up would have led to an
unacceptable order-of-magnitude increase in trigger rate
unless thresholds were raised, and that would in turn have significantly diminished the signal efficiencies.
 
This paper describes algorithms introduced during Run~2 that provide
greater pile-up resilience and background rejection while maintaining a signal acceptance similar to
that in Run~1.
These algorithms were able to keep output rates
within a tolerable 100\,Hz even at $\langle\mu\rangle = 56$.
The design of these algorithms is described in detail, and comparative studies
of their performance using data and simulation are provided.
 
The paper is organized as follows.
Section~\ref{sec:detector} describes the ATLAS detector.
The \met{} trigger algorithms are introduced in Section~\ref{sec:alg}.
The offline \met{} algorithm against which the trigger
is compared is defined in Section~\ref{sec:offlineobjects}.
The trigger performance studies and their results
are described in Section~\ref{sec:performance}.
The conclusions are presented in Section~\ref{sec:conclusion}.
 
\section{ATLAS detector}
\label{sec:detector}
 
\newcommand{\AtlasCoordFootnote}{
ATLAS uses a right-handed coordinate system with its origin at the nominal interaction point (IP)
in the centre of the detector and the \(z\)-axis along the beam pipe.
The \(x\)-axis points from the IP to the centre of the LHC ring,
and the \(y\)-axis points upwards.
Cylindrical coordinates \((r,\phi)\) are used in the transverse plane,
\(\phi\) being the azimuthal angle around the \(z\)-axis.
The pseudorapidity is defined in terms of the polar angle \(\theta\) as \(\eta = -\ln \tan(\theta/2)\).
Angular separation is measured in units of \(\Delta R \equiv \sqrt{(\Delta\eta)^{2} + (\Delta\phi)^{2}}\).}
 
The ATLAS detector~\cite{PERF-2007-01} at the LHC covers nearly the entire solid angle around the collision point.
It consists of an inner tracking detector surrounded by a thin superconducting solenoid, electromagnetic and hadronic calorimeters,
and a muon spectrometer incorporating three large superconducting toroidal magnet systems.
 
The inner-detector system is immersed in a \SI{2}{\tesla} axial magnetic field
and provides charged-particle tracking in the range \(|\eta| < 2.5\).\footnote{\AtlasCoordFootnote}
The high-granularity silicon pixel detector covers the collision vertex region~\cite{PIX-2018-001}.
It is followed by the silicon microstrip tracker.
These silicon detectors are complemented by the transition radiation tracker.

The calorimeter system has approximately 188,000 cells and
covers the pseudorapidity range \(|\eta| < 4.9\).
Within the region \(|\eta|< 3.2\), electromagnetic calorimetry is provided by barrel and
endcap high-granularity lead/liquid-argon (LAr) sampling calorimeters (\text{ECAL}),
with an additional thin LAr presampler covering \(|\eta| < 1.8\)
to correct for energy loss in material upstream of the calorimeters.
The ECAL is between 24 and 27 radiation lengths ($X_0$) deep,
and its granularity in the barrel in terms of $\Delta\eta\times\Delta\phi$
is typically $0.025\times\pi/128$, with variations in segmentation with layer and $|\eta|$
as described in Ref.~\cite{PERF-2014-07}.
 
Hadronic calorimetry is provided by the steel/scintillator-tile calorimeter (\text{HCAL}),
segmented into three barrel structures within \(|\eta| < 1.7\),
and two copper/LAr hadronic endcap calorimeters.
The solid angle coverage is completed with forward copper/LAr and tungsten/LAr calorimeter modules (\text{FCAL})
optimized for electromagnetic  (\text{FCAL1}) and hadronic  (\text{FCAL2} and \text{FCAL3}) measurements respectively.
The combined depth of the calorimeters for hadronic energy measurements
is more than 10 nuclear interaction lengths
nearly everywhere across the full detector acceptance ($|\eta| < 4.9$).
The granularity is as fine as $0.1\times\pi/32$, again with variations
in segmentation with layer and $|\eta|$ as described in Ref.~\cite{PERF-2014-07}.

The muon spectrometer comprises separate trigger and
high-precision tracking chambers measuring the deflection of muons in a magnetic field generated by the superconducting air-core toroids.
The field integral of the toroids ranges between \num{2.0} and \SI{6.0}{\tesla\metre}
across most of the detector.
A set of precision chambers covers the region \(|\eta| < 2.7\) with three layers of monitored drift tubes,
complemented by cathode strip chambers in the forward region, where the background is highest.
The muon trigger system covers the range \(|\eta| < 2.4\) with resistive plate chambers in the barrel and thin gap chambers in the endcap regions.

A two-level trigger system is used to select interesting events~\cite{TRIG-2016-01}.
It consists of a hardware-based first-level trigger
(Level-1, \lonealg) and a software-based high-level trigger (HLT)
running on a farm of approximately 50\,k processing units. The L1 trigger decision is
formed by the Central Trigger Processor, which receives inputs from the L1
calorimeter (L1Calo) ~\cite{L1CaloPaper} and L1 muon triggers as well as several other
subsystems. The L1 trigger decision is formed with a latency of \SI{2.2}{\micro\second}.
The HLT has access to the full event and a decision is made within an average time of \SI{500}{\milli\second}.
 
\section{Description of the \met{} trigger algorithms}
\label{sec:alg}
 
The operational demands of the trigger prioritize low latency, rapid processing, and large background rejection while making use of limited detector information. Thus the online \met trigger algorithms are specifically designed for this purpose and so differ from the offline \met reconstruction algorithms used in subsequent physics analyses~\cite{PERF-2016-07,ATLAS-CONF-2018-023}.
 
The ATLAS HLT processes approximately 100\,kHz of \lonealg{} accepted events,
of which about 5 to 10\,kHz come from the \lonealg{}~\met{} trigger.
The HLT~\met{} algorithms accept events at a rate of about \SI{1200}{\Hz} averaged over a typical LHC fill~\cite{ATL-DAQ-PUB-2019-001}.
The requirement that the \met{} algorithms utilize
not more than $\mathcal{O}$(100\,ms) makes the use of inner-detector tracking information generally too computationally expensive,
since the corresponding evaluation time can take $\mathcal{O}$(1--5\,s).
Thus, all of the algorithms described below use only the calorimeter.\footnote{Here and in what follows it should be understood that the singular `calorimeter' refers to the calorimeter system as a whole.}
 
For all algorithms the energy measured by the calorimeter is associated with some
set of energy depositions, generally referred to as elements.
The definition of the set of elements is algorithm-dependent. For example,
the set of elements could be all of the calorimeter cells or the reconstructed jets.
In each case, the individual elements
characterize the local energy deposits, while the complete set
captures the overall distribution of energy in the calorimeter.
Elements are indexed by the label $i$;
the energy $E_i$ deposited in each element is also associated with a polar angle $\theta_i$
(or equivalently a pseudorapidity $\eta_i$) and an azimuthal angle $\phi_i$.
 
The components of the missing transverse momentum two-vector \METvec{}
are calculated from the energy in the elements
in the approximation of massless particles
\begin{equation}\label{eq:calculate_met_from_elements}
\begin{split}
\MEX  = -\sum\limits_{i=1}^{|\mathrm{Elements}|}  E_i \sin{\theta_i} \cos{\phi_i} \;, \\
\MEY  = -\sum\limits_{i=1}^{|\mathrm{Elements}|}  E_i \sin{\theta_i} \sin{\phi_i} \;,
\end{split}
\end{equation}
where $|\mathrm{Elements}|$ indicates the number of elements.
The magnitude $\met{}=\sqrt{(\MEX)^2 + (\MEY)^2}$ of this two-vector
is used in the selection of candidate events for further study.
The quantity $E_{{\textrm T}_i} = E_i \sin \theta_i$
is conventionally known as the transverse energy, and is useful in characterizing events.
The total transverse energy in the calorimeter is given by the scalar sum
\[
\SumEt{} =\sum\limits_{i=1}^{|\mathrm{Elements}|}  E_i \sin{\theta_i}.
\]

The algorithms used are presented in the following sections. They differ in how they select the elements
which enter into the sums and in how they make corrections to the energies of the elements.
 
\subsection{Level-1 trigger }
\label{sec:L1Alg}
The ATLAS L1 trigger is implemented in firmware running on custom-made electronics~\cite{L1CaloPaper}.
Analogue sums of the input signals from calorimeter cells forming projective towers are digitized, with the granularity in the projective coordinates $\eta$ and $\phi$ being
approximately $\Delta\eta\:\times\:\Delta\phi$ = $0.1\:\times\: 0.1$ for the detector region
$|\eta|<2.5$ and both larger and less regular for $|\eta|>2.5$,
as described in Ref.~\cite{L1CaloPaper}.
The digitization results in counts that nominally correspond to 1\,\GeV{} in \ET{}.
A fixed threshold that depends on $\eta$ is then applied per tower:
the energy $E_{i}$ of any tower which is below this
threshold is set to zero in the subsequent calculations.
The threshold is adjusted to provide a fixed occupancy
of 0.5--1\% based on data unbiased by a trigger selection.
This occupancy threshold is optimized to give an acceptable rate
for a trigger that efficiently selects events with \met $>$ 150\,\GeV. As the LHC luminosity increased,
the occupancy tended to grow, leading to higher thresholds as described in Ref. \cite{nMCM}.
 
The calorimeter noise thresholds vary from 1 to 9\,\GeV{} depending on the pseudorapidity and whether
the calorimeter layer is electromagnetic or hadronic.
The noise thresholds were periodically reoptimized during the period under study,
particularly when the collider parameters, and as a result the pile-up, were varied.
In the performance studies that follow,
particular attention is paid to three periods during 2017
that have different pile-up distributions;
these periods are labelled
with the symbols $\alpha$, $\beta$ and $\gamma$.
The pile-up at the start of the LHC fill increased from around $\langle\mu\rangle = 40$ for period $\alpha$ to  $\langle\mu\rangle = 60$ for period $\gamma$.
The largest changes in threshold occurred for the towers with ~4.0~$< |\eta| < $~4.9 in the electromagnetic layer, and the thresholds were 6, 7 and 9\,\GeV{} for periods $\alpha$, $\beta$ and $\gamma$ respectively.
After the threshold is applied, the towers are summed into larger projective towers
which have an approximate granularity of $\Delta\eta\:\times\:\Delta\phi$ = $0.2\:\times\: 0.2$
and are referred to as jet elements.
The $\METvec$ is then computed by summing the $x$ and $y$ projections of the jet elements
using Eq.~(\ref{eq:calculate_met_from_elements}).
 
Events that are accepted by the L1 trigger are transferred to the HLT
where the \met{} is recalculated using one or more of the algorithms described in
Sections~\ref{sec:CellAlg} to \ref{sec:PufitAlg}.

\subsection{Trigger using calorimeter cell signals (\cellalg)}
\label{sec:CellAlg}
The most basic HLT algorithm, \cellalg, determines $\METvec{}$ from a sum over the full set of
188\,k calorimeter cells to determine $\EX$ and $\EY$,
without adjusting for hadronic vs electromagnetic
calibrations or for pile-up corrections.
To reduce the effect of noise from electronics and pile-up,
only cells satisfying  $|E_i|>2\sigma_i$ are included in this sum.
Here $\sigma_i$ is the expected energy-equivalent noise in cell $i$ described in Ref.~\cite{PERF-2014-07}.
Its value is based on expectations for electronic noise and pile-up prior to data taking.
Negative energy cells are included because the LAr electronics are designed so that
signals from pile-up in later bunch crossings appear as negative energy and so tend to cancel energy deposits from earlier pile-up signals~\cite{LARG-2009-01}.
For the 2015 and 2016 run periods, the noise thresholds were configured for an average
number of interactions $\langle\mu\rangle=30$.
For 2017 and 2018 they were configured for $\langle\mu\rangle=40$.
In addition, the requirement $E_i>-5\sigma_i$ is
used to protect against spurious large negative cell signals.
 
\subsection{Trigger using topological clusters of calorimeter cells (\tcalg)}
\label{sec:TCAlg}
The topological clustering~\cite{PERF-2014-07} of calorimeter cells
forms an early stage of many ATLAS reconstruction algorithms.
It offers the possibility of identifying clusters
as either electromagnetic or hadronic in origin,
and thus allows appropriate calibration
(`local cell weighting')
before using them as inputs for jet reconstruction and calculation of $\met{}$.
 
Topological clusters are formed in a multistage process.
First the algorithm identifies calorimeter seed cells
each with $|E_i|>4\sigma_i$.
All cells neighbouring a seed cell are collected in all three spatial dimensions
and added to the cluster.
If any of those neighbouring cells satisfy $|E_i|>2\sigma_i$,
then their neighbours are collected as well, and the process continues iteratively until no further neighbours satisfying the requirement can be identified.
Finally, all neighbouring cells are added to the cluster, regardless of their energy.
After this initial cluster formation, an algorithm is run which splits clusters between local signal maxima (again, in three dimensions).
The energies of these clusters are corrected for the type of energy deposit after each one has been classified as being either electromagnetic or hadronic in origin.
 
These energy-calibrated clusters can be used directly in an \met{} calculation,
which is denoted \tcalg{}. These topological clusters also form the inputs to all of the following algorithms.
 
\subsection{Trigger based on jets (\mhtalg)}
In most events of interest, hadronic jets
tend to dominate the visible momentum.
Since these jets can be calibrated accurately~\cite{PERF-2014-04},
there is good motivation to use them as the basis of an \met{} calculation.
In addition, the calculation of the \met from the calorimeter signals described
previously includes energy from pile-up,
while jets are corrected on-average for pile-up effects. Using only calibrated jets for \met reconstruction yields
a representation that is referred to as \mhtalg{}.
 
The \mhtalg{} algorithm calculates $\met{}$  from the negative transverse
momentum vector sum of all jets above a threshold of 7\,\GeV{} before calibration.
The HLT jets are reconstructed from calibrated topological clusters
(defined in Section~\ref{sec:TCAlg}) using the \antikt{} jet algorithm
with a radius parameter of $R = 0.4$~\cite{Cacciari:2008gp} implemented in the {\tt FastJet} toolkit~\cite{Fastjet}.
 
These jets are calibrated in a procedure similar to that used for offline physics analysis~\cite{PERF-2014-04}.
First, the estimated pile-up contribution to jets is removed using the jet-area-based pile-up suppression method \cite{CacciariJetArea,PERF-2014-03}.
After pile-up subtraction, jets are calibrated using the simulation-based calibration described in Ref.~\cite{PERF-2016-04}.
The energy deposits that arise from photons, electrons or hadronically decaying $\tau$-leptons,
are included in the jet reconstruction.
 
\subsection{Trigger implementing local pile-up suppression (\pufitalg)}
\label{sec:PufitAlg}
The \pufitalg algorithm corrects for pile-up effects on high-\Et{} calorimeter
signals contributing to \met. It employs a pile-up estimate obtained from a fit to lower-\Et{} signals.
It takes as inputs the topological clusters defined in Section~\ref{sec:TCAlg} and
combines them into $\eta$--$\phi$ patches that correspond approximately to the size of a jet with $R = 0.4$.
A fit is then performed which estimates the energy contribution to each patch from pile-up,
based on the energy deposited and its spatial fluctuations across the calorimeter.
Finally, the pile-up-subtracted patches are used to determine the \met{}.
 
The strategy is based on the assumption that high-\Et{} energy deposits
are associated with a hard-scatter collision of interest whereas the low-\Et{} deposits
are the result of pile-up.
The \pufitalg algorithm proceeds by performing a fit that constrains to zero
(within fluctuations) the summed transverse momentum components $E_x$ and $E_y$
from the pile-up energy deposits. The \MET\ vector is then determined by
summing the $E_x$ and $E_y$ of the high-\Et{} deposits after subtracting
the estimated pile-up contributions.
 
The \pufitalg{} algorithm uses the measured structure of the
energy deposition in each event. This contrasts with other approaches
such as that of Ref.~\cite{Cacciari:2007fd}
which estimate pile-up contributions by defining a median
transverse energy density $\langle\rho\rangle$ that is then used in
subtracting pile-up from high-\Et{} deposits.
The \pufitalg{} algorithm is observed to outperform the standard pile-up-density algorithms in the context of the HLT,
so these other algorithms are not described further. The full definition of the algorithm, and the event-by-event fit performed,
can be found in Appendix~\ref{sec:PufitAlgDetail}.
 
 
\section{Offline object and \met{} reconstruction}
\label{sec:offlineobjects}
 
When defining selections of events for which performance characteristics are desired,
standard ATLAS offline algorithms are used to reconstruct and identify
electrons, muons, $\tau$-leptons, jets and $b$-tagged jets, as described in Appendix~\ref{sec:selection_detail}.
 
The offline $\met{}$ is also computed using these reconstructed objects since they tend to have better resolution than
individual tracks or clusters in the calorimeter.
First, the contributions from high-\pt{} electrons, photons, $\tau$-leptons and jets are summed, following the procedure described in Ref.~\cite{PERF-2016-07}.
To account for the activity from the underlying event, tracks not associated with one of the above objects are
also included in the $\met{}$ calculation. The \met definition described above is referred to as `tight' in the following.
 
In some cases, the so-called `tenacious' offline \met definition is used
in order to make the jet selections less sensitive to pile-up.
With this algorithm, jets that have $|\eta| > $ 2.4 and \pt $<$ 35\,\GeV{} are vetoed,
along with jets with \pt $<$ 120\,\GeV{} that fail the forward jet vertex tagger (JVT) requirement
that utilizes jet correlations to reject pile-up jets in a
region without a tracking detector~\cite{PERF-2016-06}.
The working point used corresponds to an efficiency of 92\% for hard-scatter jets.
Jets with $|\eta| < $ 2.4 and \pt{} within 20--40\,\GeV{}
are used only if they satisfy a JVT requirement that yields an 85\% efficiency for hard-scatter jets.
Jets with \pt{} within 40--60\,\GeV{} and 60--120 \GeV{} are used only if they satisfy a similar requirement with an efficiency of
92\% and 97\% respectively.
 
For all purposes considered in this paper,
the offline \met{} is computed without any contribution
to the visible momentum from any muon(s).
This method of computing \met{} facilitates comparison with the \met{} trigger algorithms
which use calorimeter information only.

 
\section{\met trigger performance}
\label{sec:performance}
The figures of merit used to characterize the performance of the \met trigger algorithm include:
CPU time, trigger rate,
efficiencies with respect to well-defined references, stability of the efficiencies
for several different kinds of events, and the instantaneous luminosity dependence of these characteristics.
Depending on the characteristics under study, the L1 and HLT algorithm performances, both individually and when used consecutively, are of interest.
Good performance is characterized by a trigger which has stable high efficiency
for signal events of interest, and, at the same time, a stable low output rate.
 
The trigger efficiency is defined by:
\begin{equation}\nonumber
\varepsilon (\mathcal{S}_i) = \frac{N(\mathrm{trigger}|\mathcal{S}_i)}{N(\mathcal{S}_i)} ,
\end{equation}
where $N(\mathcal{S}_i)$ is the size of the sample of events satisfying some
selection $\mathcal{S}_i$ which is typically designed to isolate events
within a narrow range of \met{}.
To assess the efficiency of the trigger, $\mathcal{S}_i$
is relaxed to capture events that satisfy some lower \met{} threshold.
In either case, the numerator $N(\mathrm{trigger}|\mathcal{S}_i)$ is the size of the
subset of events that also satisfies the $\met$ trigger requirement.
 
\subsection{Background model based on detector resolution}
\label{sec:bgmodel}
 
The \met trigger rate behaviour in the absence of pile-up corrections can be studied
with the \cell algorithm.
This algorithm does not attempt to correct for the effects of pileup,
other than via adjustments to the cell noise thresholds.
A model has been constructed that captures the dependencies
of the unbiased event acceptance (and hence trigger rate) of the \cell trigger algorithm on pile-up.
The model is sufficient for the purpose of understanding the
behaviour of the rate and demonstrates the need for more-sophisticated algorithms
to deal with the large increase in pile-up through the period under study.
 
The \cell{} distribution is modelled with two components. The first is due to
calorimeter energy resolution effects. This resolution is assumed to depend on
the instantaneous number $\mu$ of $pp$ interactions per bunch crossing only through their combined contribution to the total calorimeter transverse energy \SumEt,
upon which the resolution in turn depends.
The second component is the high \met\ tail of the distribution,
which is assumed to arise from events with rarer measurement fluctuations and events
containing non-interacting particles (such as semileptonic decays of $b$- or $c$-hadrons).
The probability of the second class of fluctuations is assumed to scale linearly with instantaneous luminosity.
The two components are combined to form the vector sum of the two \met values, with the azimuthal angle difference between the two components
randomly oriented with respect to each other.
By modelling the dependencies in this way, and by measuring the parameters of the model at low luminosity
(and hence low pile-up),
predictions of the \met\ distribution and trigger rates can be obtained for higher $\mu$.
The detailed description of the model may be found in Appendix~\ref{sec:bgmodeldetail}.
 
\begin{table}[]
\centering\renewcommand\arraystretch{1.1}
\begin{tabular}{llll} \hline \hline
Trigger & Random triggers & $L1 \met > 30\,\GeV{}$ & $L1 \met>50\,\GeV{}$ \\
Prescale & $\mathcal{O}(10^{6})$ & $\mathcal{O}(10^{5})$ & $\mathcal{O}(10^{3})$\\
\hline\hline
\end{tabular}
\caption{\label{tab:prescaletriggers}Triggers used for comparing the \cell{} acceptance model with data.}
\end{table}
 
To compare the calculation with measurements, data are selected by combining events obtained with several triggers, shown in Table~\ref{tab:prescaletriggers}.
For low \met values, events obtained with an unbiased random trigger (zero bias) are used. The background events, which dominate the rate, were selected using a zero bias trigger,
weighted to the instantaneous luminosity per bunch by requiring that an electron trigger fired in the previous LHC
orbit of this bunch.
Such triggers are prescaled, meaning that only one in $N$ events is accepted for some number~$N$. Since the prescale factor $N$ for random triggers is high, there are not enough recorded events at high \met{} for the study of the trigger background. These events are therefore supplemented with samples collected by a suite of triggers which require \lonealg{} $\met{}$ to be greater than a set of thresholds in the range 30\,\GeV{} to 50\,\GeV{},
as shown in Table~\ref{tab:prescaletriggers}.
The efficiencies for selecting events with higher L1 $\met$ thresholds (and lowest prescales) are found successively from those selected at lower thresholds (and correspondingly higher prescale), until those with the lowest $\met$ have their efficiencies determined using the random trigger.
 
\begin{figure}[htbp]
\centering
\subfloat[]{\includegraphics[width=0.49\textwidth]{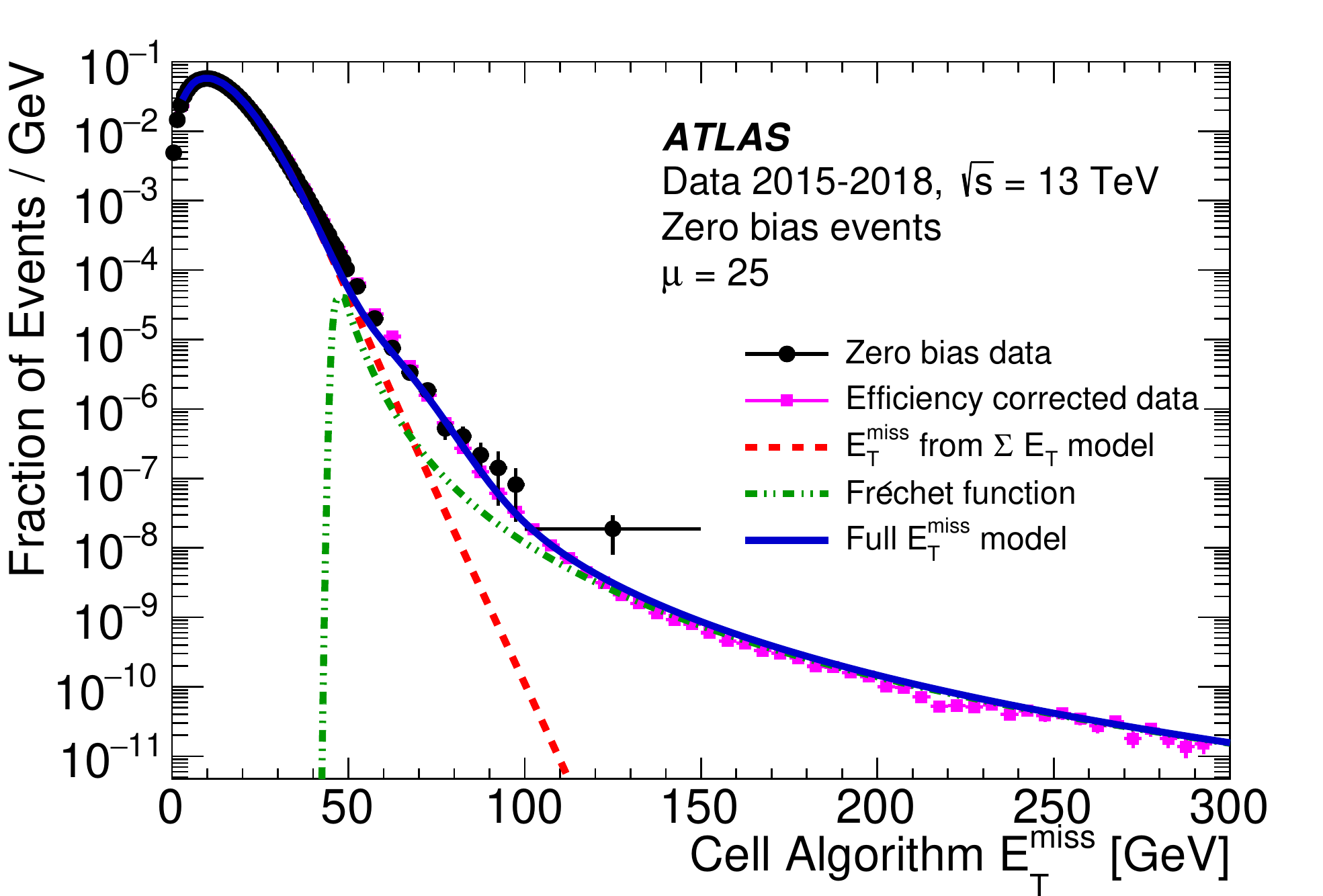}\label{fig:FitModel-a}}
\subfloat[]{\includegraphics[width=0.49\textwidth]{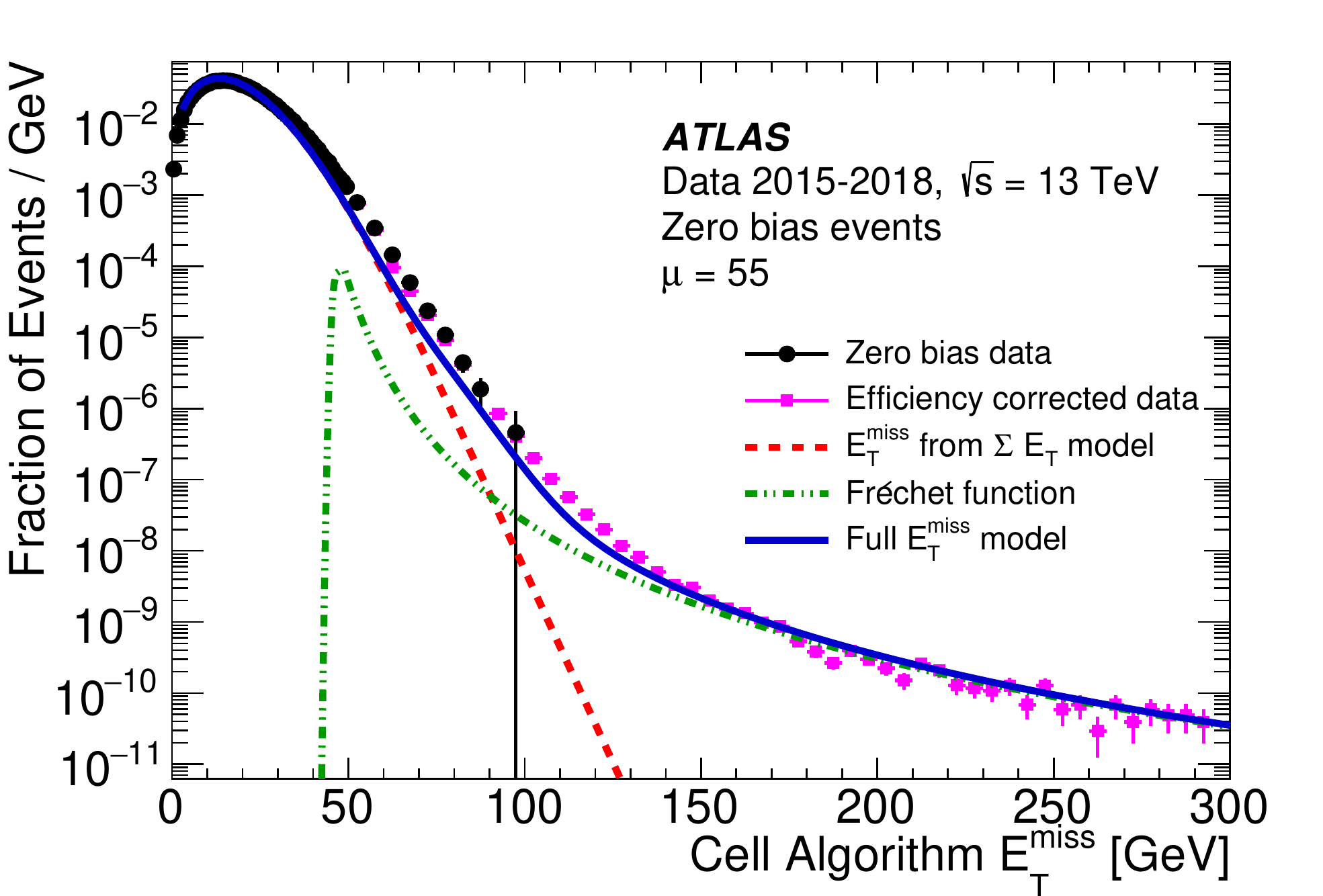}\label{fig:FitModel-b}}
 
\caption{
A comparison of the measured \cell\ distribution with
that predicted by the two-component model for two pile-up scenarios compared with data.
The circular points show the data collected using zero bias triggers, but have insufficient
luminosity to probe the higher \met portion of the distribution.
The square points extend the measured distribution using L1 $\met > 30\,\GeV$  and L1 $\met > 50\,\GeV$ data.
The uncertainties for the data points are statistical only, and much larger for the zero bias data due to the limited luminosity.
The dashed (red) curve is the prediction from the calorimeter-resolution part of the model.
The dash-dotted (green) curve is the high \met tail's probability distribution for the mean number of $pp$ interactions $\mu$ in each figure.
The solid (blue) curve is the full model prediction computed by combining the \met{} from these two individual sources shown in red and green, each calculated for $\mu=\langle\mu\rangle$. The black points show the unbiased \met\ distribution measured in data. \protect\subref{fig:FitModel-a} corresponds to a prediction for $\langle\mu\rangle = 25$ while \protect\subref{fig:FitModel-b} corresponds to $\langle\mu\rangle = 55$.
}
\label{fig:FitModel}
\end{figure}
 
Figure~\ref{fig:FitModel} compares the two-component model and its individual components with the full \met\
distribution measured in data.
When comparing data with the model, it is assumed that the instantaneous mean number $\mu$ of interactions per bunch crossing in the model is equal to its time-average $\langle\mu\rangle$ as measured over short periods in data. The data are also expected to have sensitivity to details that are not modelled, such as changes of calorimeter settings and
the LHC bunch structure. The lower-luminosity data from earlier years of Run~2 were recorded under conditions different from those for the higher-luminosity data recorded in later years, giving differences of up to an order of magnitude in rates depending on threshold and luminosity.
As is described in Appendix~\ref{sec:bgmodeldetail}, the model parameters were extracted from the full data set, and therefore are averaged over these effects. Nonetheless, as can be seen in Figure~\ref{fig:FitModel}, the model reproduces the key features of the data over the approximately nine orders of magnitude range of each distribution.
Comparisons performed for values of average pile-up in the range $15 \lesssim \langle\mu\rangle \lesssim 60$
show that the model accounts for all qualitative features of the data in this range.
Beyond these values it is found to somewhat underestimate (overestimate) the \met{} tail for higher (lower) values of $\langle\mu\rangle$.
 
Three regions can be seen in Figure~\ref{fig:FitModel}. For low \met{}, the resolution term dominates, and the rate grows exponentially with increasing $\mu$.  At high \met{}, the tail term dominates, and the rate is linear in $\mu$. Both of these terms contribute at intermediate \met{} values.  In this region there is a transition from exponential to linear behaviour with increasing \met{} threshold.  As $\mu$ increases, this transition region moves to higher values of $\met{}$.  For a fixed \met{} threshold trigger, the rate dependence on $\mu$ varies from linear to exponential with increasing $\mu$. The value of $\mu$ at which this transition occurs will vary according to the \met{} threshold applied.
 
\begin{figure}[htbp]
\centering
\includegraphics[width=0.7\textwidth]{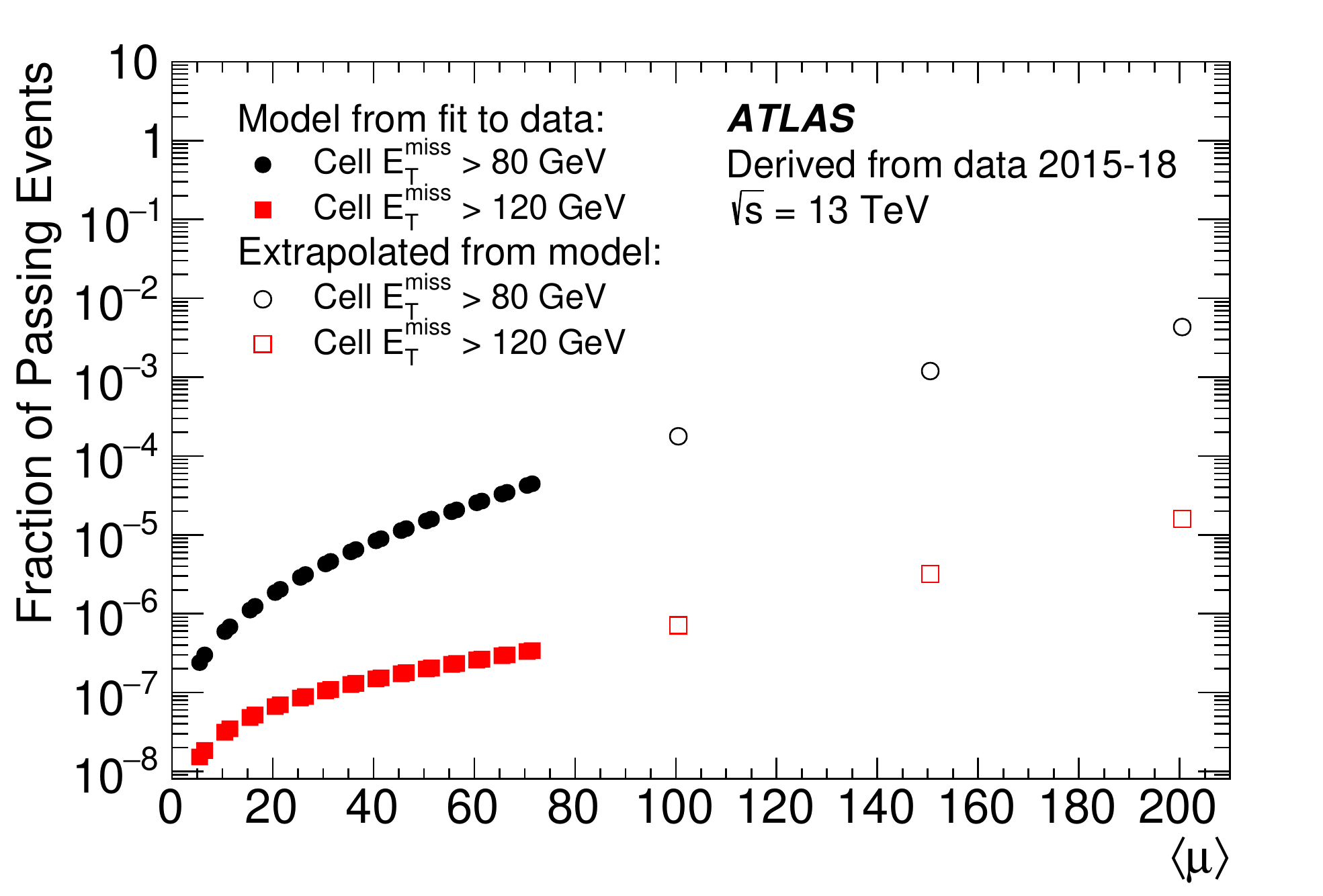}
\caption{
The \met\ model predicted trigger rate as a function of $\mu$ for the \cell\ algorithm with a threshold of $80\,\GeV$
and $120\,\GeV$, assuming no additional pile-up mitigation.
}
\label{fig:FitPrediction}
\end{figure}
 
Figure~\ref{fig:FitPrediction} shows the prediction for the \cell{} algorithm pass-fraction at fixed threshold as a function of $\mu$.
LHC Run-2 luminosities produced instantaneous $\mu$ as high as about 70, although the figure also shows extrapolated predictions up to $\mu=200$. If the \cell{} algorithm had been the primary \met{} trigger during Run~2, the threshold would have been
raised considerably to keep the trigger rate within affordable limits. This increase in threshold
would have significantly decreased the efficiency for signal events.
Algorithms which better correct for pile-up were therefore introduced for Run 2 and used either in conjunction with or in place of the \cell\ algorithm.
 
\subsection{Level-1 trigger performance}
\label{sec:L1performance}
 
The efficiency of the  \lonealg{} \met{} trigger is determined
using a $\Zboson\rightarrow \mu\mu$  events.
The muons have little interaction with the calorimeter, so
the transverse momentum \ptmumu{} of the dimuon system provides a good estimate of the \met{}
expected in the trigger calculations.
 
\begin{figure}[htbp]
\centering
\captionsetup{width=.45\linewidth}
\subfloat[]{\includegraphics[width=0.49\textwidth]{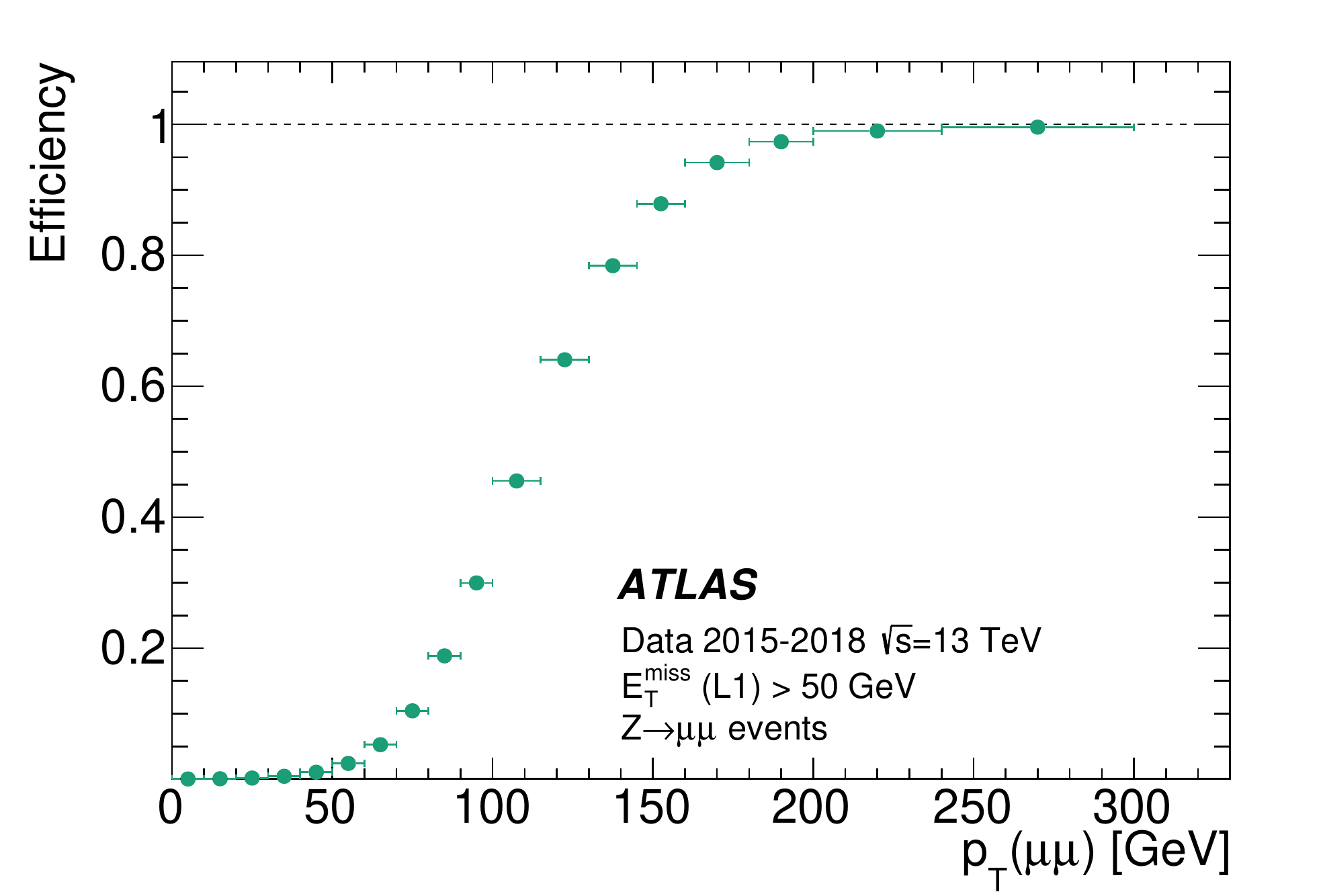}\label{fig:L1Summary:efficiency}}
\subfloat[]{
\includegraphics[width=0.49\textwidth]{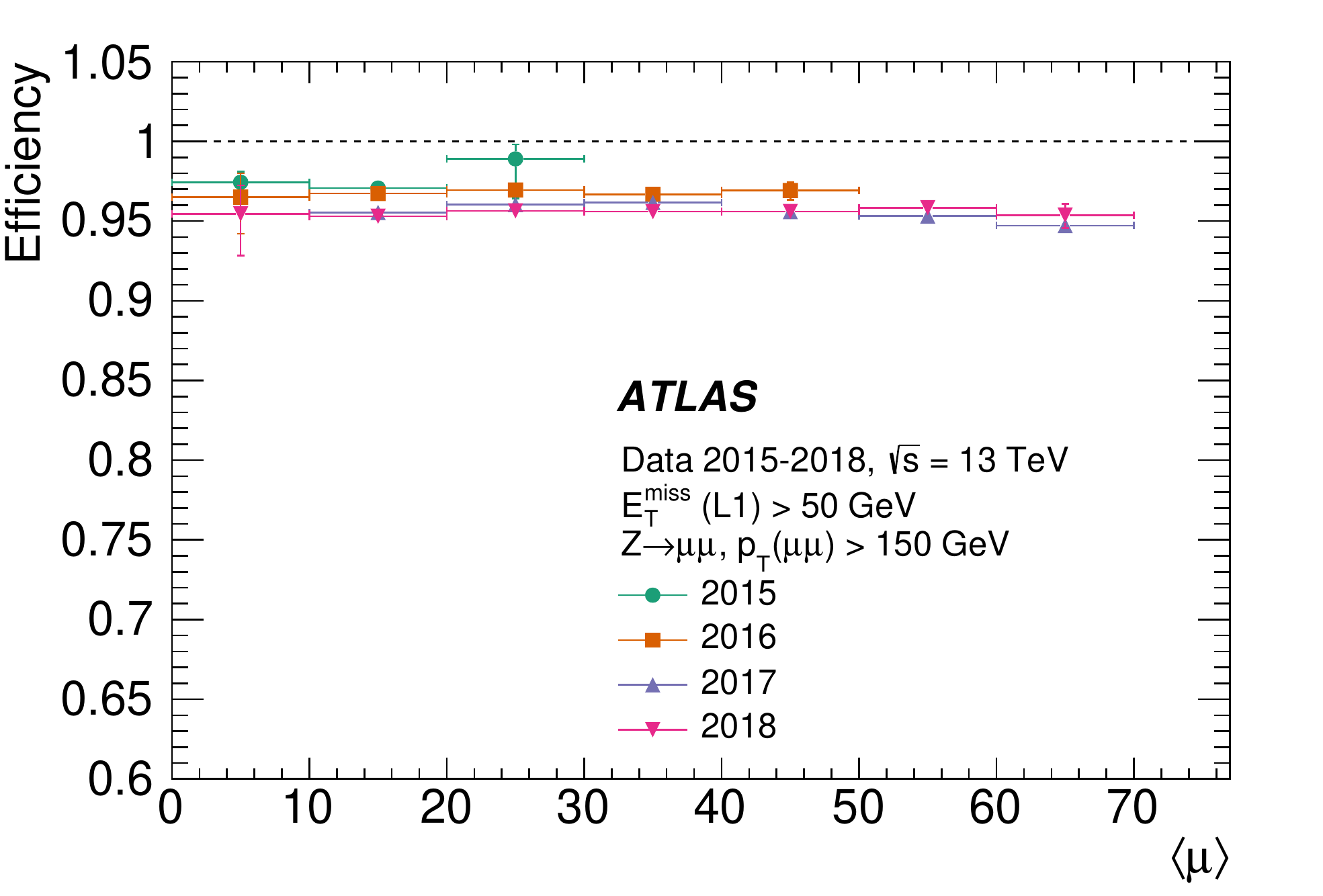}\label{fig:L1MuDep:overall}}
\captionsetup{width=\linewidth}
\caption{\protect\subref{fig:L1Summary:efficiency} The \lonealg{} \met{} trigger efficiency, shown as a function of \ptmumu{} in $\Zboson\rightarrow\mu\mu$ events. \protect\subref{fig:L1MuDep:overall} The efficiencies in the plot are shown for events satisfying a \Zmumu{} selection and with \ptmumu{} larger than 150\,\GeV~vs pile-up for each of the four years of data taking. The uncertainties are statistical.}
\end{figure}
 
To select events with two muons, a trigger
requiring either two muon candidates each with $\pt >$ 14\,\GeV, or an asymmetric threshold of
22\,\GeV{} for the leading muon and 8\,\GeV{} for the sub-leading muon was used.
The offline selection then requires that each of the two muons has $\pt>25\,\GeV$,
and that the dimuon invariant mass be in the range
$66.6\,\GeV<m(\mu\mu)<116.6\,\GeV$.
 
The efficiency is shown as a function of \ptmumu{}
in Figure~\ref{fig:L1Summary:efficiency}
for an \lonealg{} nominal threshold of 50\,\GeV,
at which the algorithm was generally run without prescaling.
It can be observed that the algorithm achieves an efficiency
of approximately 90\% for a dimuon \pt{} of 150\,\GeV. The \lonealg{} \met{} trigger efficiency for a $\Zmumu$
selection is shown as a function of $\langle\mu\rangle$ for different years
in Figure~\ref{fig:L1MuDep:overall}.
A threshold of $\ptmumu>150\,\GeV$ is used for the efficiency calculation
since for \met{} values in the range 150--175\,\GeV{},
the L1 trigger is sufficiently close to fully efficient to be interesting for many physics analyses.
It is observed that the same efficiency was maintained to within a few percent as the pile-up increased,
 
\begin{figure}[htbp]
\centering
\captionsetup{width=.45\linewidth}
\subfloat[]{\includegraphics[width=0.49\textwidth]{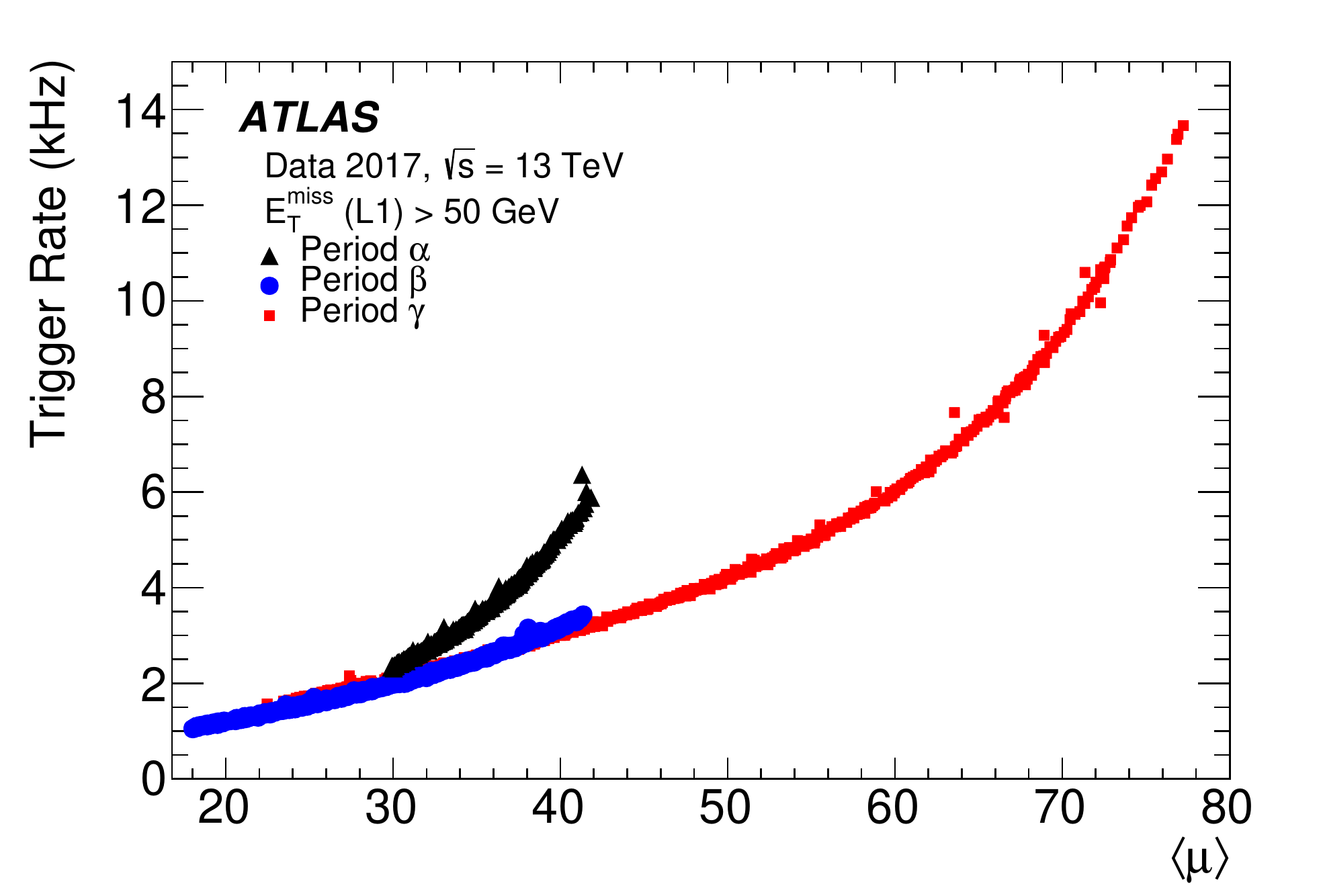}\label{fig:L1Summary:rates}}
\subfloat[]{
\includegraphics[width=0.49\textwidth]{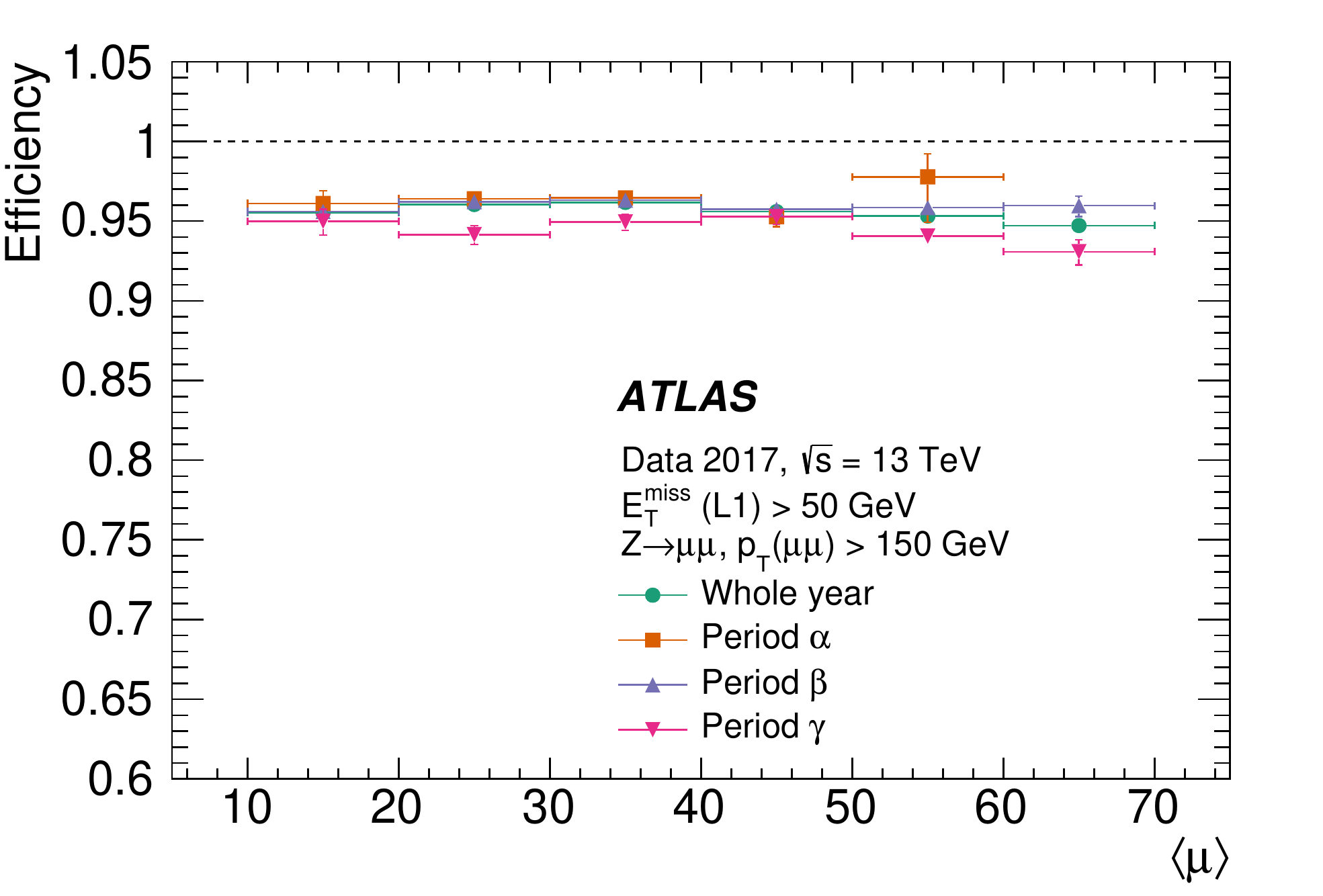}\label{fig:L1MuDep:2017}}
\captionsetup{width=\linewidth}
\caption{\protect\subref{fig:L1Summary:rates} The \lonealg{} \met{} trigger rate as a function of $\langle\mu\rangle$ for runs in three different periods ($\alpha$, $\beta$, $\gamma$) in the year 2017. \protect\subref{fig:L1MuDep:2017} The \lonealg{} \met{} trigger efficiency is shown as a function of mean pile-up for
events satisfying a \Zmumu{} selection and with \ptmumu{} larger than 150\,\GeV~in three periods during the year 2017. The uncertainties are statistical.
}
\end{figure}
 
Figure~\ref{fig:L1Summary:rates} shows the corresponding typical trigger rate
as a function of the mean pile-up $\langle\mu\rangle$,
which rises with increasing luminosity.
Each of the three periods shown has its own set of values of the \lonealg{} calorimeter noise thresholds,
which increase with increasing $\langle\mu\rangle$ as the period changes from $\alpha$ to $\beta$ to $\gamma$.
The effect of the different noise thresholds used during the periods in 2017
(labelled $\alpha$, $\beta$ and $\gamma$), can be observed.
As anticipated, higher calorimeter noise thresholds lead to much reduced trigger rates,
particularly at  higher $\langle\mu\rangle$.
 
The \lonealg{} \met{} trigger efficiency for a $\Zmumu$
selection is shown as a function of $\langle\mu\rangle$ is shown for the three periods with different noise thresholds
during 2017 in Figure~\ref{fig:L1MuDep:2017}. Even though the calorimeter noise thresholds increase to moderate the trigger rate,
the efficiency remains stable.
 
\subsection{High-level trigger performance}
 
The HLT background acceptance, which is proportional to the trigger rate,
is defined as the fraction of events that have \met computed by the HLT algorithm above a given threshold.
It is determined using events collected by a dedicated set of L1 triggers, unbiased by the HLT,
as described in Section~\ref{sec:bgmodel}.
 
The signal efficiency is determined by events collected using
the $\Zmumu$ selection described in Section~\ref{sec:L1performance}.
A subsample is selected with an additional requirement that the L1 trigger satisfy
$\met>50\,\GeV$, in order to determine the efficiency of the HLT algorithms alone.
 
\begin{figure}[htbp]
\centering
\subfloat[]{\includegraphics[width=0.49\textwidth]{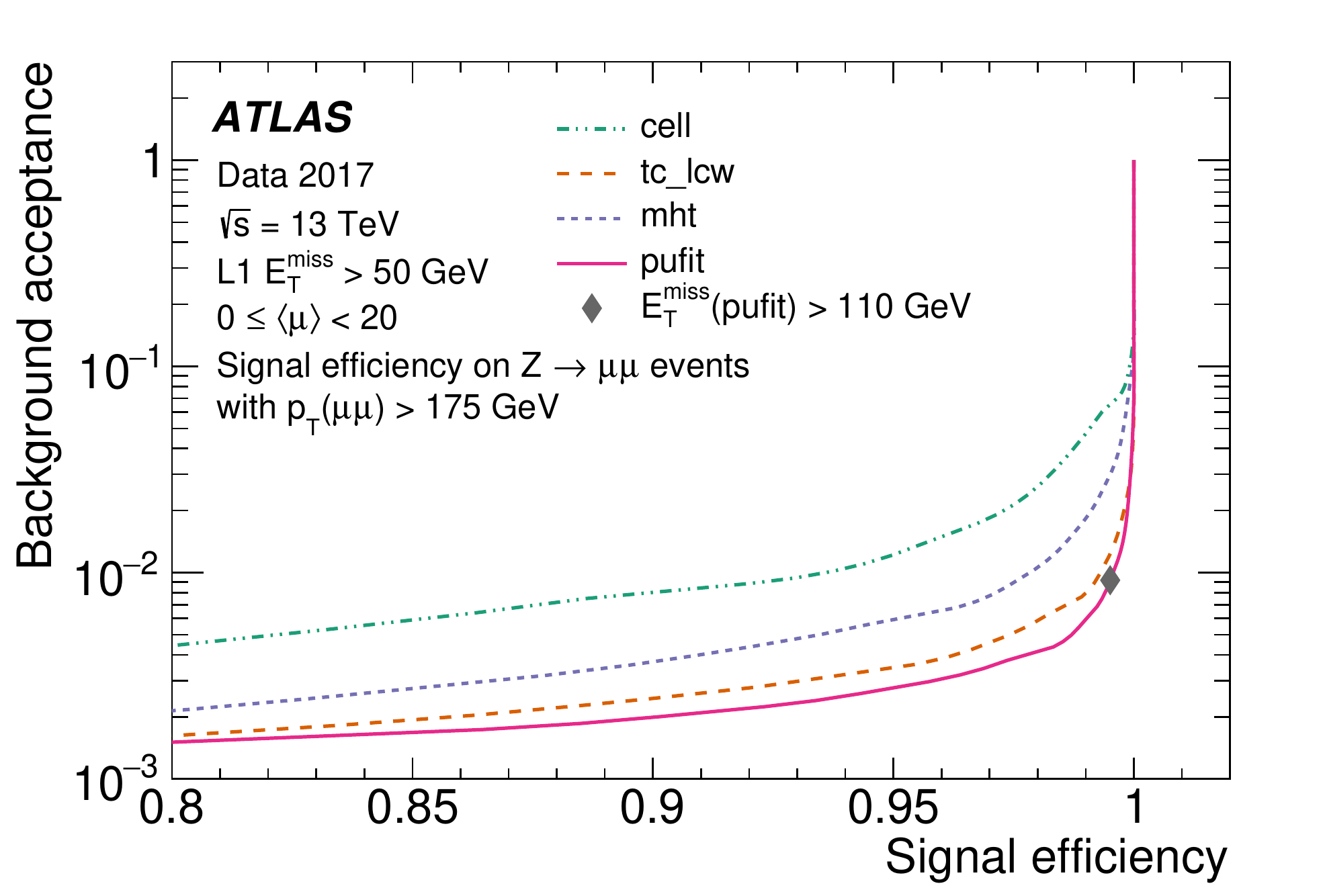}    \label{fig:RatevsEffBinned-a}}
\subfloat[]{\includegraphics[width=0.49\textwidth]{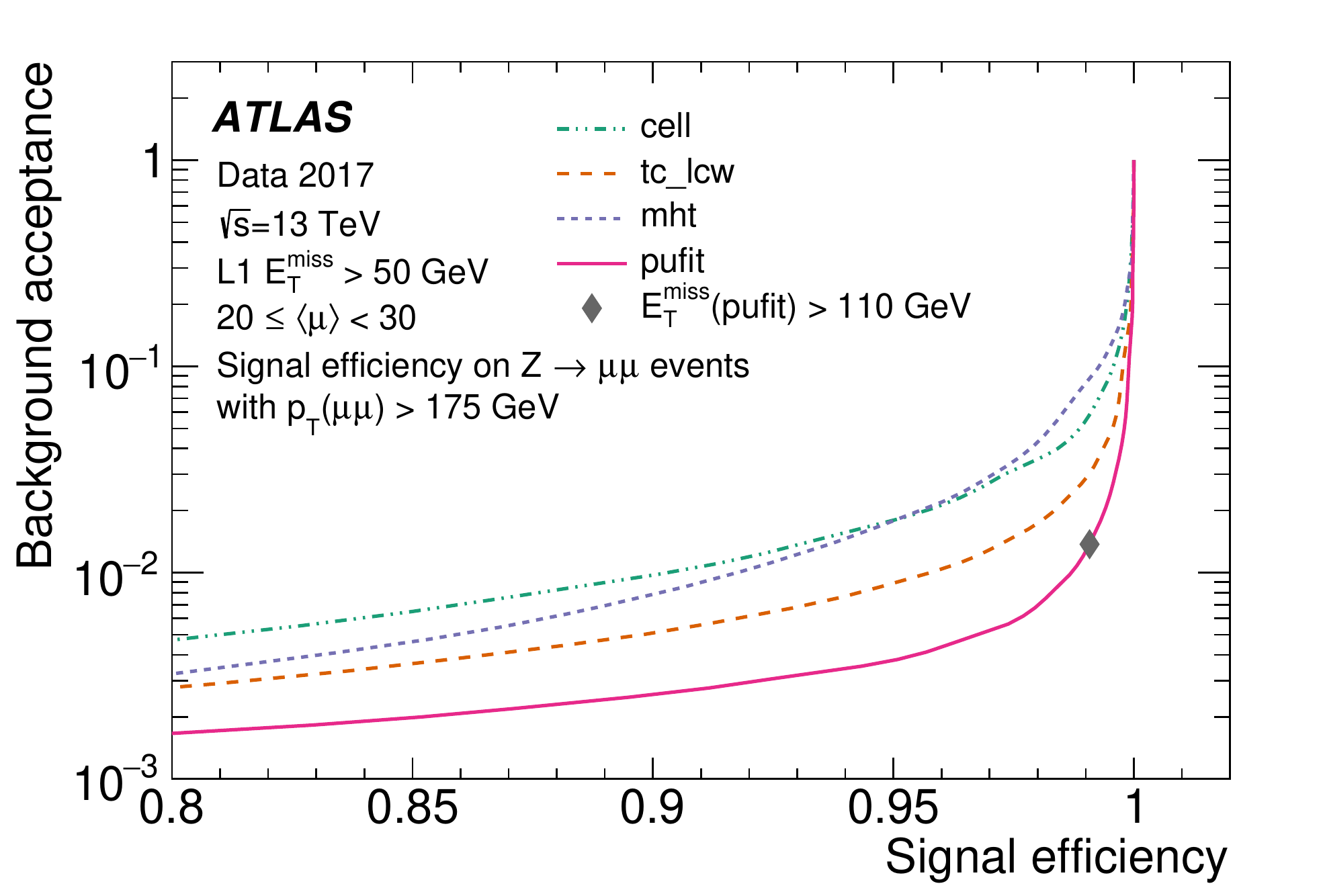} \label{fig:RatevsEffBinned-b}} \\
\subfloat[]{\includegraphics[width=0.49\textwidth]{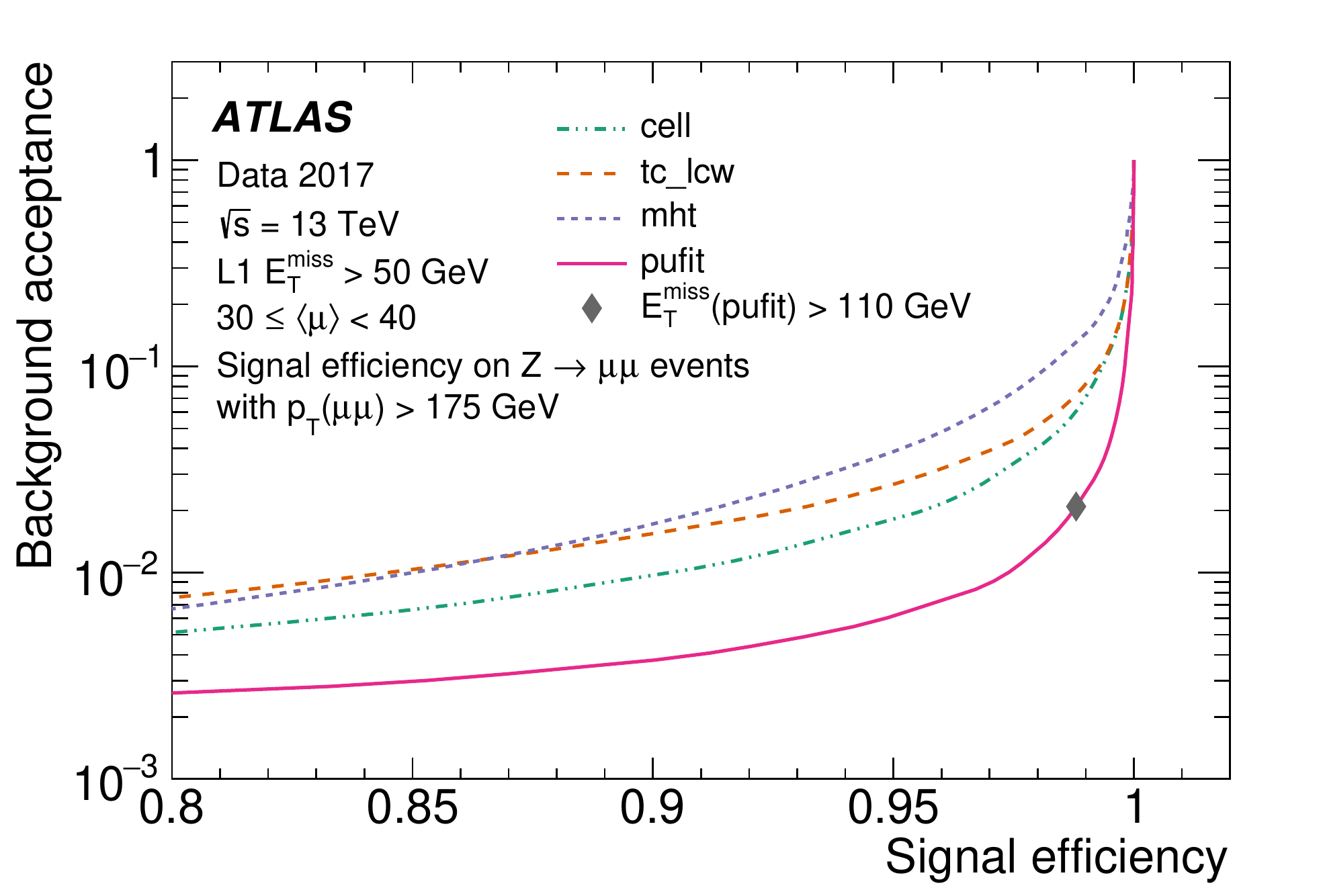}   \label{fig:RatevsEffBinned-c}}
\subfloat[]{\includegraphics[width=0.49\textwidth]{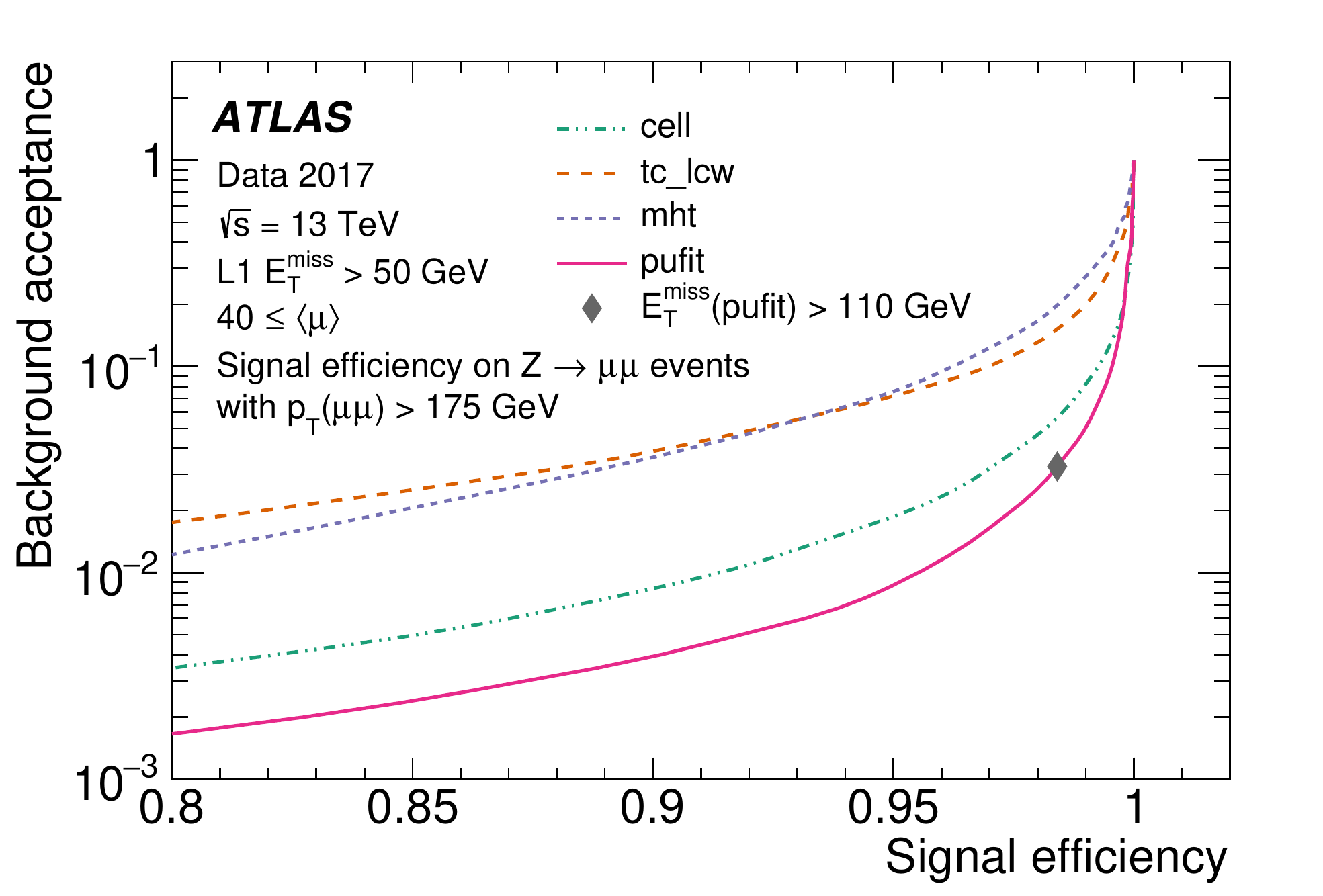}      \label{fig:RatevsEffBinned-d}}
\caption{Background acceptance vs signal efficiency for each of four individual HLT \met{} algorithms for a \Zmumu{} selection with \ptmumu{} $>$ 175\,\GeV~for data recorded in the year 2017.
The diamond indicates the performance of the $\pufit>110\,\GeV{}$ trigger.
Each of the four lower panels shows a different range of $\langle\mu\rangle$: \protect\subref{fig:RatevsEffBinned-a} $0\leq\langle\mu\rangle < 20$, \protect\subref{fig:RatevsEffBinned-b} $20\leq\langle\mu\rangle < 30$, \protect\subref{fig:RatevsEffBinned-c} $30\leq\langle\mu\rangle < 40$ and \protect\subref{fig:RatevsEffBinned-d} $40\leq\langle\mu\rangle$.\label{fig:RatevsEffMuBinned} }
\end{figure}

Curves of background rejection versus signal efficiency are obtained by varying the
HLT trigger threshold.
Figure~\ref{fig:RatevsEffMuBinned} compares such curves for the four \met algorithms
defined in Section~\ref{sec:alg}, for different amounts of pile-up.
For low pile-up ($\langle\mu\rangle<20$) the efficiencies at which the
\tcalg{} and the \mht{} algorithms have equal-efficiency rejection power within a factor of three to that of \pufit.
The \cell{} algorithm has lower corresponding efficiency.
As pile-up increases,  \tcalg{} and \mhtalg{} suffer the most degradation in their performance,
whereas the \pufit{} trigger, which was designed to be robust against increasing pile-up,
continues to simultaneously achieve good signal efficiency and large background rejection.
 
\begin{figure}[htbp]
\centering
\subfloat[]{\includegraphics[width=0.49\textwidth]{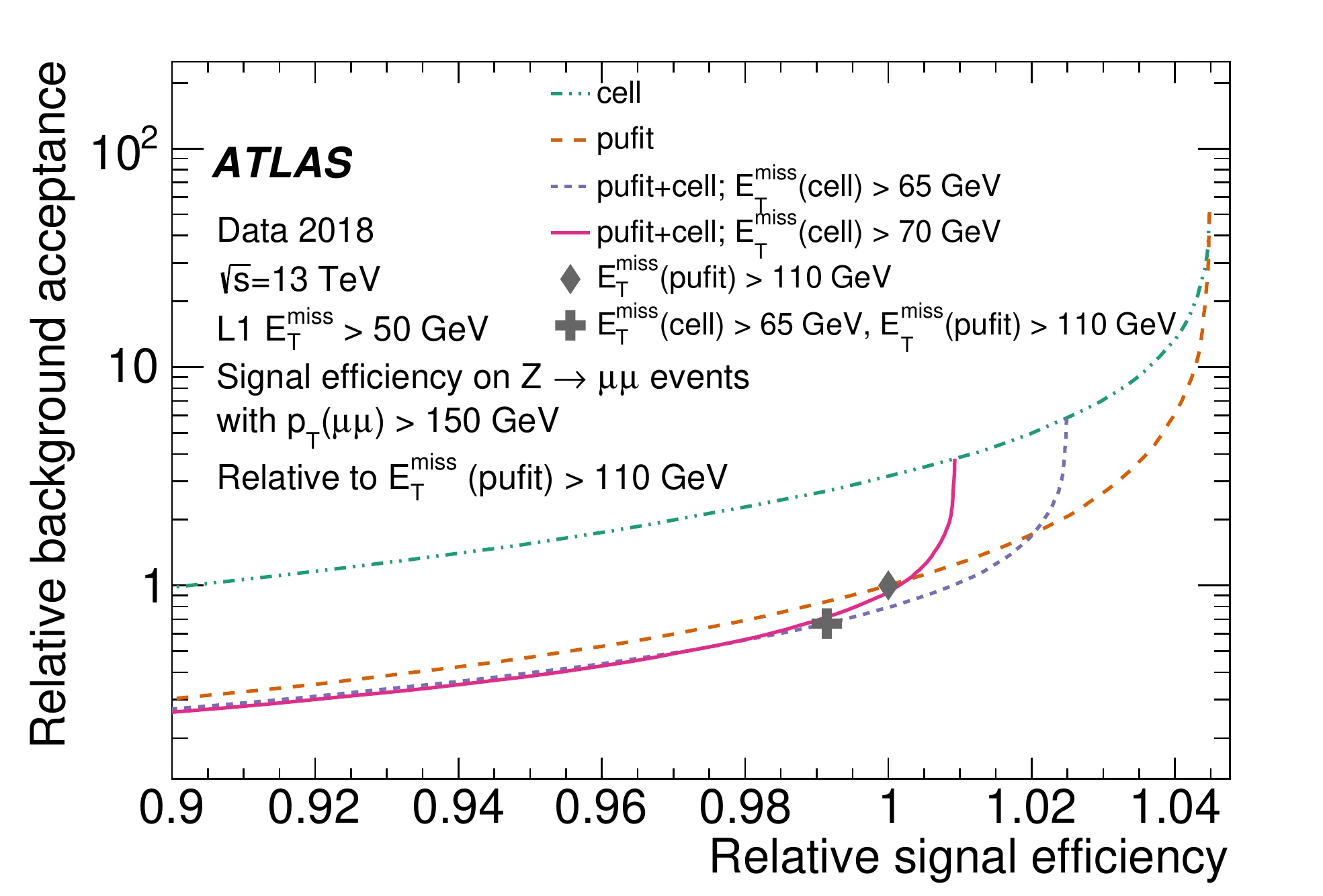}\label{fig:AccVsEffHLTCellPufit-a}}
\subfloat[]{\includegraphics[width=0.49\textwidth]{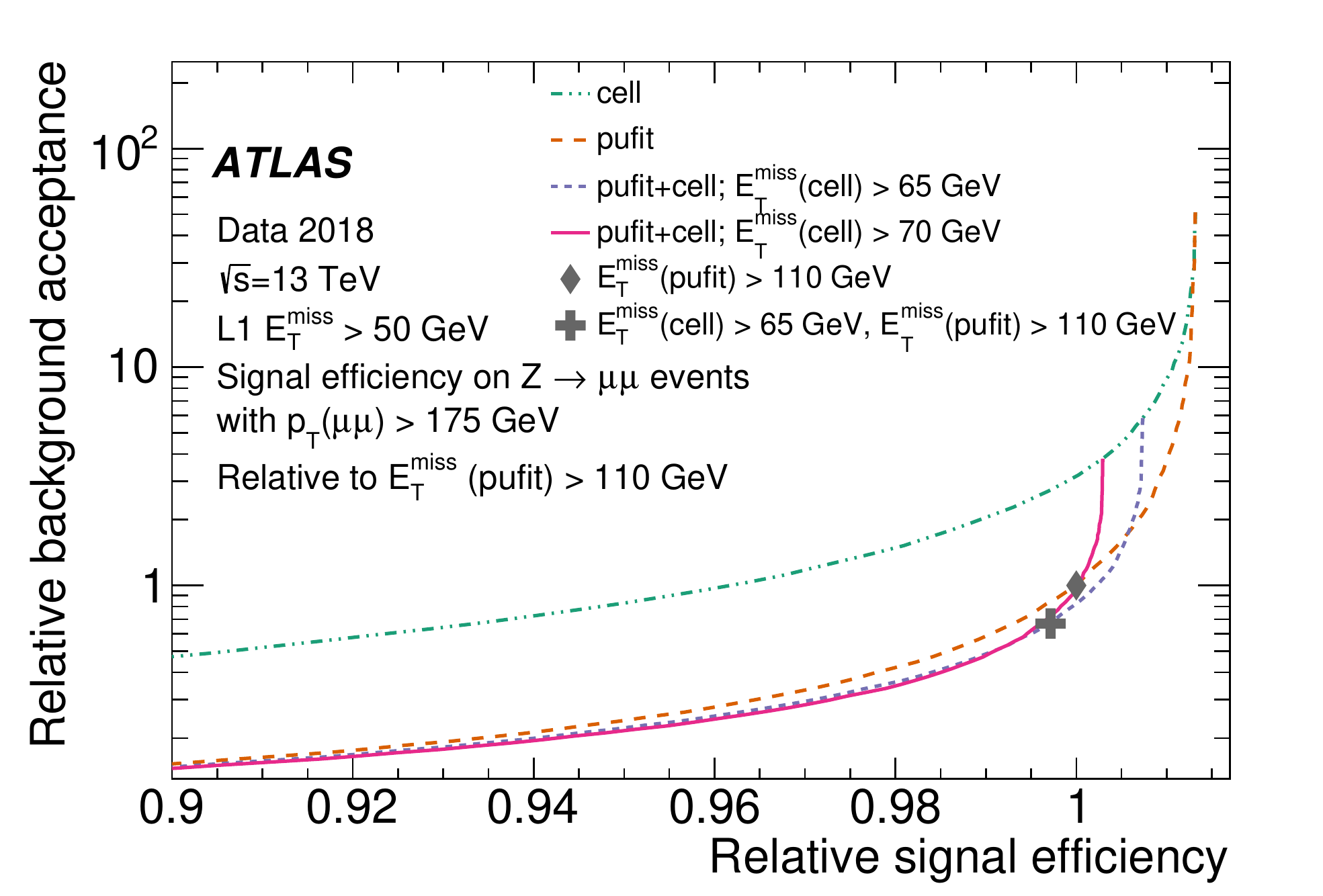}\label{fig:AccVsEffHLTCellPufit-b}}
\caption{Relative background acceptance fraction vs. relative efficiency for
two different \ptmumu{} thresholds: \protect\subref{fig:AccVsEffHLTCellPufit-a} $\ptmumu{}>150\,\GeV$ and \protect\subref{fig:AccVsEffHLTCellPufit-b} $\ptmumu{}>175\,\GeV$ ~for data recorded in the year 2018.
Two of the curves show the performance of the stand-alone
\cellalg{} algorithm and the stand-alone \pufitalg{} algorithm.
The other two show combined algorithms each formed by requiring that the event
satisfy both a fixed threshold (either 65\,\GeV{} or 70\,\GeV{} as shown in the legend)
for the \cellalg{} algorithm and a \pufit{} threshold which varies along the curve.
In each plot the background acceptance fractions and the efficiencies are relative to those of the $\pufit\ > 110\,\GeV$ trigger and thus can be greater than one.
The diamond indicates the performance of the $\pufit>110\,\GeV{}$ trigger while the cross indicates the performance of the combined \big($\pufit>110\,\GeV{}$ and $\cell>50\,\GeV{}$\big) trigger.
\label{fig:AccVsEffHLTCellPufit}
}
\end{figure}
 
\begin{figure}[htbp]
\centering
\captionsetup{width=.49\linewidth}
\subfloat[]{\includegraphics[width=0.49\textwidth]{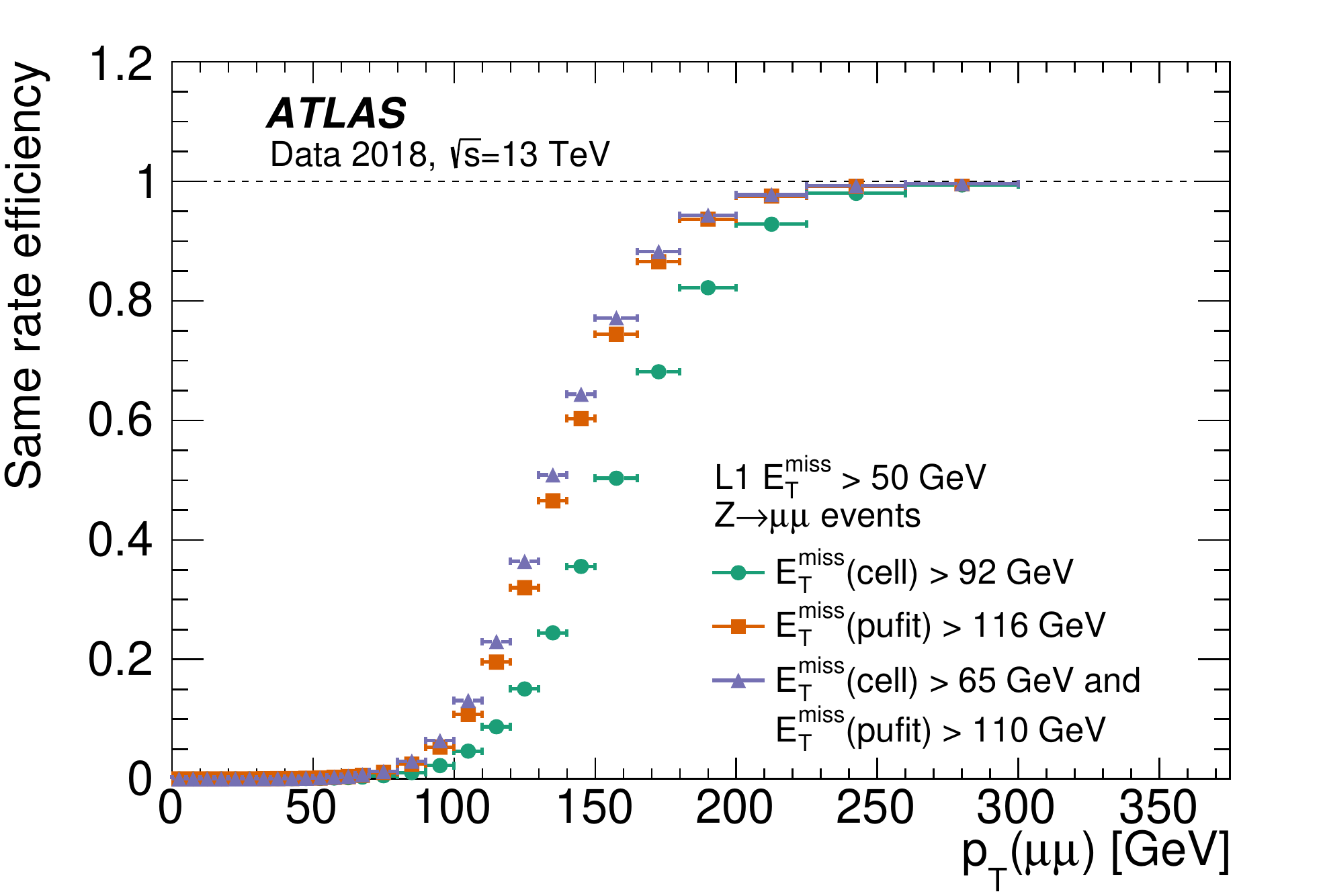}\label{fig:HLTCellPufitEqualRate-a} }
\subfloat[]{\includegraphics[width=0.49\textwidth]{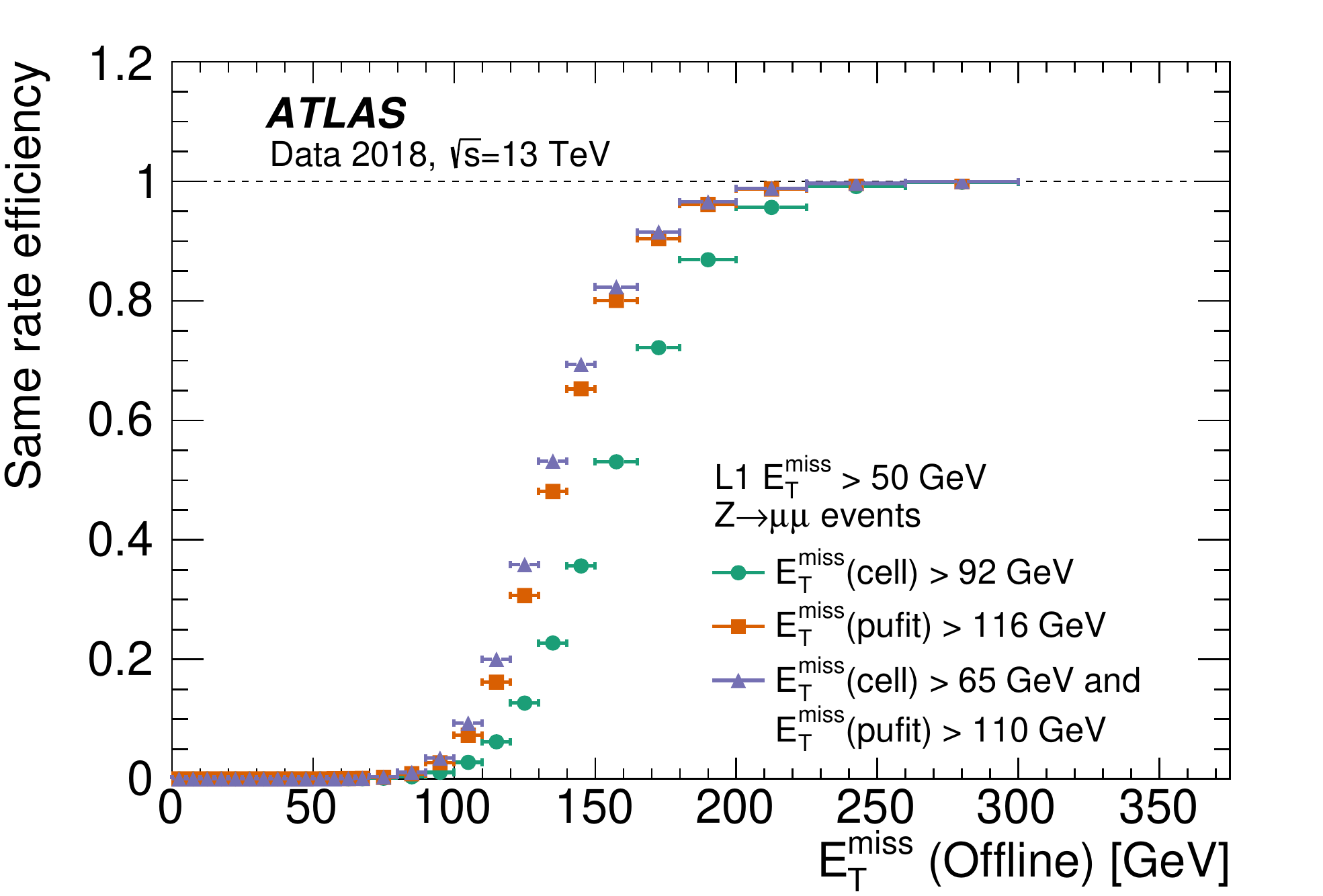} \label{fig:HLTCellPufitEqualRate-b}}
\captionsetup{width=\linewidth}
\caption{
Turn-on efficiency curves are shown for \Zmumu{} events for three algorithms: the
\cellalg{} algorithm alone, the \pufitalg{} algorithm alone and the combined \cellalg{}+\pufitalg{} algorithm.
The thresholds are set such that the algorithms have equal rates, and the data were recorded in the year 2018.
\protect\subref{fig:HLTCellPufitEqualRate-a} The trigger efficiency with respect to \ptmumu. \protect\subref{fig:HLTCellPufitEqualRate-b} The trigger efficiency with respect to the offline \met\ calculation with muons treated as being invisible.
}
\label{fig:HLTCellPufitEqualRate}
\end{figure}
 
By combining different high-level triggers it was found to be
possible to further improve the overall HLT performance.
The simplest way to achieve this is by demanding that more than one
\met{} algorithm indicates that the event has high \met{}.
The rationale for such a combination is as follows.
The trigger rate of each algorithm for \met greater than about 50\,\GeV{} is typically
dominated by contributions from the resolution tails of poorly measured events
which often contain little true \met{}.
Since these tails depend on the details of the algorithm,
populations of poorly reconstructed events in the high $\met{}$ tails
differ between algorithms.
By contrast, events with large true $\met{}$ caused by invisible particles
tend to produce a large $\met{}$ with all algorithms.
Therefore, requiring events to have large $\met{}$ in more than one algorithm,
with appropriate thresholds for each,
can result in reduced trigger rates for a similar overall efficiency.
 
The joint use of two \met\ algorithms was found to be particularly useful
when combining the \pufitalg\ and \cellalg{} algorithms.
Figure~\ref{fig:AccVsEffHLTCellPufit} shows the relative
signal acceptance and background rejection curves of the combined \pufitalg{}+\cellalg{} algorithm
compared with those of \pufitalg alone or \cellalg alone.
With suitable thresholds, the combinations can have a higher rejection
at the same efficiency than does either algorithm used alone.
 
The efficiencies of the \cellalg, \pufitalg{} and combined \pufitalg{}+\cellalg{}
algorithms are shown as a function of \ptmumu{} and as a function of the offline \met{}
in Figure~\ref{fig:HLTCellPufitEqualRate}.
In order to have a fair comparison between the algorithms,
each algorithm's trigger threshold has been set such that their background rejections
(and hence trigger acceptance rates) are equal.
The combined \pufitalg{}+\cellalg{} algorithm is again observed to have higher efficiency
for signal events throughout the turn-on region than does either of the individual algorithms.
The behaviour is consistent regardless of whether the
efficiency is calculated as a function of \ptmumu{} or the offline \met{} (with muons treated as invisible).
 
The improved acceptance for physics analyses that results
from using the new algorithms can be considerable.
Figure~\ref{fig:HLTCellPufitEqualRate} shows that the
new \pufitalg{}+\cellalg{} trigger reaches its plateau efficiency at a $Z$ boson \pt{}
about 25\,\GeV{} below that of the \cellalg{} trigger operating at the same rate.
This permits a correspondingly 25\,\GeV{} lower offline \met{} threshold to be used
for any analysis using an \met{} trigger.
This has a particularly important effect for physics channels in which the \met{}
distribution falls rapidly.
For those analyses that select events on the \met{} turn-on region
where the trigger is not fully efficient,
the \pufitalg{}+\cellalg{} algorithm
recovers up to double the number of events of interest compared to \cellalg{} alone.
An example is for the search for Higgs to $b$-quarks~\cite{HIGG-2018-04} associated with the decay \Znunu. For this analysis the acceptance decreases from 12\% with \met $>$ 150\,\GeV to only 5\% with \met $>$ 200 \GeV~\cite{TDR029}. If the offline threshold were to increase to 225\,\GeV the acceptance would have been only 3.5\%.
 
\begin{figure}[htbp]
\centering
\captionsetup{width=0.48\linewidth}
\subfloat[]{\includegraphics[width=0.49\textwidth]{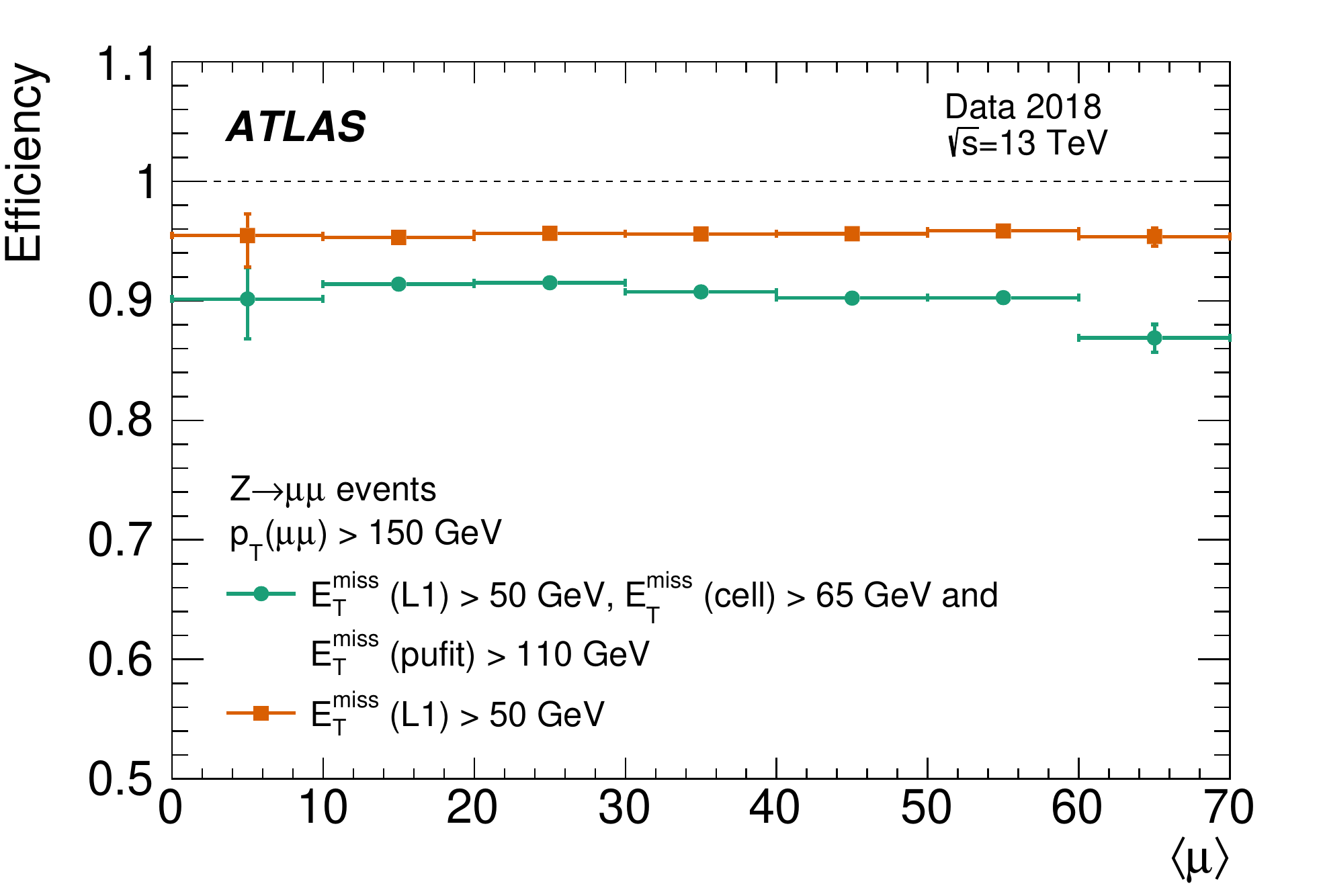}\label{fig:ZvsOffline-a}}
\subfloat[]{\includegraphics[width=0.49\textwidth]{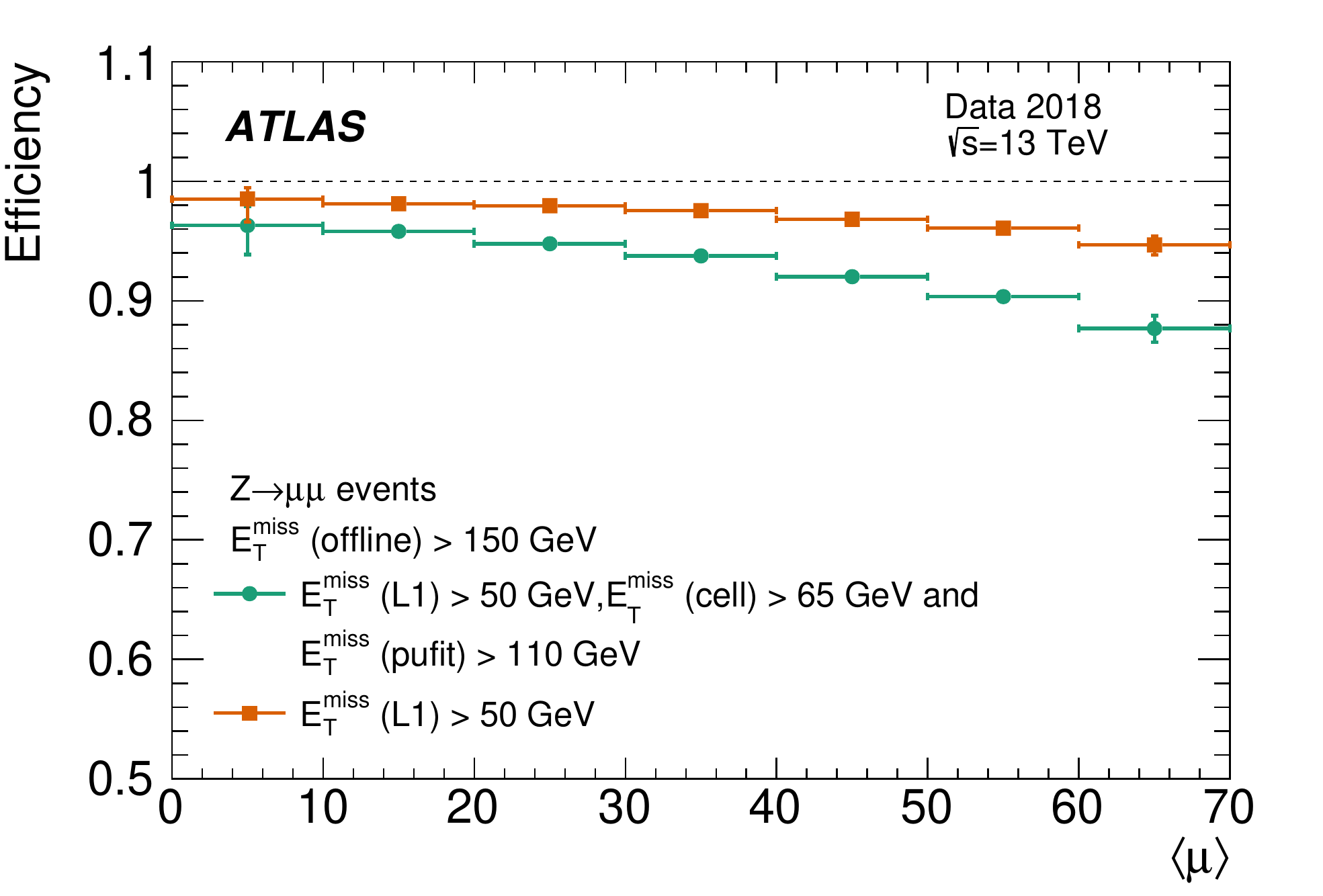}\label{fig:ZvsOffline-b}} \\
\subfloat[]{\includegraphics[width=0.49\textwidth]{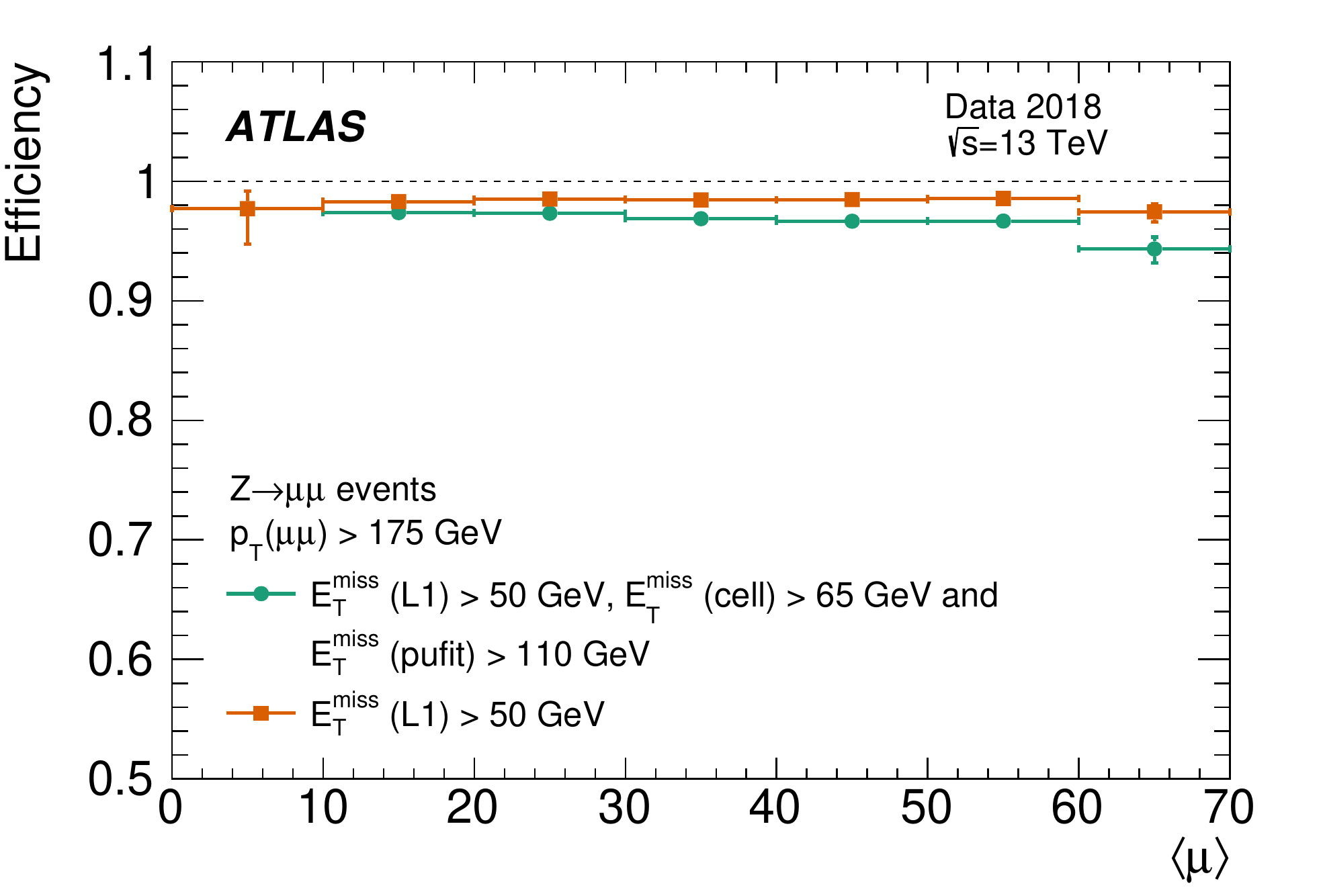}\label{fig:ZvsOffline-c}}
\subfloat[]{\includegraphics[width=0.49\textwidth]{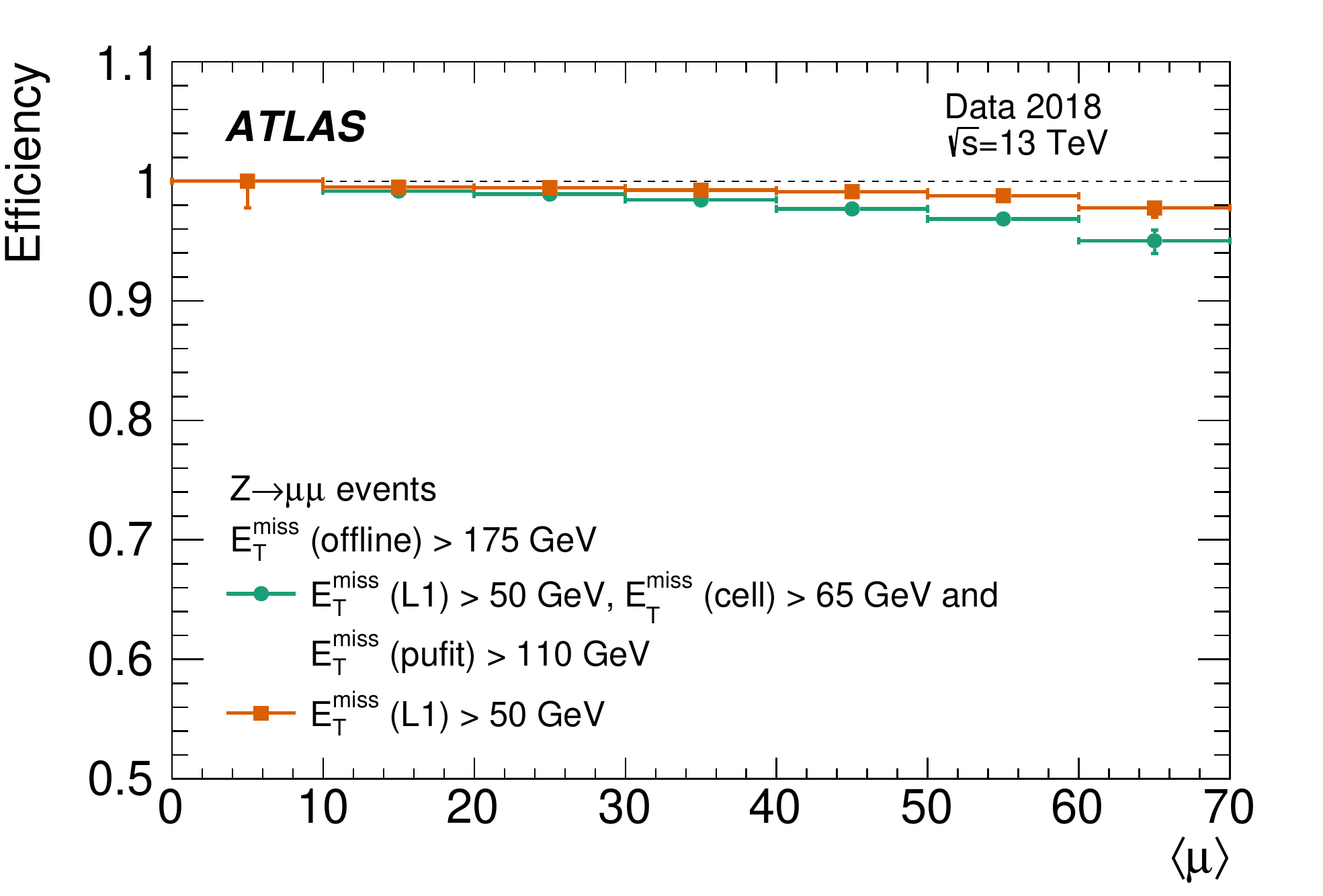}\label{fig:ZvsOffline-d}}
\captionsetup{width=\linewidth}
\caption{Efficiencies for \Zmumu{} events are shown for the \lonealg{} $\met{}>50\,\GeV{}$ trigger (square)
and for the complete L1+HLT trigger chain (circle) that also requires $\pufit>110\,\GeV{}$. The uncertainties are statistical.
Each is shown as a function of $\langle\mu\rangle$,
either for a \ptmumu{} threshold as in the left plots: \protect\subref{fig:ZvsOffline-a} and \protect\subref{fig:ZvsOffline-c} or for an offline \met{} threshold as shown in the right plots: \protect\subref{fig:ZvsOffline-b} and \protect\subref{fig:ZvsOffline-d}. The upper two plots \protect\subref{fig:ZvsOffline-a} and \protect\subref{fig:ZvsOffline-b} show thresholds of 150\,\GeV, while the lower two plots \protect\subref{fig:ZvsOffline-c} and \protect\subref{fig:ZvsOffline-d} correspond to thresholds of 175\,\GeV.
}
\label{fig:ZvsOffline}
\end{figure}

To further examine the efficiency of the trigger algorithms with respect to the offline \met{},
the \met{} trigger efficiency is calculated
after applying either an additional offline $\met>150\,(175)\,\GeV$ requirement
or an offline $\ptmumu >150\,(175)\,\GeV$ requirement.
Figure~\ref{fig:ZvsOffline}~(left) shows efficiencies
for both the L1 trigger and
the full (L1+HLT) trigger chain for data recorded at the end of 2018.
The trigger efficiencies for a fixed \ptmumu{} threshold show no significant decrease,
even for the highest values of $\langle\mu\rangle$.
However, when compared with an offline \met{} threshold in Figure~\ref{fig:ZvsOffline}~(right),
an apparent degradation of the trigger efficiency is observed for high $\langle\mu\rangle$.
This indicates that the difference between the online and offline \met{} definitions
has a larger effect at higher $\langle\mu\rangle$.
 
\subsection{Trigger menu evolution and performance}

\begin{table}[]
\centering\renewcommand\arraystretch{1.1}
\begin{tabular}{llllll} \hline \hline
Year &Trigger name & HLT algorithm & L1 threshold & HLT threshold & $\int$$\mathcal{L}\,\text{\rmfamily d}t$ \\
&          & [\GeV] & [\GeV]  & [\GeV] & [\fbinv]\\ \hline
2015 &\texttt{HLT\_xe70\_mht\_L1XE50} & \mhtalg & 50 & ~~70 &  ~~3.5 \\
2016 &\texttt{HLT\_xe90\_mht\_L1XE50} &\mhtalg & 50 & ~~90 &  12.7 \\
2016 &\texttt{HLT\_xe110\_mht\_L1XE50} &\mhtalg & 50 & 110 &  30.0 \\
\hline
2017 &\texttt{HLT\_xe90\_pufit\_L1XE50} &\pufitalg,\,\cellalg & 50 & ~~90,\,50 &  21.8 \\
2017 &\texttt{HLT\_xe100\_pufit\_L1XE50} &\pufitalg,\,\cellalg & 50 & 100,\,50 &  33.0 \\
2017 &\texttt{HLT\_xe110\_pufit\_L1XE50(55)} &\pufitalg,\,\cellalg & 50\,(55) & 110,\,50 &  47.7 \\
\hline
2018 &\texttt{HLT\_xe110\_pufit\_xe65\_L1XE50} &\pufitalg,\,\cellalg & 50 & 110,\,65 &  57.0 \\
2018 &\texttt{HLT\_xe110\_pufit\_xe70\_L1XE50} & \pufitalg,\,\cellalg & 50 & 110,\,70 &  62.6 \\  \hline \hline
\end{tabular}
\caption{The evolution of the primary \met{} physics triggers through the years of the LHC physics Run 2 from 2015 to 2018.
For each year the table shows the algorithms used, the L1 and HLT thresholds
applied and the integrated luminosity collected.
Where two HLT thresholds are given, the first corresponds to the \pufitalg{} algorithm and the second to the \cellalg{} algorithm.
In 2017, the \pufitalg{} algorithm was used in conjunction
with an additional requirement that $\cell>50\,\GeV{}$, which is not explicit in its name.
The integrated luminosities are not exclusive and cannot be summed to obtain a total integrated luminosity.}
\label{table:triggermenu}
\end{table}
 
\begin{figure}[htbp]
\centering
\includegraphics[width=0.7\textwidth]{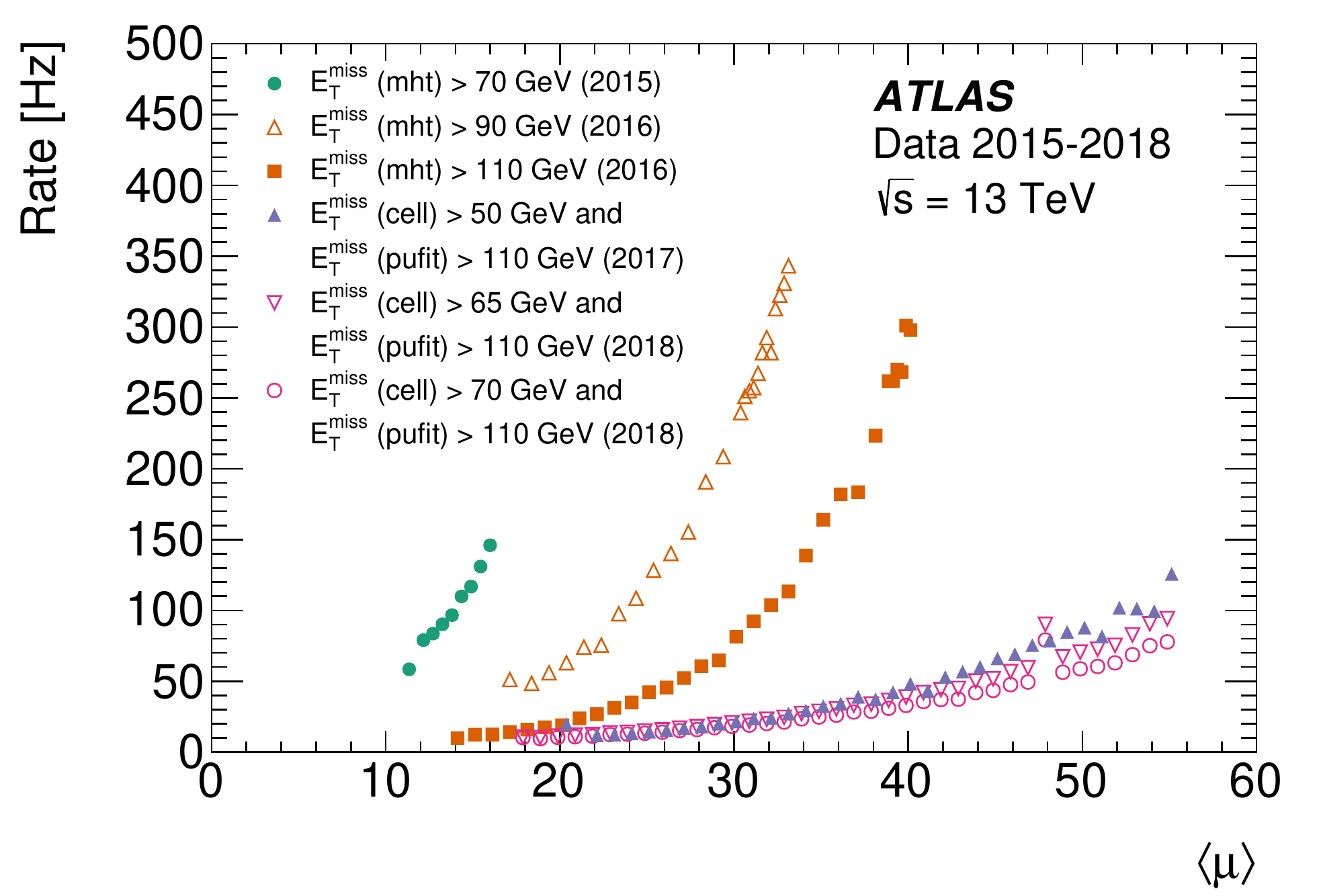}
\caption{High-level trigger output rates, as a function of $\langle\mu\rangle$, shown separately for example runs in each year 2015--2018, for triggers \texttt{HLT\_xe70\_mht} (2015), \texttt{HLT\_xe90\_mht} and \texttt{HLT\_xe110\_mht} (2016), \texttt{HLT\_xe110\_pufit} (2017), \texttt{HLT\_xe110\_pufit\_xe65} and \texttt{HLT\_xe110\_pufit\_xe70} (2018). The \texttt{HLT\_xe110\_pufit} trigger used during 2017 also included an implicit requirement of $\cell>50\,\GeV$.
}
\label{fig:hltrate}
\end{figure}
 
Due to the dependence of the algorithm efficiencies and trigger rates upon luminosity, it was necessary to
update the primary physics triggers to cope with the increasing pile-up levels.
Since the \lonealg{} rate was reduced by adjusting calorimeter noise thresholds,
only small adjustments needed to be made to the overall \lonealg{} threshold.
Table~\ref{table:triggermenu} summarizes the algorithms
and trigger thresholds used during Run-2 data taking. In 2015--2016, the \mht was used.
From 2016, the \pufit was combined with \cell, thereby mitigating the effect of pile-up.
 
The trigger names carry information about the algorithms and thresholds used.
For example \texttt{L1\_XE50},
denotes that the requirement is placed upon the first-level trigger (\texttt{L1}), that
the requirement is on the value of \met (\texttt{XE}), and provides
the value of the \texttt{L1} trigger threshold (\text{50}\,\GeV).
The naming convention for full trigger paths can be parsed to give the trigger algorithms and
their thresholds. For example, for the trigger path named \texttt{HLT\_xe110\_pufit\_L1XE50},
the prefix \texttt{HLT} indicates that the event must satisfy the high-level trigger requirement;
\texttt{xe110} indicates that the HLT threshold used was 110\,\GeV; \texttt{pufit} refers to the HLT algorithm
used (except in the special case of the \cell{} algorithm, where the additional algorithm name is omitted),
and \texttt{L1XE50} refers to the L1 item used and its threshold.
 
Typical output rates for various HLT algorithms are shown year-by-year in Figure~\ref{fig:hltrate}.
The reduction in rate obtained by using the \pufitalg-based algorithms is
a factor of ten or more for higher values of $\langle\mu\rangle$.
 
\begin{figure}[htbp]
\centering
\captionsetup{width=0.48\linewidth}
\subfloat[]{\includegraphics[width=0.49\textwidth]{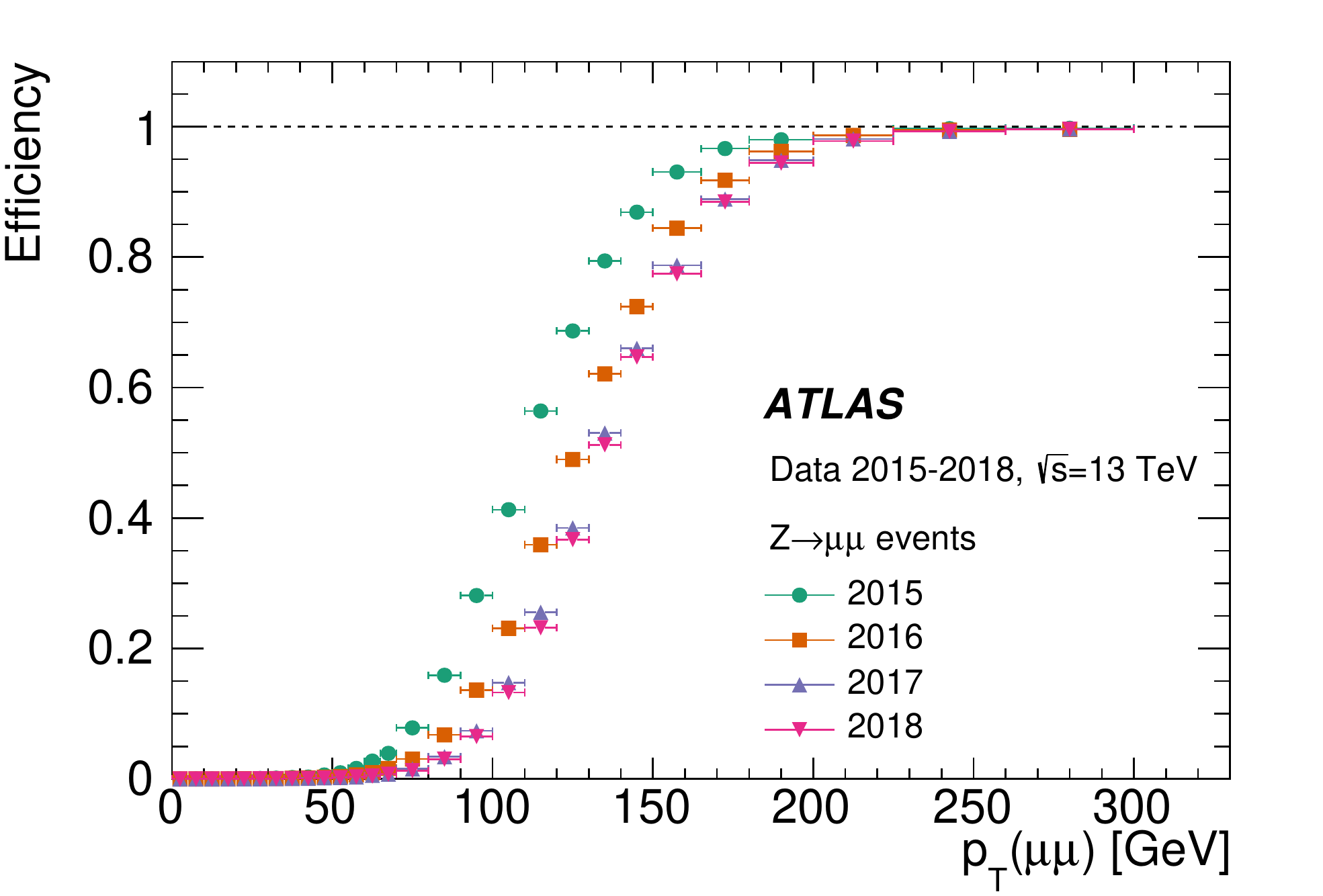}\label{fig:SummaryEff:turnon}}
\subfloat[]{\includegraphics[width=0.49\textwidth]{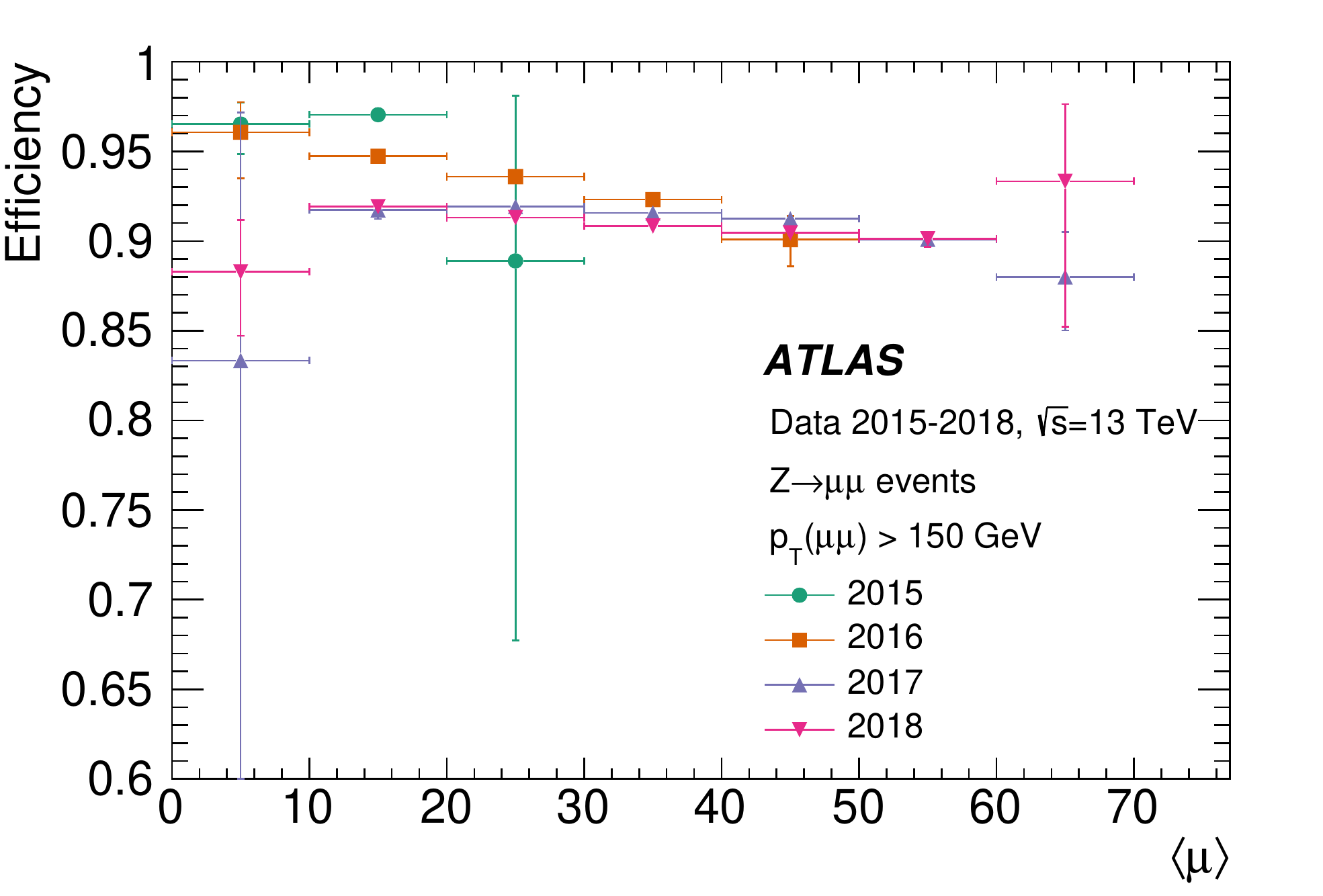}\label{fig:SummaryEff:mu}}
\captionsetup{width=\linewidth}
\caption{Full-chain trigger efficiencies for each year \protect\subref{fig:SummaryEff:turnon} as a function of \ptmumu and \protect\subref{fig:SummaryEff:mu} as a function of $\langle\mu\rangle$ for $\ptmumu>150\,\GeV$.
The efficiency corresponds to that of the lowest unprescaled trigger that is adjusted throughout each year (Table~\ref{table:triggermenu}). The uncertainties are statistical.}
\label{fig:SummaryEff}
\end{figure}

The overall (\lonealg+HLT) \met\ trigger efficiency is shown year-by-year in Figure~\ref{fig:SummaryEff}.
The efficiency is shown both as a function of \ptmumu{} and as a function of  pile-up.
The latter demonstrates that the efficiency remained stable within a few percent even at the
highest pile-up values recorded.
 
\subsection{Algorithm computation times}
 
Average CPU times for the various steps used in the HLT \met{} algorithms are given in
Table~\ref{table:HLTalgCPUtime}.
For all algorithms except \cellalg{}, the fraction of the computation time needed
for evaluating the final \met{} from previously determined input elements is negligible,
and most of the CPU time is spent reconstructing cells and topological clusters.
All steps satisfy the requirement described in Section~\ref{sec:alg}
that the CPU time does not exceed $\mathcal{O}$(100\,ms).
 
\begin{table}[ht]
\centering\renewcommand\arraystretch{1.2}
\begin{tabular}{lccccc}\hline \hline
\multicolumn{1}{c}{Algorithm step} & Time per step [ms]& \multicolumn{4}{c}{Algorithm} \\
\multicolumn{2}{l}{}  &  \tcalg & \pufitalg & \cellalg & \mhtalg \\ \hline
Calorimeter cell reconstruction                 & 20      & $\bullet$  & $\bullet$     & $\bullet$    & $\bullet$   \\
Topological cluster reconstruction               & 75       & $\bullet$  & $\bullet$    &  -    & $\bullet$   \\
Jet reconstruction                  & 15        &  -  &   -    &   -   & $\bullet$   \\\hline
\met evaluation time [ms]     &      -       & -  & -     & 40   & -   \\ \hline
Total time (ms)           &  - & 95         & 95   &  60     &   110   \\\hline\hline
\end{tabular}
\caption{The average execution time of each step in computing \met
in the HLT online farm. The dot ($\bullet$) indicates the required steps for each algorithm.
The time to evaluate the \met is shown as well, with the total time per step
added to the evaluation time.
\label{table:HLTalgCPUtime}
}
\end{table}
 
\FloatBarrier
\subsection{Dependence on event characteristics}
 
All of the previous efficiencies are computed
using the clean reference sample selected with two muons in the final state. This sample is dominated by \Zmumu{} events produced with additional jets.
Because the detector response is not identical for events selected according to different criteria,
the computation of \met{} also depends on the event characteristics,
for example whether jets or electrons are required to be present. In this section the trigger
efficiency is evaluated and compared for a variety of offline event selections.
 
\begin{table}[]
\centering\renewcommand\arraystretch{1.2}
\begin{tabular}{llll}
\hline\hline
Physics process & Offline \MET definition & Lepton(s)                                                                                                       & Kinematics    \\ \hline
\Zmumu  & \ptmumu         & $\mu\mu$ 						& $66.6 < \Mmumu < 116.6\,\GeV$       \\
$\Wboson \rightarrow e\nu$      & Tight       & $e$ & -   \\
VBF        & Tight  & $\mu$, \pt $>$ 30\,\GeV  							& Exactly two jets, $\pt > 80\,(50)\,\GeV$, \\
&&& $|\Delta\phi(\text{jj})| < 1.8$, $|\Delta \eta(\text{jj})| > 4.9$ \\
\ttbar    & Tight     &  $e$, $\mu$, $e^{\pm}\mu^{\mp}$     				& $\geq$ 2 $b$-tagged jets  \\ \hline\hline
\end{tabular}
\caption{\label{table:selections}Definition
of offline analysis selections used for efficiency measurements, labelled by the physics process being examined.
All indicated lepton requirements implicitly require $\pt(\ell) > 25\,\GeV$ unless specified. The offline \met definitions correspond
to different working points. When multiple jets are required with different \pt thresholds, the threshold for the subleading jet is listed in parentheses, e.g. 80 (50)\,\GeV.}
\end{table}
 
To complement the \Zmumu{} events, four other selections are defined, as
shown in Table~\ref{table:selections}.
The \ttbar{} selections target the pair production of top quarks, and are particularly
useful for examining performance in events with a large number of jets.
These selections require that the event contain either (i) exactly one electron and no muons or
(ii) exactly one muon and no electrons or
(iii) exactly one electron and one muon.
The vector boson fusion (VBF) selection targets events characterized by two energetic jets,
where typically at least one jet is in the forward calorimeter.
The selection requires two jets separated by a large pseudorapidity difference
and not back-to-back in azimuth.
The $W$ boson sample targets the $W\rightarrow e\nu$ process, and samples events with electromagnetic energy deposits
that can be larger than 50\,\GeV.
 
The $W$ boson, VBF and \ttbar{} samples were each collected using a trigger that
selects events containing a single isolated electron or muon.
The required lepton transverse momentum thresholds were
in the range 24--26\,\GeV, where the \pt value corresponds to the lowest threshold
single lepton trigger available for a given luminosity.
 
\begin{figure}[htbp]
\centering
\captionsetup{width=0.45\linewidth}
\subfloat[]{\includegraphics[width=0.49\textwidth]{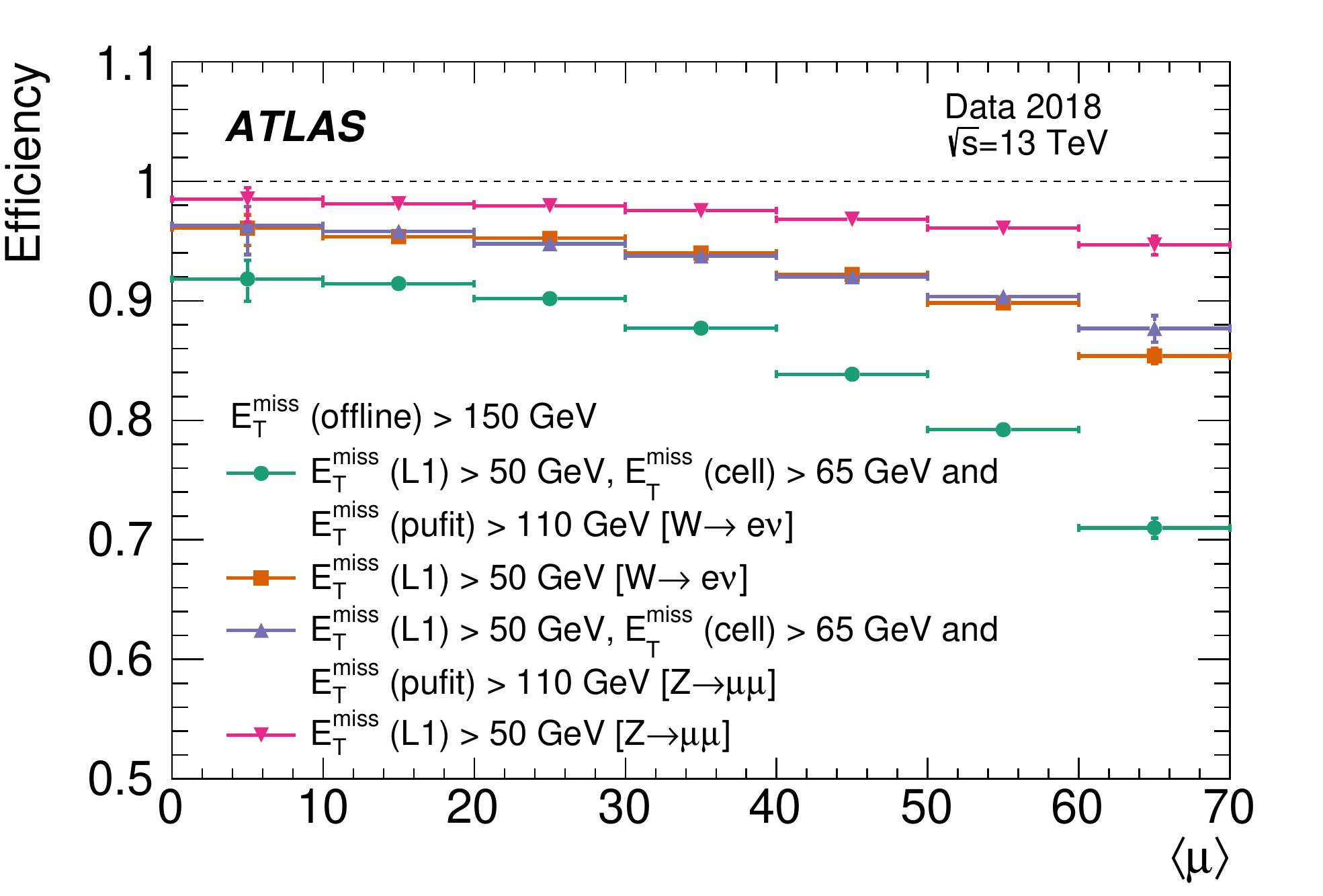}\label{fig:EffEventChar-a}}
\subfloat[]{\includegraphics[width=0.49\textwidth]{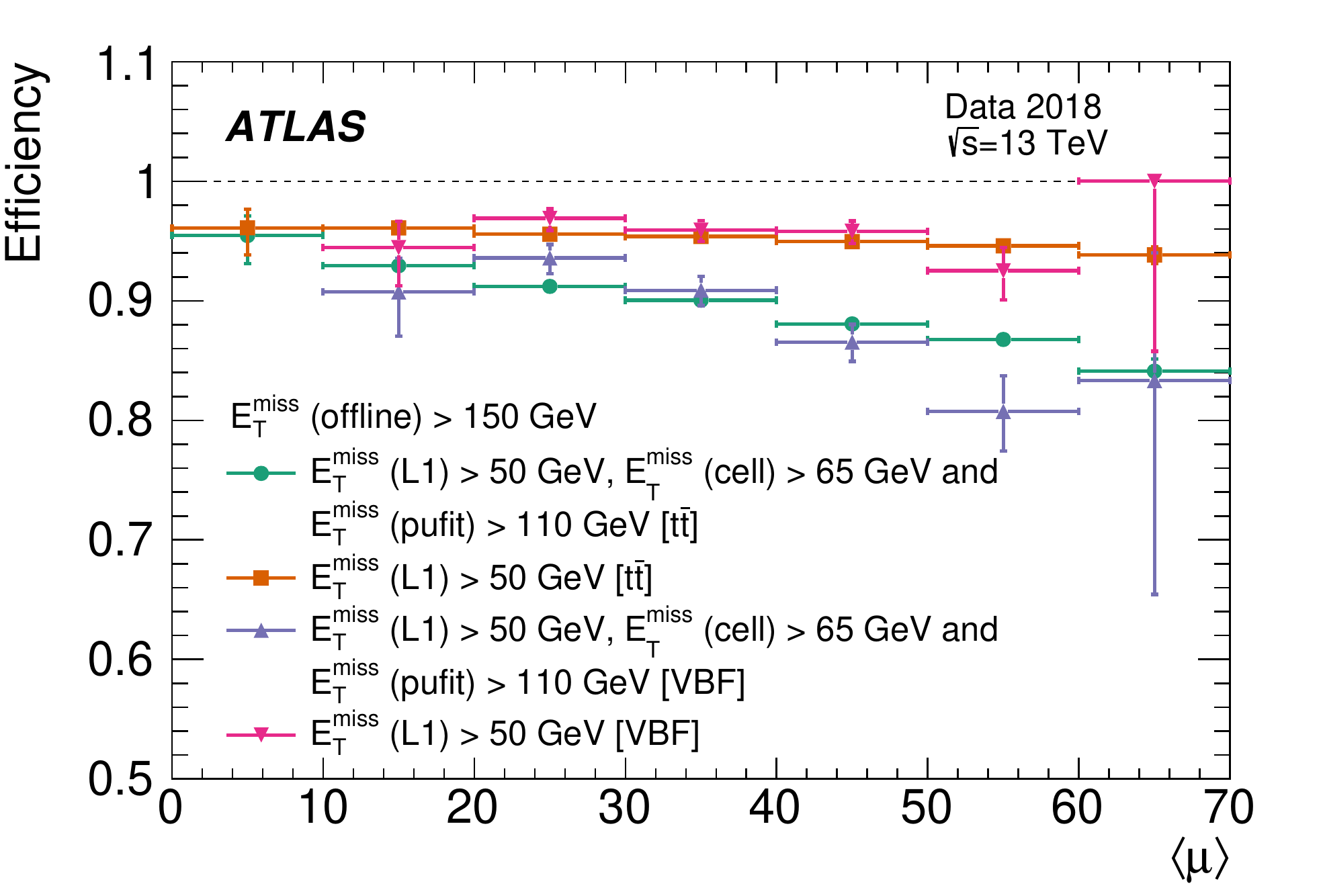}\label{fig:EffEventChar-b}} \\
\subfloat[]{\includegraphics[width=0.49\textwidth]{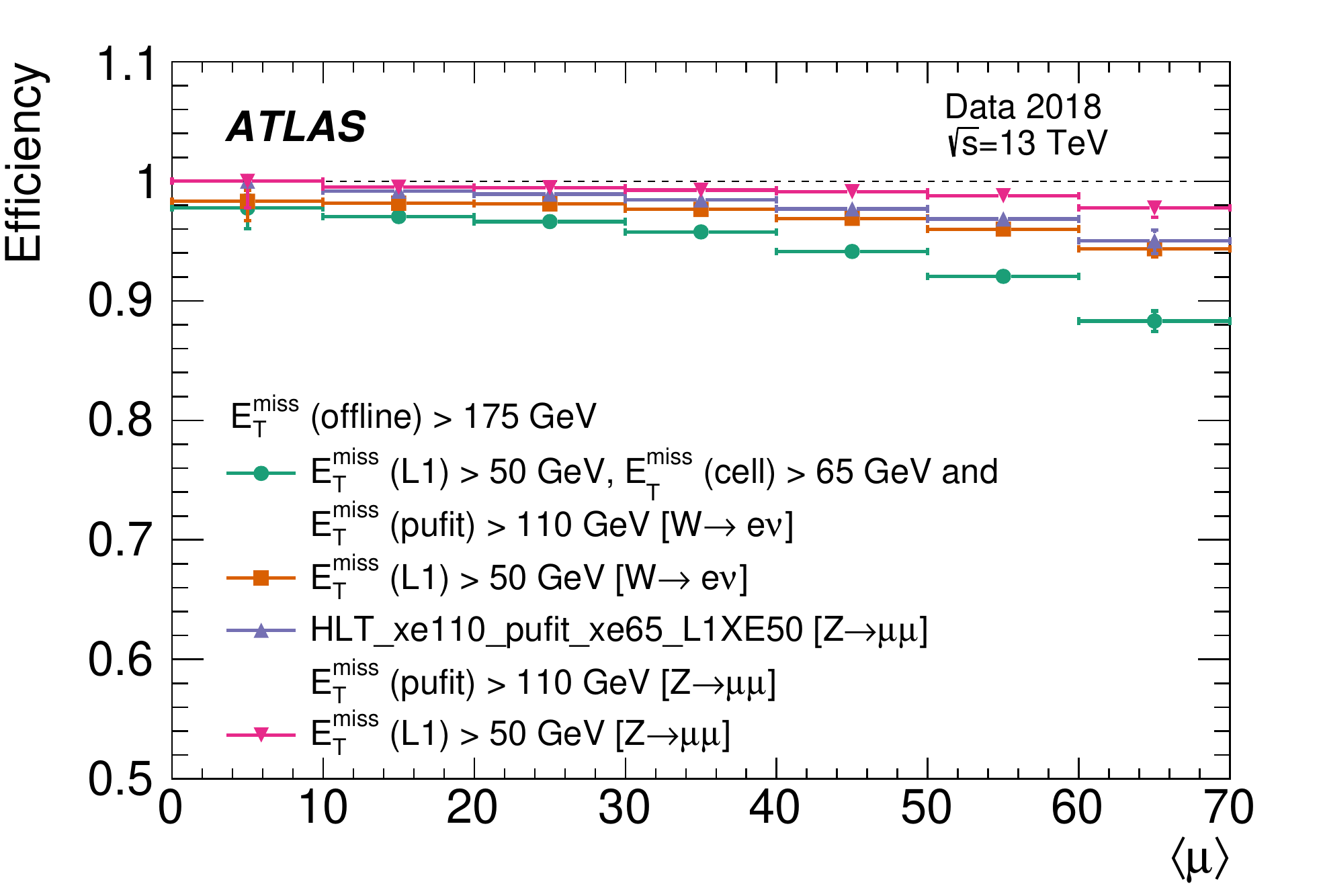}\label{fig:EffEventChar-c}}
\subfloat[]{\includegraphics[width=0.49\textwidth]{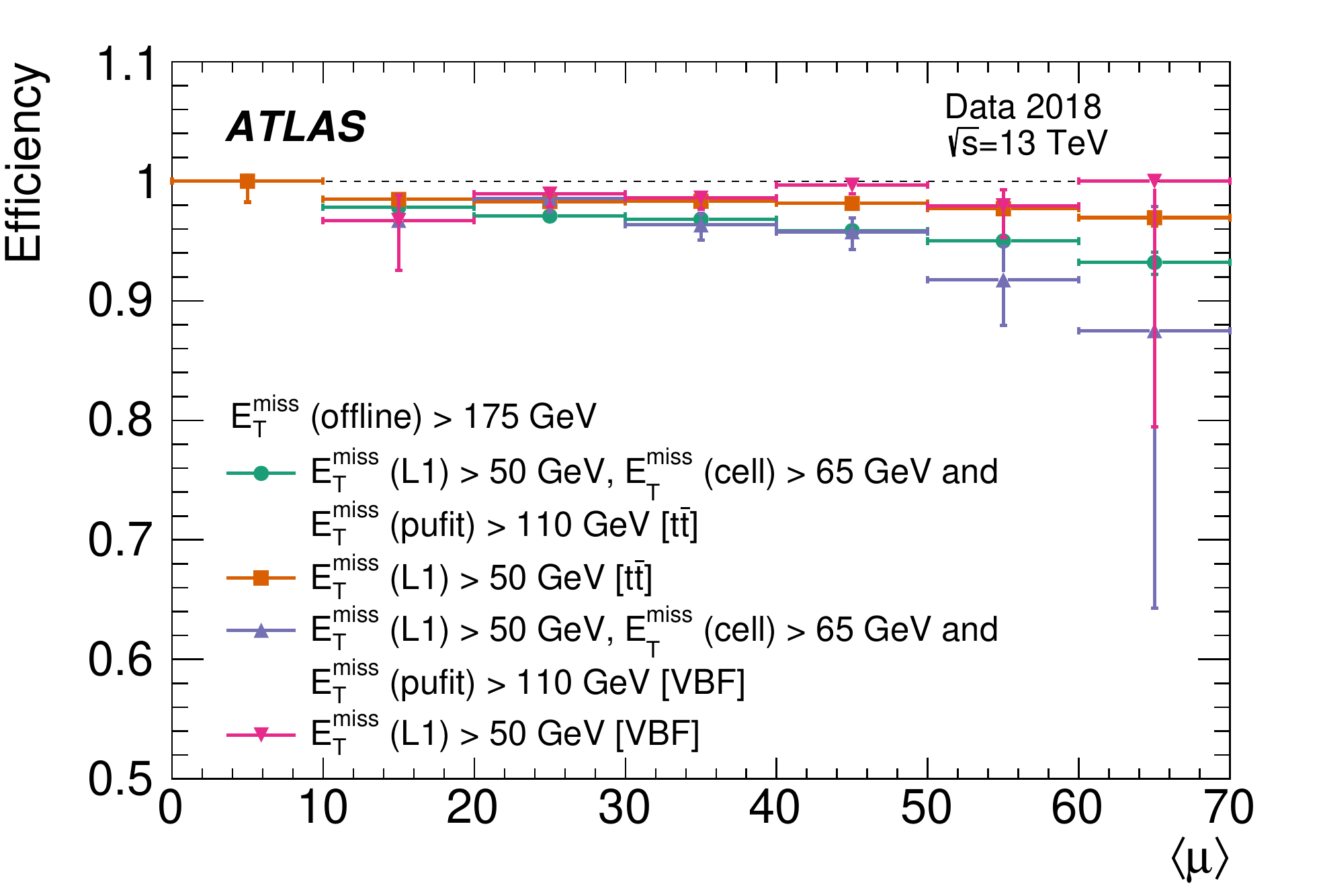}\label{fig:EffEventChar-d}}
\captionsetup{width=\linewidth}
\caption{Efficiencies for the first-level trigger \texttt{L1XE50} and the combined L1+HLT trigger chain \texttt{HLT\_xe110\_pufit\_xe65\_L1XE50} in data recorded in the year 2018 are shown as a function of $\langle\mu\rangle$ for two different offline \met{} thresholds and four different physics selections: \protect\subref{fig:EffEventChar-a} $\Wboson\rightarrow e\nu$ and $\Zmumu$ selections with offline \met $>$ 150\,\GeV~\protect\subref{fig:EffEventChar-b} $\ttbar$ and vector boson fusion selections with offline \met $>$ 150\,\GeV~\protect\subref{fig:EffEventChar-c} $\Wboson\rightarrow e\nu$ and $\Zmumu$ selections with offline \met $>$ 175\,\GeV~\protect\subref{fig:EffEventChar-d} $\ttbar$ and vector boson fusion selections with offline \met $>$ 175\,\GeV. The uncertainties are statistical.
}
\label{fig:EffEventChar}
\end{figure}
 
Figure~\ref{fig:EffEventChar} shows the stability of the efficiencies with
respect to pile-up after
requiring that the offline \met{} be larger than
150\,(175)\,\GeV{} for these four different physics selections.
In general, the efficiency for $\langle\mu\rangle > 50$
tends to be approximately 10--20\% lower for events containing an electron
rather than a muon.
A difference in behaviour is not unexpected given that electrons
are included in the calculation of the visible momentum, whereas muons are not.
The offline \met calculation uses offline electrons that have better resolution
compared to the trigger algorithm.
It can also be seen that events containing forward jets in the VBF selections, or containing jets from top quark decays (right)
have a somewhat different behaviour than do those events selected without jet requirements (left).
Such variations are also to be expected, since the \met{} resolution and scale change if any jets are present
and depend on their energies and the region(s)
of the calorimeter in which they are found.
 
\subsection{Comparison with Monte Carlo simulation}
It is important for most physics analyses to quantify the efficiency with which the trigger selects the events of interest.
Depending on the details of the analysis, the \met trigger efficiency may be determined from data alone, from a Monte Carlo simulation, or from a combination of the two.
A concern with using Monte Carlo simulation to derive the \met trigger efficiency is
the effect of any residual difference between data and simulation.
For example, as described in Section~\ref{sec:L1Alg},
the noise thresholds of the \lonealg{} \met{} trigger algorithm are adjusted periodically during data taking, but it is generally impractical to include
such changes in the simulation.
Because the \met{} triggers use information from the full calorimeter,
the efficiency determined using \met\ triggers is more sensitive to changes in noise thresholds
than the efficiency of other triggers.
 
Many ATLAS physics analyses render residual \met{} trigger inefficiencies
largely immaterial by requiring that the offline $\met$ be larger than $200\,\GeV$. This requirement means that selected events are in
a region in which the trigger efficiency is greater than 99\% and therefore inefficiencies are negligible.
However, some analyses, particularly those in which the
number of events falls rapidly with increasing
\met{} (such as those in Refs.~\cite{HIGG-2018-04,EWKSusy,EXOT-2016-37})
motivate the use of \met thresholds below the trigger plateau in order to maintain high signal efficiency.
For these and similar cases the \met trigger efficiency needs to be determined, often by using Monte Carlo simulations.
Any differences between the simulation and the data may therefore
lead to an incorrect
calculation of the efficiency if the simulation alone were to be relied upon.
 
To account for residual differences between data and simulation,
corrections referred to as scale factors are determined by measuring the ratio of the trigger efficiency using data to that expected from simulation.
These are subsequently applied to correct the signal and background simulation.
In the case of the \met{} trigger the values of the scale factors vary with properties that include, e.g.,
the value of the trigger threshold, the cell noise thresholds,
the definition of offline \met\ and the details of the offline selection.
Given that the \met\ trigger is used for a large range of offline selections with widely varying final states,
no single scale factor suitable for all cases can be found.
Instead, analysis-specific corrections must be employed.
 
\begin{figure}[htbp]
\centering
\captionsetup{width=0.45\linewidth}
\subfloat[]{\includegraphics[width=0.49\textwidth]{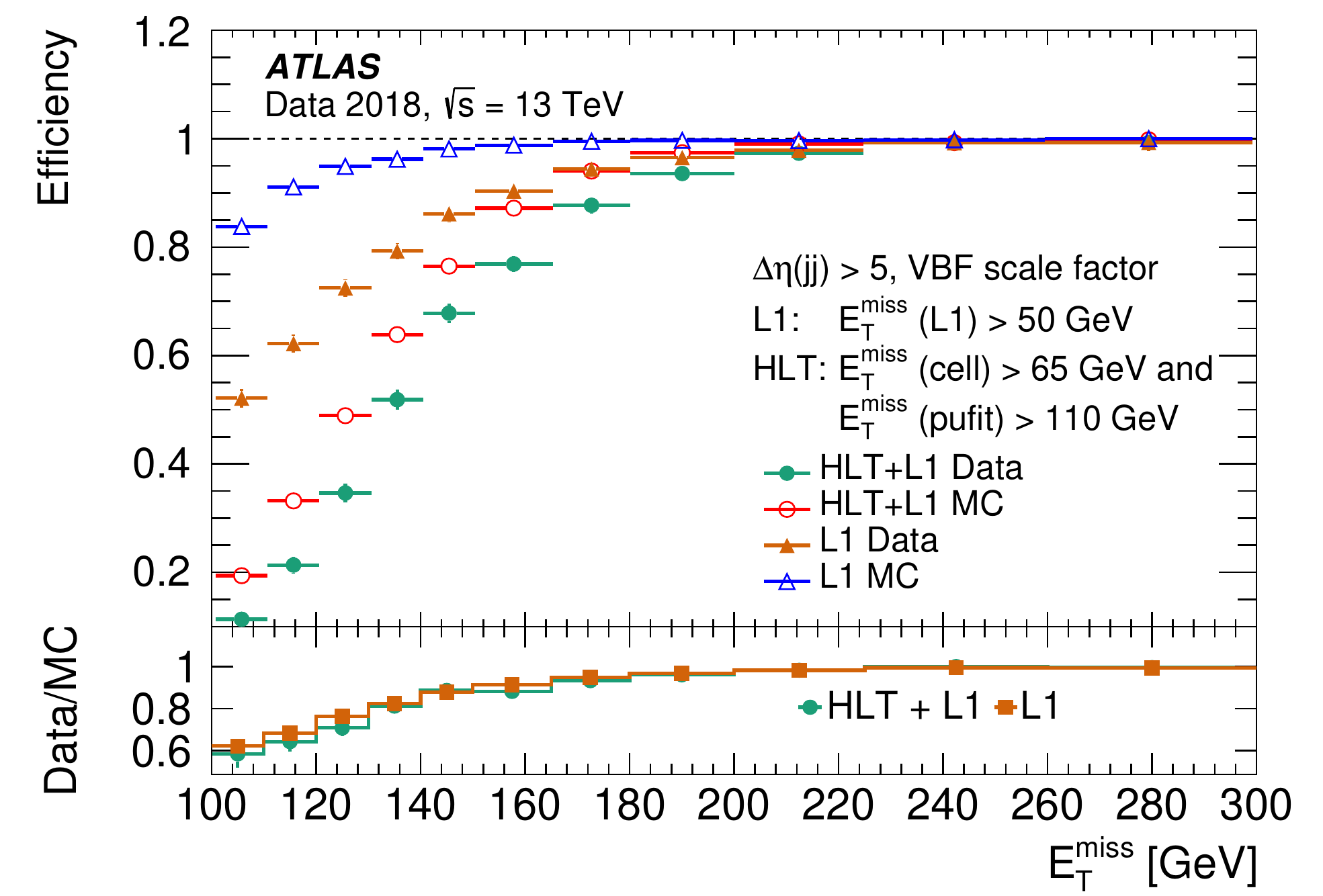}\label{fig:ScaleFactors:turnon}}
\subfloat[]{\includegraphics[width=0.49\textwidth]{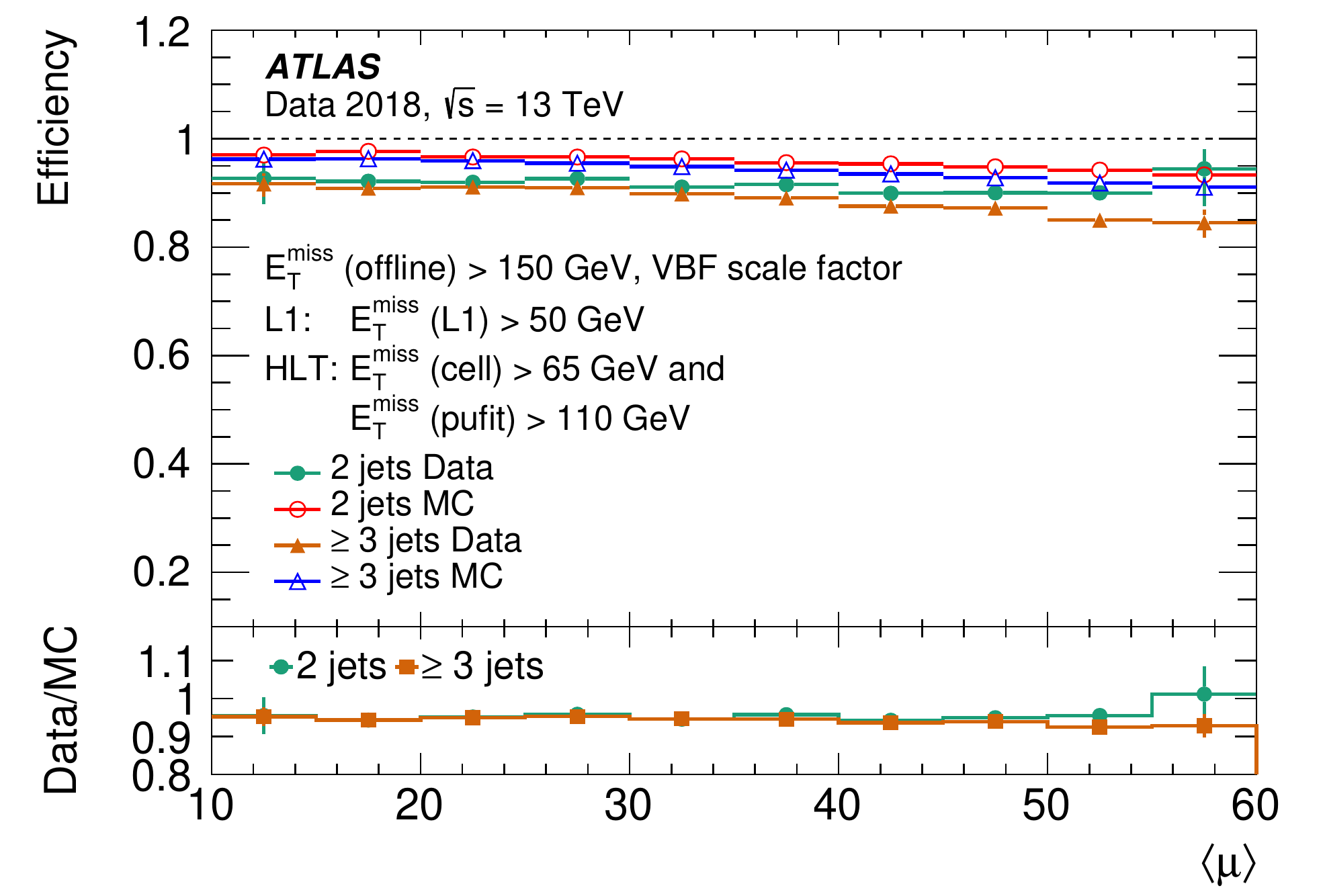}\label{fig:ScaleFactors:mu}}
\captionsetup{width=\linewidth}
\caption{Efficiencies from data and from simulation for L1 (\texttt{L1XE50}) and the combined L1+HLT chain (\texttt{HLT\_xe110\_putfit\_xe65\_L1XE50}) triggers are shown for a VBF selection which requires at least two jets that are well separated in rapidity. The uncertainties are statistical only.
The lower panels show the ratios between efficiencies in data and simulation. \protect\subref{fig:ScaleFactors:turnon} Efficiencies as a function of the offline \met{}. In this plot the jet rapidity difference requirement has been tightened to $\Delta \eta(\text{jj}) > 5$ \protect\subref{fig:ScaleFactors:mu} efficiencies as a function of $\langle\mu\rangle$ for an offline requirement of $\met > 150\,\GeV$ using the `tenacious' working point; the efficiencies in data and MC simulation are compared for two different selections, one requiring exactly two and the other $\geq$3 jets.
}
\end{figure}
 
Comparisons between the trigger efficiency as predicted by Monte Carlo simulation
and as measured in data were performed for both the L1 and the
combined L1+HLT trigger chain, using
data recorded during 2018.
The trigger employed for those data is the combined \pufitalg{}+\cellalg{} algorithm with
thresholds as indicated in Table~\ref{table:triggermenu}.
The variant of the offline \met{} used is that referred to as `tenacious',
and is described in Section~\ref{sec:offlineobjects}.
The efficiency is measured using events containing a single muon,
which, as elsewhere, is treated as being invisible in the \met{} calculation.
The selection requirements are otherwise similar to the row labelled 'VBF' in Table~\ref{table:selections}, except that:
the requirements on the VBF jets are changed such that $|\Delta\phi(\text{jj})| < 2.0$ and \mjj $>$ 200\,\GeV{}. In addition, for Figure~\ref{fig:ScaleFactors:turnon}, exactly two jets are required and $|\Delta \eta(\text{jj})| > 5.0$, while Figure~\ref{fig:ScaleFactors:mu} is binned in jet multiplicity and $|\Delta \eta(\text{jj})| > 3.5$.
This selection is similar to that used in the VBF Higgs-to-invisible analysis~\cite{EXOT-2016-37}.
 
Monte Carlo simulated samples of the \Znunu, \Wlnu{} and \Zll{} processes
were generated at next-to-leading order (NLO) in strong coupling constant $\alpha_{\mathrm s}$ using \SHERPAV{2.2.1}~\cite{Gleisberg:2008ta}.
These calculations use the Comix~\cite{Gleisberg:2008fv} and OpenLoops~\cite{Cascioli:2011va} matrix element generators, and merging was
done with the \SHERPA{} parton shower~\cite{Schumann:2007mg} using the ME+PS@NLO prescription~\cite{Hoeche:2012yf}.
The NNPDF3.0 parton distribution function (PDF) set~\cite{Ball:2014uwa} computed at next-to-next-to-leading order in $\alpha_{\mathrm s}$ was used, along with dedicated parton shower tuning parameters developed for \SHERPAV{2.2.1}~\cite{ATL-PHYS-PUB-2017-006}.
After the events were generated, the response of the detector~\cite{SOFT-2010-01} is simulated using $\GEANT4$~\cite{Agostinelli:2002hh}. Each
\Wboson{} or \Zboson{} boson event was overlaid with 15--70 pile-up collisions to match the distribution in data. The pile-up events were
simulated using \PYTHIAV{8.1}~\cite{Sjostrand:2007gs} with the MSTW2008 PDF set~\cite{Ball:2014uwa} and A3 set of parameters tuned to data~\cite{ATL-PHYS-PUB-2016-017,Skands:2014pea}.
The efficiency as a function of offline \met is shown
in Figure~\ref{fig:ScaleFactors:turnon}.
In the trigger turn-on region at lower values of offline \met{}, a difference can be observed
between data and simulation, an effect largely attributable to the L1 trigger.
The lower panel shows the ratio of data-determined to simulated efficiency as a function of the offline \met.
This ratio is an example of the scale factor that can be applied to simulated Monte Carlo events to correct
their \met{} trigger efficiency.
Figure~\ref{fig:ScaleFactors:mu} shows the efficiency with respect to $\langle\mu\rangle$ for a selection that requires offline $\met > 150\,\GeV$ and either exactly 2 or $\geq$3 jets.
It can be seen that the Monte Carlo simulations overestimate the efficiency by a few percent.
Such considerations show the need both to correct for the differences between data and simulation when working in the turn-on region, and to understand the behaviour of the resulting scale factors for appropriate selections.
 
\FloatBarrier
 
\section{Conclusion}
\label{sec:conclusion}
Despite the considerable increase in luminosity during Run~2 of the LHC (2015--2018),
it was possible to maintain the excellent performance of the ATLAS \met{} trigger.
This was achieved through a dedicated programme of developing, testing,
evaluating and optimizing various pile-up mitigation algorithms.
 
For triggers without any pile-up correction, a steep increase in trigger rate
with pile-up is observed.
This behaviour is consistent with expectations from a two-component
background \met{} distribution model.
 
Several \met trigger algorithms were introduced in ATLAS during LHC Run~2.
Both the first-level and high-level trigger algorithms were improved to maintain a similar level of efficiency throughout the
data-taking period.
These included a new high-level trigger algorithm which uses a fit to determine pile-up-induced local energy deposits
in individual events to reduce the impact of increasing luminosity on the \met\ trigger rate.
In addition, it was found that combining algorithms related to different sources of high-\met\ tails could,
with an appropriate choice of thresholds,
help maintain high efficiency while keeping trigger rates under control.
The new algorithms enabled the offline \met{} selection to become efficient
about 25\,\GeV{} earlier than would have previously been possible.
In the trigger turn-on region the new algorithms recovered up to double the number of events
that would have been possible previously.
 
A study of the \met\ trigger performance for different signal samples shows only a small degradation of the efficiency despite the factor of four increase in instantaneous luminosity during the four-year LHC Run-2 period.
The \met\ trigger behaviour agrees in general with predictions from Monte Carlo simulations.
However, the low-\met\ region, where the trigger is not fully efficient, is more difficult to model precisely.
Since different \met\ algorithms respond differently depending on the characteristics of the event,
it is necessary for analyses working in this low-\met{} region to determine the specific
corrections appropriate to their particular event selection.
 
The predictions from the background model extend up to $\mu=200$,
the value anticipated for the HL-LHC \cite{TDR029}. The predicted rates are those that would be obtained if no
changes were made to the algorithm to mitigate pile-up.
The equal-threshold acceptance fractions predicted for this level of pile-up are about two orders of magnitude higher than in Run~2,
making that environment even more challenging for the \met\ trigger.
The upgrades planned for ATLAS from 2021 to 2025 will
enable a factor of ten increase in trigger rate and a first-level trigger that takes advantage of the full calorimeter granularity.
As found in the results presented in this paper, mitigation of pile-up is possible,
by use of specifically-designed algorithms which can achieve lower rates for the same signal efficiency.
Use of tracking information can also mitigate the pile-up effect, as can increasing the cell noise thresholds.
However, from the rate predictions it is clear that further development of the high-level \met\ trigger algorithms will also be required.
 
\clearpage
\appendix
\part*{Appendix}
\addcontentsline{toc}{part}{Appendix}
 
\section{Full definition of the trigger implementing local pile-up suppression}
\label{sec:PufitAlgDetail}
 
The details of the \pufitalg{} algorithm, introduced in Section~\ref{sec:PufitAlg}, are as follows.
 
To estimate the contribution from high-\Et{} energy deposits,
the calorimeter is divided into equal
area `patches'. Using 8 divisions in $\phi$ and 14 in $\eta$
yields a total of 112 patches, each with an area
similar to that of an  $R = 0.4$ jet. The energy for each patch is computed from the
sum of all the clusters that fall within that patch. To prevent
a jet from being split into two low-\ET patches, four sets of patches are constructed
in parallel, with each set shifted by half of a patch size along $\eta$ and/or $\phi$.
From the four candidate sets of patches,
the algorithm selects the set that yields the largest scalar sum of \Et{} from the high-\Et{} patches.
The transverse energy in each
patch is then used to select the hard-scatter and pile-up patches. Finally, a fit
is performed to determine the pile-up contribution to each hard-scatter patch.
 
The fit is constrained by two assumptions:
first that the vector sum of \METvec{} over all pile-up contributions should be zero,
and second that the pile-up should be relatively evenly distributed throughout the detector.
 
\paragraph{Stage 1: selection of hard-scatter patches}
The hard-scatter patch threshold is determined for each event from the trimmed mean
and variance of the distribution of the patch \ET values in the event.
The patch \ET values are sorted in ascending order
and  a trimmed mean is determined by discarding the 5\% of patches with the lowest \ET and the 5\% of patches with the highest \Et.
The trimmed mean is given by:
\begin{equation}\nonumber
\langle \text{\ET} \rangle_{\mathrm{patch}} = \frac{1}{N} \sum_{j=0.05N}^{0.95N} E_{{\textrm T}_j},
\end{equation}
where $N$ is the total number of patches.
The variance of the patch \ET{} values is estimated using the sample variance, calculated as
\begin{equation}\nonumber
V_{\mathrm{patch}} = \frac{1}{N} \left\{\sum_{j=1}^{0.95N} \left( E_{{\textrm T}_j} - \langle E_{\textrm T} \rangle_{\mathrm{patch}} \right)^2
+ \sum_{j=1}^{0.05N} \left( E_{{\textrm T}_j} - \langle E_{\textrm T} \rangle_{\mathrm{patch}} \right)^2 \right\}~,
\end{equation}
where the second term uses the lowest $5\%$ of patches (already included in the first term) to estimate
the contribution to the variance from pile-up in the highest $5\%$. The highest 5\% are excluded because the sample is biased by hard-scatter jets.
The patch threshold is then set to
\begin{equation}\nonumber
E_{\textrm T}^{\mathrm{thres}} = \langle E_{\textrm T} \rangle_{\mathrm{patch}} + n_{\sigma} \sqrt{V_{\mathrm{patch}}}~,
\end{equation}
with $n_{\sigma}$ chosen to be 5, based on rate considerations.
The typical value for $E_{\textrm T}^{\mathrm{thres}}$ after the L1 requirement is $30\,\GeV$.
 
With the threshold determined, patches are categorized as high-\ET or low-\ET if they have transverse
energy above or below $E_{\textrm T}^{\mathrm{thres}}$, respectively.
Any event that has no patch above threshold
has \pufitalg $\met{}$ set to zero and is rejected by the trigger. Approximately 97\% of background events
that pass the \texttt{L1XE50} trigger have at least one hard-scatter patch.
 
\paragraph{Stage 2: performing the fit}
The fit of the pile-up contribution to each of the high-\Et{} patches
is performed by minimizing the $\chi^2$ function
\begin{equation}\nonumber
\chi^2({\mathcal E}_{{\textrm T}_1},\ldots,{\mathcal E}_{{\textrm T}_m}) = \Delta^T V^{-1} \Delta~.
\end{equation}
In this expression  $V$ is the associated covariance matrix, and
the variables ${\mathcal E_{{\textrm T}_k}}$ are the pile-up contributions to the transverse
energies of the $m$ high-\ET patches.
The values of ${\mathcal E_{{\textrm T}_k}}$ are determined by minimizing this $\chi^2$ function.
The $(m+2)$-dimensional vector $\Delta$ is given by
\begin{equation}\nonumber
\Delta =
\left(
\begin{aligned}
\sum_{i=1}^{N_{\mathrm{low}}} E_{x_i} & + \sum_{k=1}^{m} {\mathcal E}_{{\textrm T}_k} \cos\phi_{k} \\
\sum_{i=1}^{N_{\mathrm{low}}} E_{y_i} & + \sum_{k=1}^{m} {\mathcal E}_{{\textrm T}_k} \sin\phi_{k} \\
\left(A_1 / A_{\mathrm{low}} \times \sum_{i=1}^{N_{\mathrm{low}}} E_{{\textrm T}_i}\right) & - {\mathcal E}_{{\textrm T}_1} \\
& \vdots \\
\left(A_m / A_{\mathrm{low}} \times \sum_{i=1}^{N_{\mathrm{low}}} E_{{\textrm T}_i}\right) & - {\mathcal E}_{{\textrm T}_m}
\end{aligned}
\right)
\end{equation}
where the $A_k$ are the areas of each of the $m$ high-\ET patches,
and $A_{\mathrm{low}}$ is the summed area of the low-\ET{} patches. The first two lines
impose the transverse momentum conservation constraints $\sum E_x=0$ and $\sum E_y=0$ on the pile-up
contributions. On each of those two lines, the first sum (for which $N_{\mathrm{low}}=N-m$)
runs over the low-\ET patches, and the
second over the estimated contributions of pile-up in the high-\ET{} patches.
The quantities
$\cos\phi_{k}$ and $\sin\phi_{k}$ are given by $E_{x_k}/E_{{\textrm T}_k}$ and
$E_{y_k}/E_{{\textrm T}_k}$, respectively.
Each of the remaining $m$ lines corresponds to one of the high-\ET{} patches, and penalizes any difference between the to-be-fitted pile-up estimate in that patch ${\mathcal E_{{\textrm T}_k}}$
and the amount of pile-up transverse energy that would be expected in it, based
on the event-wide average transverse energy density
(as calculated from the low-\ET{} patches).
The event-wide transverse energy is given by the term
$A_k / A_{\mathrm{low}} \times \sum_{i=1}^{N_{\mathrm{low}}} E_{{\textrm T}_i}$,
where the ratio  $\left(\sum_{i=1}^{N_{\mathrm{low}}} E_{{\textrm T}_i}\right) / A_{\mathrm{low}}$ is
the average energy density from the low-\Et{} patches.
 
For the determination of the covariance matrix, the fluctuations
of all of the low-\Et{} patches are calculated from the contribution to the calorimeter energies due to detector resolution.
The uncertainty in each measured patch transverse energy $E_{{\textrm T}_i}$ is taken as
\begin{equation}\nonumber
\sigma_i^2 = r_0^2 + r^2 E_{{\textrm T}_i}~,
\end{equation}
where $r$ is the resolution scale, set to $0.5\,\GeV^{1/2}$ and determined from calorimeter energy resolutions.
The $r_0$ term is a resolution floor ($0.05\,\GeV$), introduced to avoid numerical problems for
events where most of the patches have no deposited \ET.
The patch sample variance $V_{\mathrm{patch}}$ calculated
above is used for the lower $m$ diagonal elements.
The full covariance matrix is
\begin{equation}\nonumber
V=
\left( \begin{array}{cccccc}
V_{11} & V_{12} & 0 & 0 & \hdots & 0\\
V_{12} & V_{22} & 0 & 0 & \hdots & 0\\
0 & 0 & s V_{\mathrm{patch}}  & 0 & \hdots & 0\\
0 & 0 & 0 & \ddots & \hdots  & 0 \\
\vdots & \vdots & \vdots & \vdots & \ddots  & \vdots \\
0 & 0 & 0 & 0 & \hdots & s V_{\mathrm{patch}} \\
\end{array} \right)
\end{equation}
where the upper $2\times 2$ block is given by
\begin{equation}\nonumber
V_\mathrm{cov} \equiv
\left( \begin{array}{ll}
V_{11} & V_{12} \\
V_{12} & V_{22} \\
\end{array} \right) =
\left( \begin{array}{ll}
\sum_i^{N_{\mathrm{low}}} \sigma_i^2 \cos^2\phi_i &  \sum_i^{N_{\mathrm{low}}}  \sigma_i^2 \cos\phi_i \sin\phi_i \\
\sum_i^{N_{\mathrm{low}}} \sigma_i^2 \cos\phi_i \sin\phi_i &  \sum_i^{N_{\mathrm{low}}} \sigma_i^2 \sin^2\phi_i  \\
\end{array} \right )~,
\end{equation}
and $s=1$ is determined by optimizing the fit.
 
The minimization
of the $\chi^2$ function with respect to the ${\mathcal E}_{{\textrm T}_k}$
is performed analytically,
avoiding computationally expensive numerical minimization.
The solution involves solving an $m\times m$ linear system, where $m$, the number of above-threshold patches, is typically 4 or 5.
 
Minimizing the function gives
\begin{equation}\nonumber
0=  \frac{1}{2} \frac{\partial \chi^2}{\partial {\mathcal{E}_\textrm{T}}_k} =
(\cos\phi_k,\,\sin\phi_k)V^{-1}_\mathrm{cov}
\begin{pmatrix}\sum_{i=1}^{N_\mathrm{low}}E_{x_i} + \sum_{j=1}^m \mathcal{E}_{{\rm T}_j}\cos\phi_j\\
\sum_{i=1}^{N_\mathrm{low}}E_{{y}_i} + \sum_{j=1}^m \mathcal{E}_{{\rm T}_j}\sin\phi_j\\\end{pmatrix}
- \frac{1}{sV_\mathrm{patch}} \left(\left(\frac{A_k}{A_\mathrm{low}}\sum_{i=1}^{N_\mathrm{low}} {E_{\rm T}}_i \right)- {\mathcal{E}_{\rm T}}_k\right) \, .
\end{equation}
 
The solution to this equation is given by the matrix equation
\begin{equation}\nonumber
\mathcal{E}_{{\rm T}_k} = \left[X^{-1}\right]_{ki}c_i~,\end{equation}
where the $m \times m$ matrix
\begin{equation}\nonumber
X_{ik} = (\cos\phi_i,\,\sin\phi_i)V^{-1}_\mathrm{cov}\begin{pmatrix}\cos\phi_k\\\sin\phi_k\end{pmatrix} + \frac{\delta_{ik}}{sV_\mathrm{patch}}
\end{equation}
and
\begin{equation}\nonumber
c_i = \frac{A_i}{A_\mathrm{low}}\frac{\sum_{j=1}^{N_\mathrm{low}}E_{{\rm T}_j}} {sV_\mathrm{patch}} -
(\cos\phi_i,\,\sin\phi_i)V^{-1}_\mathrm{cov}\begin{pmatrix}\sum_{j=1}^{N_\mathrm{low}}E_{x_j}\\\sum_{j=1}^{N_\mathrm{low}}E_{y_j}\end{pmatrix}\,.
\end{equation}
 
The \MET\ for the event is then
calculated from the $m$ high-\ET patches and the fit results for the pile-up contribution to
the energy in these patches, where now
\begin{equation}\nonumber
\begin{aligned}
\MEX & = - \sum_{k=1}^m \left( E_{x_k} - {\mathcal E}_{{\textrm T}_k} \cos\phi_k \right) \ , \\
\MEY & = - \sum_{k=1}^m \left( E_{y_k} - {\mathcal E}_{{\textrm T}_k} \sin\phi_k \right) \ ,
\end{aligned}
\end{equation}
and finally $\met{}=\sqrt{(\MEX)^2+(\MEY)^2}$.
\section{Details of the offline reconstruction algorithms}
\label{sec:selection_detail}
In the offline analyses, tracks with $\pt > 400\,\MeV$ are reconstructed to identify a common vertex~\cite{PERF-2015-01} and
an event is required to have at least one such vertex.
If more than one vertex is found in the event, the primary vertex is defined
as the one with the largest $\sum_{t} p_{\text{T},t}^{2}$,
where the sum runs over all tracks associated with the vertex.
 
Electrons are reconstructed by matching tracks that originate from the primary vertex
with clusters of energy deposits in the calorimeter~\cite{EGAM-2018-01}. They are required to
have $\pt > 25\,\GeV$ and $|\eta| < 2.47$. To suppress electrons that originate from the hadronic showers of jets,
the electron is also required to be isolated using information from the calorimeter and tracks within a cone of size $\Delta R = 0.2$ around the electron direction.
The total efficiency of electron reconstruction and isolation is 85\%
for an electron with $\pt \approx 30\,\GeV$.

Muons are reconstructed by matching tracks from the muon spectrometer with inner-detector tracks that are associated with the primary
vertex~\cite{PERF-2015-10}. Muons are required to have $\pt >25\,\GeV$ and $|\eta| < 2.47$, and they
are required to be isolated. For the isolation working point used, the efficiency for a muon with \pt in the range 20--100\,\GeV{} is 96\%.
 
Jets are reconstructed starting with information from the calorimeter.
Cells are grouped into topological clusters~\cite{PERF-2014-07},
described in Section~\ref{sec:TCAlg},
which in turn are used to reconstruct jets using the \antikt{} algorithm
with a radius parameter of $R$ = 0.4. Jets are required to have $\pt >  20\,\GeV$ and
$|\eta|< 4.5$. To suppress jets that come from pile-up, those that have $|\eta| < 2.4$ are
required to exceed the jet vertex tagger (JVT) threshold~\cite{PERF-2014-03}.
The JVT uses tracks and vertices to assign a likelihood for a jet to be associated with the
primary vertex.
Jets that fail the JVT requirement are removed, as they are likely to be due to pile-up.
The efficiency to select a jet from a signal process is 92\%, while jets originating from
pile-up collisions are rejected 99\% of the time in the \pt range 20--50\,\GeV{}.
Each jet with
$|\eta|< 2.5$ is evaluated for the likelihood that it originated from a
$b$-hadron~\cite{PERF-2016-05}.
The efficiency for tagging a jet containing a $b$-hadron is 77\%
for jets originating from $\ttbar$ events~\cite{FTAG-2018-01}.
 
When selecting events based on the number of jets and leptons (electrons or muons), it is
necessary to resolve the ambiguity for cases where a reconstructed jet and a lepton
both result from the same detector signals.
To remove such ambiguities, a sequential overlap removal procedure is defined.
If the jet is within a cone of size $\Delta R$ = 0.2 around any electron,
the jet is removed. If a jet is found within a conical annulus defined by $0.2 < \Delta R < 0.4$, then the electron is removed.
If there is a muon within $\Delta R = 0.4$ of a jet, and the jet has at least three charged tracks with $\pt > 500\,\MeV$,
then the muon is removed; otherwise the jet is removed.
\section{The \cell{} background distribution model}
\label{sec:bgmodeldetail}
\renewcommand\SumEtShort{\SumEt}
The details of the background \cell{} distribution model discussed in Section~\ref{sec:bgmodel} are as follows.
The model has two components. The first covers the bulk of the distribution, which is assumed to originate from resolution effects.
The second component, which dominates in the tail, comes from rare occurrences which scale linearly with the luminosity,
and hence, for fixed beam parameters, linearly with the instantaneous average number $\mu$ of interactions per bunch crossing.
 
\paragraph{Component 1: \SumEtShort \ resolution-dependent bulk distribution}
The bulk resolution effects are assumed to depend on the total transverse
energy deposited in the calorimeter, $\SumEtShort$.
It is assumed that the two perpendicular components of \met
are independently identically Gaussian distributed with a width
$\sigma(\SumEtShort)$ that depends
on \SumEtShort,
such that events with a given \SumEtShort\ will obey a Rayleigh distribution
in \met,
\[
R\left(\met | \sigma(\SumEtShort)\right) = (\met/\sigma^2)\exp(-[\met]^2/2\sigma^2).
\]
In turn the probability density function of \SumEtShort\ is assumed to be a function
$P(\SumEtShort| \mu, \vec{\gamma})$, where $\mu$ is the instantaneous average number
of interactions per bunch crossing and
$\vec{\gamma}$ is used to parameterize any other dependencies.
For a given $\mu$, \met is distributed according to the probability density function
\begin{equation}\label{eq:metpdf}
P_\textrm{res}(\met|\mu) = \int_0^\infty R(\met|\sigma(\SumEtShort))\,P(\SumEtShort| \mu, \vec{\gamma})~{\rm d}(\SumEtShort).
\end{equation}
 
\begin{figure}[htbp]
\centering
\captionsetup{width=0.45\linewidth}
\subfloat[]{\includegraphics[width=0.49\textwidth]{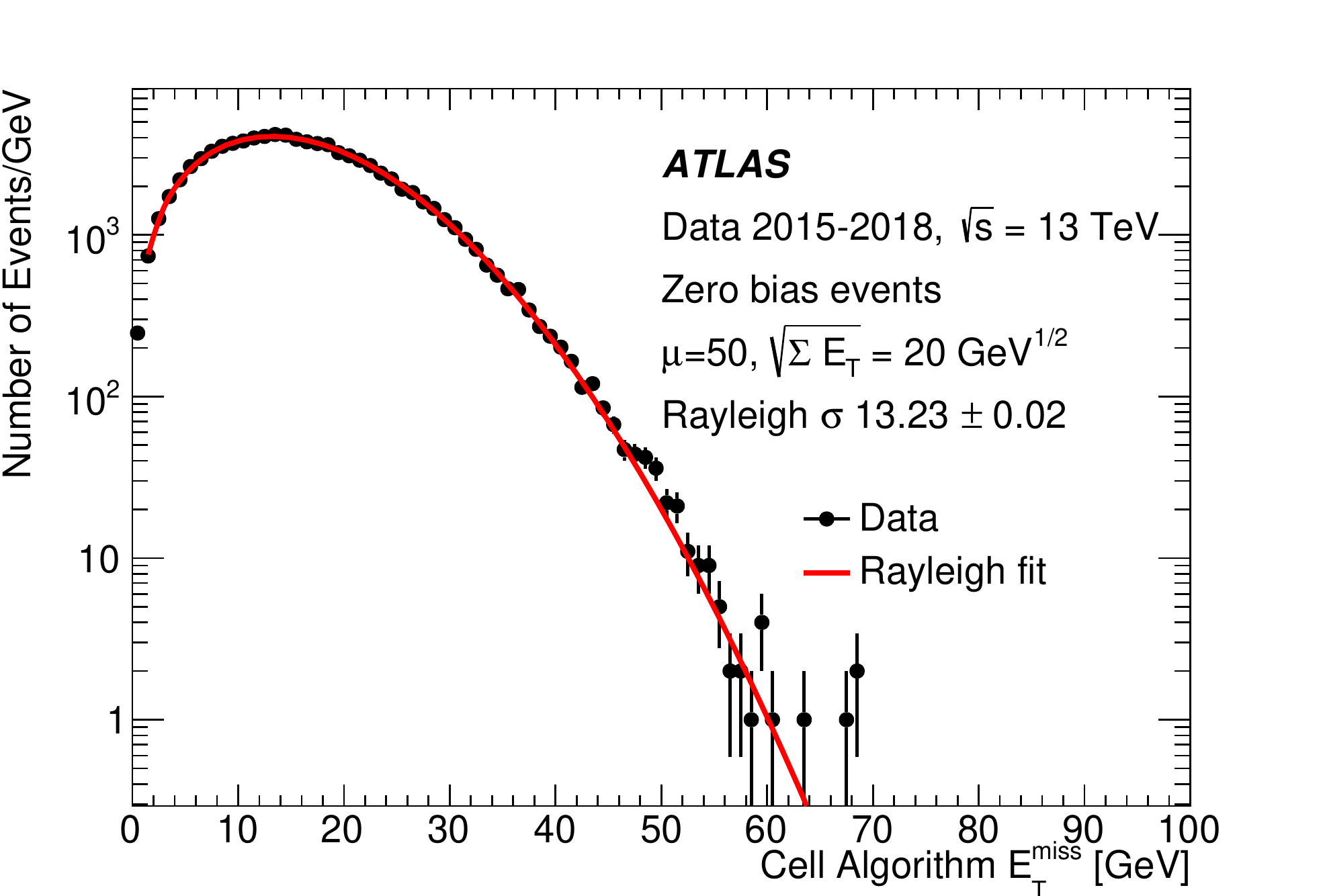}\label{fig:RayleighFit}}
\subfloat[]{\includegraphics[width=0.49\textwidth]{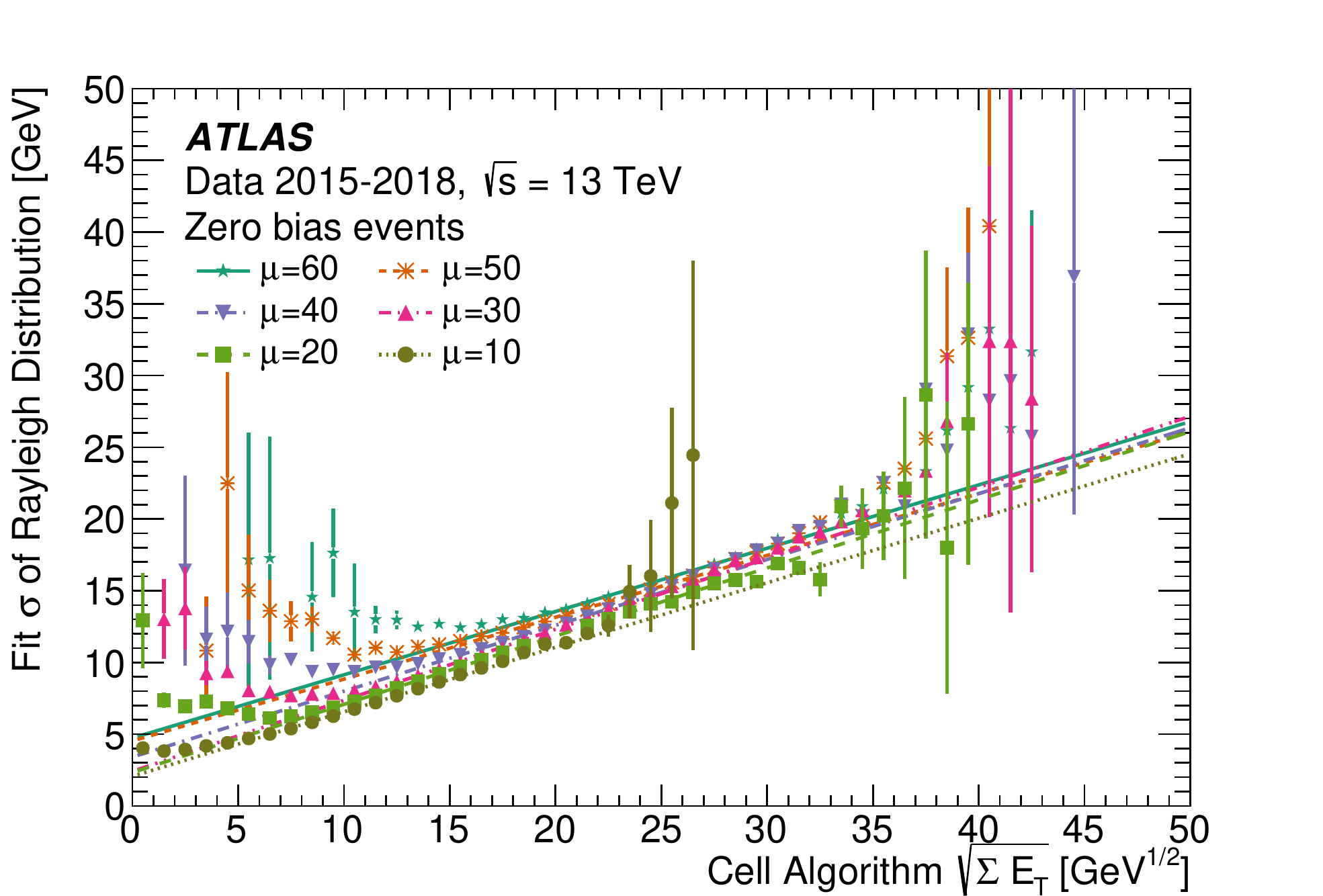}\label{fig:FitSigmas}}
\captionsetup{width=\linewidth}
\caption{Data and fits to data for \met{} model. \protect\subref{fig:RayleighFit} A comparison of the measured cell-algorithm \met\ distribution with
a Rayleigh distribution fit, for a narrow range of pile-up centred on $\langle\mu\rangle\approx 50$. The black histogram is the distribution from the unbiased data. The red curve is the result of a fit to a Rayleigh distribution. \protect\subref{fig:FitSigmas} Values of $\sigma$ are shown from Rayleigh fits to
\met\ distributions for 6 values of $\langle\mu\rangle$ and 50 values of $\sqrt{\SumEt}$ per $\langle\mu\rangle$ value, as well as the data from which the fits are obtained. Also shown are straight-line fits to the 50 $\sqrt{\SumEt}$\ bins for each $\langle\mu\rangle$.
}
\end{figure}
 
In order to sample from the distribution $P_\textrm{res}(\met|\mu)$,
expressions for both $\sigma(\SumEtShort)$ and $P(\SumEtShort| \mu, \vec{\gamma})$ are needed.
 
Fits to find  $\sigma(\SumEtShort)$ are discussed first. Figure~\ref{fig:RayleighFit} shows an example \cell distribution for
a narrow range around $\langle\mu\rangle\approx50$ and $\SumEt \approx 400\,\GeV$.
The plot also shows a Rayleigh fit to the \met{} distribution; the fit characterizes
it well except at the highest \met\ values.
The points in Figure~\ref{fig:FitSigmas} show the values of $\sigma$ obtained from
fits to the \met\ distribution for
6 values of $\langle\mu\rangle$ and 50 bins of $(\SumEtShort)^{1/2}$ for each $\langle\mu\rangle$.
Although some $\langle\mu\rangle$ dependence is visible, the model provides
a reasonable description of the data
using a parameterization of the resolution given by $\sigma(\SumEtShort) = \alpha_1+\alpha_2\sqrt{\SumEtShort}$
with $\alpha_1= 3\,\GeV$ and $\alpha_2= 0.465\,{\GeV}^{1/2}$.
Plots like those in Figure~\ref{fig:FitSigmas} show that this linear fit is a good model for the bulk of events and performs poorly only in ranges of the
total transverse energy where the Rayleigh fits do not provide a good description of the \met{}.
 
To find the $\SumEtShort$ probability density function $P(\SumEtShort| \mu, \vec{\gamma})$,
one may first consider the distribution
$S_1(\SumEtShort| \vec{\gamma})$ of the \SumEtShort\ for a single proton--proton interaction.
The corresponding distribution $S_n(\SumEtShort| \vec{\gamma})$ for $n$ simultaneous proton--proton interactions
is given by the $n$-fold convolution of $S_1(\SumEtShort| \vec{\gamma})$.
The values of $n$ that are sampled for any $\mu$ are assumed to be Poisson distributed with mean $\mu$,
so that the resulting distribution
$P(\SumEtShort | \mu, \vec{\gamma})$ is given by a Poisson-weighted sum of terms, each term being the product
of the Poisson probability for $n$ interactions given $\mu$ with the corresponding  $S_n(\SumEtShort| \vec{\gamma})$:
 
\begin{equation}\nonumber
P(\SumEtShort | \mu, \vec{\gamma}) =\sum_{n=1}^{\infty}\text{Poi}(n|\mu) S_n(\SumEtShort|\vec\gamma)~.
\end{equation}
 
\begin{figure}[htbp]
\centering
\includegraphics[width=0.49\textwidth]{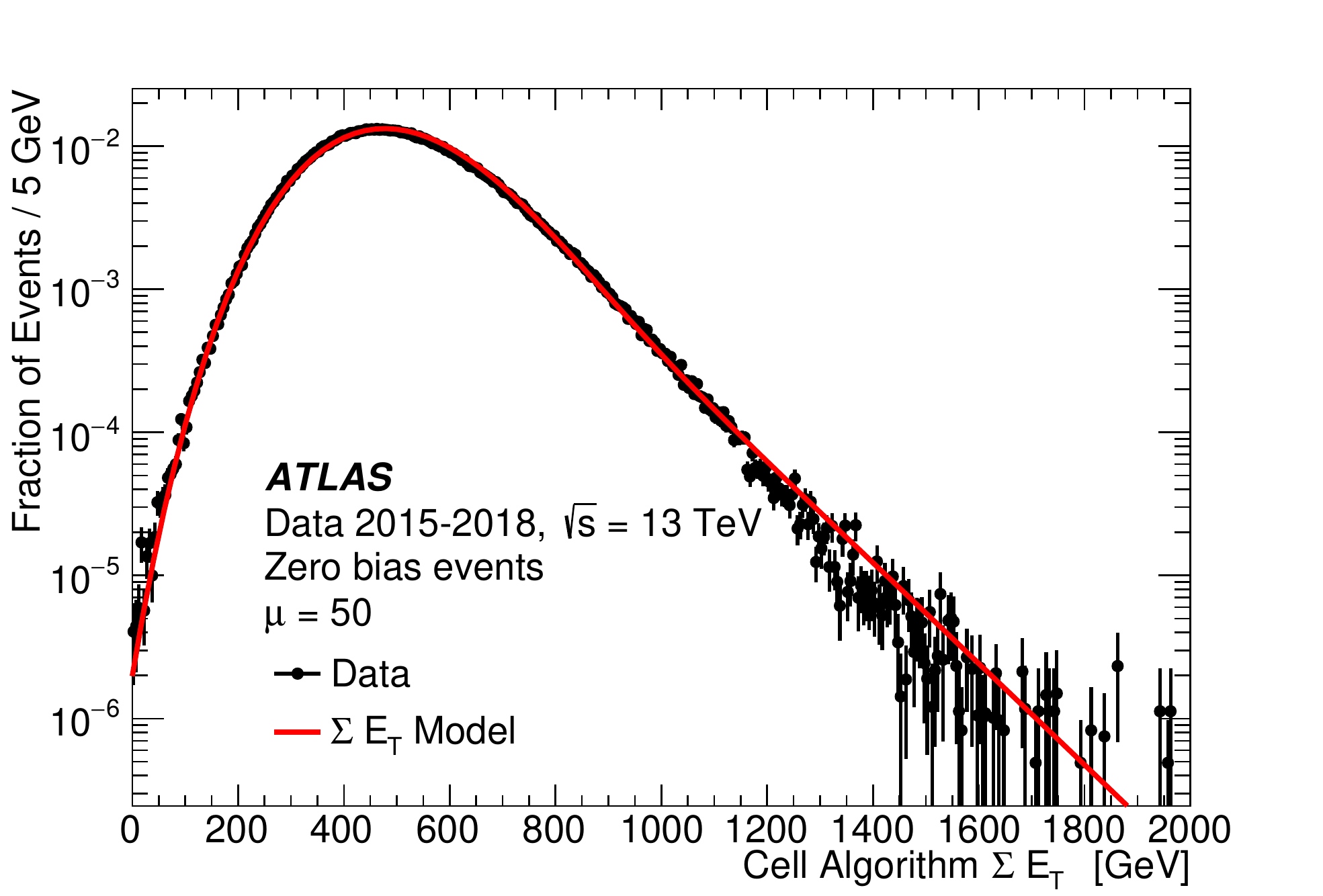}
\caption{
The probability density function $P(\SumEtShort| \mu, \vec{\gamma})$
described in the text is shown
and compared with data recorded using a zero bias trigger.
The model and the data both correspond to pile-up of $\langle\mu\rangle \approx 50$.   }
\label{fig:SumETmodel}
\end{figure}

In the absence of data taken at low $\langle\mu\rangle$ under the same beam conditions,
$S_1(\SumEtShort | \vec{\gamma})$ cannot be determined directly.
Instead, various models of $S_1(\SumEtShort| \vec{\gamma})$ are explored
by varying their parameters until the calculation gives a reasonable match to
the measured high-luminosity $P(\SumEtShort| \mu, \vec{\gamma})$ distributions.
It is found that a single-interaction \SumEt\ distribution of the form
\[
S_1(\SumEtShort| \vec{\gamma}) = \gamma_3\gamma_1\exp(-\gamma_1\Sigma^\prime)+(1-\gamma_3)\gamma_2\exp(-\gamma_2\Sigma^\prime)
\]
gives a result that matches the data well for $\mu$ of about 20 or greater. The parameters of this expression are $\gamma_1 = \SI{0.070}{\GeV^{-1}}$, $\gamma_2 = \SI{0.009}{\GeV^{-1}}$, $\gamma_3 = 0.995$, and for the
$\mu$-dependent substitution $\Sigma^\prime = \SumEtShort-(\gamma_4+\gamma_5\times \mu)$
with $\gamma_4 = \SI{25}{\GeV}$ and  $\gamma_5  =\SI{-5}{\GeV}$.
Figure~\ref{fig:SumETmodel} shows an example of the $\SumEtShort$ modelling for $\langle\mu\rangle\approx50$.
 
The $n$-fold convolution can be carried out analytically only for a limited number of functions.
For example, the single-interaction  \SumEt\ distribution for the LHC Run~1 was approximated by a single exponential function,
and the convolution was then performed as described in Ref.~\cite{mettheory}.
For the more complicated functions used in the present calculation, the convolution and Poisson sum
is computed using a fast Fourier transform
and the method described in Refs.~\cite{HongboHu} and \cite{Cranmer}.
 
The Fourier transform $\widetilde{P}(\omega| \mu, \vec{\gamma})$
of the weighted convolution sum $P(\SumEtShort| \mu, \vec{\gamma})$
is rewritten as the infinite sum of Fourier transforms
over $S_1(\SumEtShort| \vec{\gamma})$, denoted by $\widetilde{S_1}(\omega| \vec{\gamma})$.
This facilitates calculation, as the infinite sum can be performed using the Fourier transform:
\begin{equation}\label{eq:sumetcalc}
\begin{split}
\widetilde{P}(\omega| \mu, \vec{\gamma}) &=\sum_{n=1}^{\infty}\text{Poi}(n|\mu) \widetilde{S_n}(\omega| \vec\gamma) \\
&= \sum_{n=1}^{\infty}\text{Poi}(n|\mu) (2\pi)^{n-1}(\widetilde{S_1}(\omega| \vec\gamma))^n  \\
&= \frac{1}{2\pi}\mathrm{e}^{-\mu}\sum_{n=1}^{\infty} (2\pi)^n \mu^n (\widetilde{S_1}(\omega| \vec\gamma))^n /n! \\
&= \frac{1}{2\pi}\mathrm{e}^{-\mu}\left(\sum_{n=0}^{\infty} [2\pi\mu \widetilde{S_1}(\omega| \vec\gamma)]^n/n! - 1 \right)  \\
&= \frac{1}{2\pi}\mathrm{e}^{-\mu}\left(\exp\left(2\pi\mu \widetilde{S_1}(\omega| \vec\gamma)\right)-1\right).
\end{split}
\end{equation}
 
In principle, one could also take a logarithm of the exponential term in the final expression above to directly
solve for $\widetilde{S_1}(\omega| \vec{\gamma})$
from $\widetilde{P}(\omega| \mu, \vec{\gamma})$.
However, the uncertainties in the $P(\SumEtShort| \mu, \vec{\gamma})$
distribution are amplified in this process, making it impossible in practice to obtain useful results in this way.
 
\paragraph{Component 2: Luminosity-dependent tail from rare events}
The Rayleigh distribution resolution model described above is found to describe the observed \cell{} distribution well for \met{} up to about 50\,\GeV, but not beyond.
A second component, which contributes more at larger \met{}
values, is assumed to come from a combination of events with rare calorimeter measurement fluctuations,
for example very poor measurement of a single jet in a two-jet event, and from events containing non-interacting particles,
such as those in which semileptonic $b$-hadron decays produce neutrinos.
The probability of such an event is expected to scale linearly with luminosity,
and hence for fixed bunch-crossing rate, to scale linearly with $\mu$.
Indeed, a Fr\'echet function that is linearly dependent on $\mu$,
\begin{equation}\label{eq:frechet}
\mu F(\met|m,\alpha,s) = \mu A \frac{\alpha}{s} \left( \frac{\met-m}{s} \right)^{-1-\alpha}
\exp \left[ {-\left(\frac{\met-m}{s}\right)^{-\alpha}} \right]
\end{equation}
is found to describe the high \met
distribution,
with parameters $\alpha=3.54$, $s=8\,\GeV$, and $m=40\,\GeV$, and
with the probability per single $pp$ interaction of such a tail event being $A = 10^{-5}$.
 
\paragraph{Full \met distribution}
The complete model \met distribution is then obtained by vector addition of the two \met sources.
A random azimuthal angle $\phi_\text{rand}$ is assumed between the two vectors and the magnitude of the \met
is sampled from two sources. These are
(i) the \met distribution in Eq.~\eqref{eq:metpdf} from the resolution function
of the total transverse energy in Eq.~\eqref{eq:sumetcalc}, and
(ii) a tail modelled with the Fr\'echet distribution, depending linearly upon $\mu$, in Eq.~\eqref{eq:frechet}.
The combination yields
\begin{equation}\nonumber
\text{Event\ } \met = \big| \met(\text{resolution})~\hat {\bf n}_{\mathrm T}(\phi = 0) + \met(\text{tail})~\hat {\bf n}_\mathrm{T}(\phi = \phi_\text{rand}) \big |.
\end{equation}
It is the distribution of this quantity that is compared with data in Section~\ref{sec:bgmodel}.
 
\clearpage
 
\section*{Acknowledgements}

We thank CERN for the very successful operation of the LHC, as well as the
support staff from our institutions without whom ATLAS could not be
operated efficiently.
 
We acknowledge the support of ANPCyT, Argentina; YerPhI, Armenia; ARC, Australia; BMWFW and FWF, Austria; ANAS, Azerbaijan; SSTC, Belarus; CNPq and FAPESP, Brazil; NSERC, NRC and CFI, Canada; CERN; CONICYT, Chile; CAS, MOST and NSFC, China; COLCIENCIAS, Colombia; MSMT CR, MPO CR and VSC CR, Czech Republic; DNRF and DNSRC, Denmark; IN2P3-CNRS and CEA-DRF/IRFU, France; SRNSFG, Georgia; BMBF, HGF and MPG, Germany; GSRT, Greece; RGC and Hong Kong SAR, China; ISF and Benoziyo Center, Israel; INFN, Italy; MEXT and JSPS, Japan; CNRST, Morocco; NWO, Netherlands; RCN, Norway; MNiSW and NCN, Poland; FCT, Portugal; MNE/IFA, Romania; MES of Russia and NRC KI, Russia Federation; JINR; MESTD, Serbia; MSSR, Slovakia; ARRS and MIZ\v{S}, Slovenia; DST/NRF, South Africa; MINECO, Spain; SRC and Wallenberg Foundation, Sweden; SERI, SNSF and Cantons of Bern and Geneva, Switzerland; MOST, Taiwan; TAEK, Turkey; STFC, United Kingdom; DOE and NSF, United States of America. In addition, individual groups and members have received support from BCKDF, CANARIE, Compute Canada and CRC, Canada; ERC, ERDF, Horizon 2020, Marie Sk{\l}odowska-Curie Actions and COST, European Union; Investissements d'Avenir Labex, Investissements d'Avenir Idex and ANR, France; DFG and AvH Foundation, Germany; Herakleitos, Thales and Aristeia programmes co-financed by EU-ESF and the Greek NSRF, Greece; BSF-NSF and GIF, Israel; CERCA Programme Generalitat de Catalunya and PROMETEO Programme Generalitat Valenciana, Spain; G\"{o}ran Gustafssons Stiftelse, Sweden; The Royal Society and Leverhulme Trust, United Kingdom.
 
The crucial computing support from all WLCG partners is acknowledged gratefully, in particular from CERN, the ATLAS Tier-1 facilities at TRIUMF (Canada), NDGF (Denmark, Norway, Sweden), CC-IN2P3 (France), KIT/GridKA (Germany), INFN-CNAF (Italy), NL-T1 (Netherlands), PIC (Spain), ASGC (Taiwan), RAL (UK) and BNL (USA), the Tier-2 facilities worldwide and large non-WLCG resource providers. Major contributors of computing resources are listed in Ref.~\cite{ATL-SOFT-PUB-2020-001}.
 
\printbibliography

\clearpage \input{atlas_authlist}
\end{document}

%% file: atlas_authlist.tex
 
\begin{flushleft}
\hypersetup{urlcolor=black}
{\Large The ATLAS Collaboration}

\bigskip

\AtlasOrcid[0000-0002-6665-4934]{G.~Aad}$^\textrm{\scriptsize 102}$,    
\AtlasOrcid[0000-0002-5888-2734]{B.~Abbott}$^\textrm{\scriptsize 128}$,    
\AtlasOrcid[0000-0002-7248-3203]{D.C.~Abbott}$^\textrm{\scriptsize 103}$,    
\AtlasOrcid[0000-0002-2788-3822]{A.~Abed~Abud}$^\textrm{\scriptsize 36}$,    
\AtlasOrcid[0000-0002-1002-1652]{K.~Abeling}$^\textrm{\scriptsize 53}$,    
\AtlasOrcid[0000-0002-2987-4006]{D.K.~Abhayasinghe}$^\textrm{\scriptsize 94}$,    
\AtlasOrcid[0000-0002-8496-9294]{S.H.~Abidi}$^\textrm{\scriptsize 167}$,    
\AtlasOrcid[0000-0002-8279-9324]{O.S.~AbouZeid}$^\textrm{\scriptsize 40}$,    
\AtlasOrcid{N.L.~Abraham}$^\textrm{\scriptsize 156}$,    
\AtlasOrcid[0000-0001-5329-6640]{H.~Abramowicz}$^\textrm{\scriptsize 161}$
% ...
\bigskip
\\

% ...
$^{36}$CERN, Geneva; Switzerland.\\
% ...
$^{40}$Niels Bohr Institute, University of Copenhagen, Copenhagen; Denmark.\\
% ...
$^{53}$II. Physikalisches Institut, Georg-August-Universit\"{a}t G\"ottingen, G\"ottingen; Germany.\\
% ...
$^{94}$Department of Physics, Royal Holloway University of London, Egham; United Kingdom.\\
% ...
$^{102}$CPPM, Aix-Marseille Universit\'e, CNRS/IN2P3, Marseille; France.\\
$^{103}$Department of Physics, University of Massachusetts, Amherst MA; United States of America.\\
% ...
$^{128}$Homer L. Dodge Department of Physics and Astronomy, University of Oklahoma, Norman OK; United States of America.\\
% ...
$^{156}$Department of Physics and Astronomy, University of Sussex, Brighton; United Kingdom.\\
% ...
$^{161}$Raymond and Beverly Sackler School of Physics and Astronomy, Tel Aviv University, Tel Aviv; Israel.\\
% ...
$^{167}$Department of Physics, University of Toronto, Toronto ON; Canada.\\
% ...

\end{flushleft}
